%% file: Ly2015_MACT1.arxiv.v3.tex
\documentclass[numberedappendix]{emulateapj}
\usepackage{lscape,epsfig,amsmath,amssymb,graphics}
\usepackage{morefloats}
\usepackage[colorlinks=true,  citecolor=blue,  pagecolor=blue,
  linkcolor=blue,  menucolor=blue, urlcolor=blue,
  linkbordercolor={0 0 1}, frenchlinks=True, breaklinks]{hyperref}
\usepackage{multirow}
\usepackage[yyyymmdd,hhmmss]{datetime}
\begin{document}
\newcommand{\Sname}{$\mathcal{MACT}$}
\newcommand{\Ha}{H$\alpha$}
\newcommand{\Hb}{H$\beta$}
\newcommand{\Hg}{H$\gamma$}
\newcommand{\Hd}{H$\delta$}
\newcommand{\Hyd}{{\rm H}}
\newcommand{\Hae}{\Hyd\alpha}
\newcommand{\Hbe}{\Hyd\beta}
\newcommand{\Hge}{\Hyd\gamma}
\newcommand{\Hde}{\Hyd\delta}
\newcommand{\Lya}{Ly$\alpha$}
\newcommand{\NII}{[{\rm N}\,\textsc{ii}]}
\newcommand{\OIII}{[{\rm O}\,\textsc{iii}]}
\newcommand{\OII}{[{\rm O}\,\textsc{ii}]}
\newcommand{\SII}{[{\rm S}\,\textsc{ii}]}
\newcommand{\NeIII}{[{\rm Ne}\,\textsc{iii}]}
\newcommand{\OIIIa}{\OIII$\lambda$4363}
\newcommand{\OI}{[{\rm O}\,\textsc{i}]}
\newcommand{\OIl}{\OI$\lambda$6300}

\newcommand{\OIIIA}{[\textsc{O iii}]{\rm-A}}
\newcommand{\zphotf}{z_{\rm phot}}
\newcommand{\zspecf}{z_{\rm spec}}
\newcommand{\zphot}{$\zphotf$}
\newcommand{\zspec}{$\zspecf$}
\newcommand{\Rcf}{R_{\rm C}}
\newcommand{\Rc}{$\Rcf$}

\newcommand{\Pagel}{$R_{23}$}
\newcommand{\Oratio}{$O_{32}$}
\newcommand{\Te}{$T_e$}
\newcommand{\OH}{12\,+\,$\log({\rm O/H})$}
\newcommand{\OHm}{12\,+\,\log({\rm O/H})}
\newcommand{\zsun}{$Z_{\sun}$}
\newcommand{\MB}{$M_B$}
\newcommand{\fifth}{$Z$=0.004}
\newcommand{\solar}{$Z$=0.02}

\newcommand{\rNB}{{\rm NB}}
\newcommand{\nOII}{$\sim$1,300}

\newcommand{\HaHbi}{2.86}

\newcommand{\EBV}{E(B-V)}
\newcommand{\EBVa}{$E(B-V)$}
\newcommand{\cc}{cm$^{-3}$}
\newcommand{\mm}{$\mu$m}
\newcommand{\Msun}{$M_{\sun}$}
\newcommand{\Mstar}{$M_{\star}$}
\newcommand{\iyr}{yr$^{-1}$}

\newcommand{\MZ}{\Mstar--$Z$}
\newcommand{\MZR}{MZR}
\newcommand{\zmin}{0.05}

\newcommand{\za}{0.01}
\newcommand{\zb}{1.62}
\newcommand{\areaa}{870.4}
\newcommand{\areab}{788.7}

\newcommand{\Nem}{9264}
\newcommand{\Nspec}{3243}
\newcommand{\Nspecg}{1911}
\newcommand{\Nspecgz}{1493}
\newcommand{\NMMTspec}{1820}
\newcommand{\NMMTspecg}{845}
\newcommand{\NKeckspec}{1423}
\newcommand{\NKeckspecg}{1313}
\newcommand{\Ndet}{66}
\newcommand{\Nrel}{98}
\newcommand{\Ntot}{164}
\newcommand{\Ndetf}{19}
\newcommand{\NMMTf}{37}
\newcommand{\NKeckf}{33}

\newcommand{\NOIIIMMT}{67}
\newcommand{\NOIIIKeck}{119}

\newcommand{\NHa}{33}
\newcommand{\NXMPG}{16}
\newcommand{\zslope}{-1.66}
\newcommand{\DsSFR}{$\Delta({\rm sSFR})_{\rm MS}$}

\newcommand{\GALEX}{{\it GALEX}}
\newcommand{\FUV}{{\it FUV}}
\newcommand{\NUV}{{\it NUV}}

\newcommand{\TM}{\tablenotemark{M}}
\newcommand{\TA}{\tablenotemark{a}}
\newcommand{\TB}{\tablenotemark{b}}
\newcommand{\TC}{\tablenotemark{c}}
\newcommand{\TD}{\tablenotemark{d}}
\newcommand{\TE}{\tablenotemark{e}}
\newcommand{\TF}{\tablenotemark{f}}
\newcommand{\TG}{\tablenotemark{g}}

\newcommand{\pa}{\phantom{1}}

\newcommand{\SFRA}{--0.19$\pm$0.79}
\newcommand{\SFRM}{--0.18}
\newcommand{\sSFRA}{10$^{-8.3}$}
\newcommand{\sSFRt}{210 Myr}
\newcommand{\MassA}{$1.5\times10^8$}
\newcommand{\MassM}{$1.3\times10^8$}
\newcommand{\AgeA}{8.0}

\newcommand{\SAGN}{2}
\newcommand{\FMS}{2}
\newcommand{\bias}{4.1}

\defcitealias{ly07}{Ly07}
\defcitealias{ly14}{Ly14}
\defcitealias{and13}{AM13}
\defcitealias{cha03}{Chabrier}
\defcitealias{MACTII}{Paper II}

\title{THE METAL ABUNDANCES ACROSS COSMIC TIME ($\mathcal{MACT}$) SURVEY. I. OPTICAL
  SPECTROSCOPY IN THE SUBARU DEEP FIELD}

\author{Chun Ly,\altaffilmark{1} Sangeeta Malhotra,\altaffilmark{2}
  Matthew A. Malkan,\altaffilmark{3} Jane R. Rigby,\altaffilmark{1}
  Nobunari Kashikawa,\altaffilmark{4,5} Mithi A. de los Reyes,\altaffilmark{6}
  and James E. Rhoads\altaffilmark{2}}

\submitted{Received 2016 February 1; revised 2016 June 1; accepted 2016 June 7;
  published 2016 September 1}
\shorttitle{The Metal Abundances across Cosmic Time Survey}
\shortauthors{Ly et al.}
\email{astro.chun@gmail.com}

\altaffiltext{1}{Observational Cosmology Laboratory, NASA Goddard Space Flight Center,
  8800 Greenbelt Road, Greenbelt, MD 20771, USA}
\altaffiltext{2}{School of Earth and Space Exploration, Arizona State University,
  Tempe, AZ 85287, USA}
\altaffiltext{3}{Department of Physics and Astronomy, UCLA, Los Angeles, CA 90095-1547, USA}
\altaffiltext{4}{Optical and Infrared Astronomy Division, National Astronomical
  Observatory, Mitaka, Tokyo, Japan}
\altaffiltext{5}{Department of Astronomy, School of Science, Graduate University
  for Advanced Studies, Mitaka, Tokyo, Japan}
\altaffiltext{6}{Department of Physics, North Carolina State University, Raleigh, NC, USA}

\begin{abstract}
  Deep rest-frame optical spectroscopy is critical for characterizing and understanding
  the physical conditions and properties of the ionized gas in galaxies. Here, we
  present a new spectroscopic survey called ``Metal Abundances across Cosmic Time''
  or \Sname, which will obtain rest-frame optical spectra for $\sim$3000 emission-line
  galaxies. This paper describes the optical spectroscopy that has been conducted
  with MMT/Hectospec and Keck/DEIMOS for $\approx$1900 $z=0.1$--1 emission-line galaxies
  selected from our narrowband and intermediate-band imaging in the Subaru Deep Field.
  In addition, we present a sample of \Ntot\ galaxies for which we have measured the weak
  \OIIIa\ line (\Ndet\ with at least 3$\sigma$ detections and \Nrel\ with significant
  upper limits). This nebular emission line determines the gas-phase metallicity by
  measuring the electron temperature of the ionized gas. This paper presents the optical
  spectra, emission-line measurements, interstellar properties (e.g., metallicity, gas
  density), and stellar properties (e.g., star formation rates, stellar mass).
  \citetalias{MACTII} of the \Sname\ survey (Ly et al.) presents the first results on
  the stellar mass--gas metallicity relation at $z\lesssim1$ using the sample with
  \OIIIa\ measurements.
\end{abstract}


\section{INTRODUCTION}
\label{1}

The ionized gas of the interstellar medium (ISM) is a key component in galaxies and
is directly affected by (1) photo-ionizing radiation from young massive stars, (2)
heating from lower mass stars, (3) accretion of pristine and enriched gas on
$\sim$10--100 kpc scales and from nearby interacting galaxies, and (4) gas outflows from
the stellar winds of supernovae. To diagnose and understand the physical conditions of
the ISM, deep spectroscopy is necessary, preferably in the rest-frame optical range
(3000--7000 \AA) where hydrogen Balmer recombination lines and collisionally
excited ionic metal lines are available to characterize the interstellar gas.

For example, the Baldwin--Phillips--Terlevich (``BPT'') diagnostic diagrams
\citep{bal81,vei87} of \OIII$\lambda$5007/\Hb\ as a function of \NII$\lambda$6583/\Ha,
\SII$\lambda\lambda$6716, 6731/\Ha, or \OI$\lambda$6300/\Ha\ are fundamental for
classifying galaxies and understanding whether massive young stars, shocks, or active
galactic nuclei (AGNs) photoionize the gas.

The hydrogen Balmer lines directly trace high-mass star formation, as they are
recombination lines that arise from the ionizing radiation produced by short-lived
($\lesssim$10 Myr) OB stars \citep[e.g.,][]{ken98,newha,lee12}. Also, comparing the
flux ratios of the Balmer lines (e.g., \Ha/\Hb), known as the Balmer decrements
\citep{ost06} with the intrinsic ratios predicted for recombination provides estimates
of the interstellar reddening.

Finally, the gas-phase heavy-element abundance (``metallicity'') of the ISM is
primarily set by enrichment from star formation, but diluted by the accretion of new
pristine gas, and loss within winds that drive gas out of the galaxies
\citep[e.g.,][]{dal07,dave11,lil13}. For this reason, the gas metallicity, when
combined with stellar mass and star formation rate (SFR) determinations, is a
fundamental observable that allows for an understanding of how galaxies process gas.

The Sloan Digital Sky Survey \citep[SDSS;][]{york00} is the largest optical
spectroscopic study of local ($z\lesssim0.2$) galaxies, and has (1) classified galaxies
and AGNs \citep{kau03}; (2) determined SFRs and interstellar dust reddening \citep{bri04};
and (3) examined the dependence of the gas metallicity on stellar properties, such as
the stellar mass--metallicity (\MZ) relation \citep{tre04}.

At $z\gtrsim0.2$, optical spectroscopic surveys such as zCOSMOS \citep{lil09}, DEEP2
\citep{new13}, and AGES \citep{koc12}, have struggled to determine the physical properties
of the interstellar gas in \textit{individual} galaxies over a wide range of redshift
and stellar mass. These limitations are driven mainly by the design of the surveys to
determine spectroscopic redshifts, often with only one or two emission or absorption
lines.

For example, \cite{cre12} used the zCOSMOS catalog of $\sim$10,000 bright
($I_{\rm AB}<22.5$) galaxies to measure metallicity in only 3\% of the zCOSMOS sample
(334 galaxies) at $z\lesssim0.75$. The low success of obtaining metallicity is due to a
combination of the relatively short integration (one hour) and low spectral resolution
($\lambda/\Delta\lambda \sim 600$). For the DEEP2 survey, only 9\% of the DEEP2 sample
have metallicity measurements due to a combination of a limited spectral coverage
(6000--9000 \AA) and a pre-selection using broad-band colors for galaxies at
$z\gtrsim0.7$ \citep[4140 galaxies;][]{ly15}. While DEEP2 has measured the \MZ\ relation
over 1.5 dex in stellar mass \citep{zah11}, the limited spectral coverage and selection
function reduce the metallicity sample to galaxies in a small redshift range of
$z\approx0.75$--0.82.

Another limitation of these surveys is the lack of galaxies below \Mstar\ $\sim10^9$
\Msun. For example, AGES targeted $\sim$3000 galaxies at $0.05\lesssim z\lesssim0.75$.
However, they were limited to galaxies brighter than $I_{AB}\approx20.5$, and thus were
unable to study the \MZ\ relation below $M_{\star}\approx10^{10}$ \Msun\ ($\approx$10$^{11}$
\Msun) at $z\gtrsim0.35$ ($z\gtrsim0.55$) \citep{mou11}. While the magnitude-limited
selections of zCOSMOS and DEEP2 ($R_{AB}\lesssim24$) are fainter than AGES, they limit
$z\sim0.7$ galaxies to $M_{\star} \approx2\times10^9$ \Msun.

To address the lack of deep rest-frame optical spectroscopy for thousands of galaxies
at $z\lesssim1$, specifically those with lower stellar masses (\Mstar\ $\lesssim10^9$
\Msun), we have conducted a spectroscopic survey called ``Metal Abundances across Cosmic
Time'' (\Sname), which targets $\approx$1900 $z\lesssim1$ star-forming galaxies. This
survey has obtained between 2 and 12 hr of on-source integration for each galaxy with
optical spectrographs on Keck and MMT.
The primary goal of the survey is to obtain reliable measurements of the gas-phase
metallicity and other physical properties of the ISM in galaxies, such as the SFR,
gas density, ionization parameter, dust content, and the source of photo-ionizing
radiation (star formation and/or AGNs). The galaxy sample of \Sname\ encompasses
more than 3 dex in stellar mass, including galaxies with stellar masses as low as
$M_{\star}\sim3\times10^6$ \Msun\ and $3\times10^7$ \Msun\ at $z\sim0.1$ and $z\sim1$,
respectively. To effectively select low-mass galaxies over a wide range of redshift,
\Sname\ targeted galaxies in the Subaru Deep Field \citep[SDF;][]{kas04} that have
excess flux in the narrowband and/or intermediate-band filters, which is now
understood to be produced by nebular emission lines from star formation or AGNs
\citep[e.g.,][and references therein]{ly07,newha}.

Another advantage of \Sname\ compared to previous spectroscopic surveys is that it is
the first to use the electron temperature (\Te) method to measure the evolution of the
\MZ\ relation over $\approx$8 billion years. This approach uses the weak \OIIIa\ nebular
emission line, which is governed by the ability of the gas to cool from a higher oxygen
abundance, O/H \citep{all84}. Previous studies have been limited to the use of strong
nebular emission lines for metallicity estimations. Such ``strong-line'' calibrations
have raised concerns about their validity for higher redshift due to growing evidence
that the physical conditions of the ISM in $z\gtrsim1$ galaxies are different from
local galaxies \citep[e.g.,][]{ste14,san15,cow16,dop16}.

This paper (Paper I) describes the optical spectroscopy conducted for the \Sname\
survey, and discusses the sample of \Ndet\ galaxies with detections of \OIIIa\ at
$z=\zmin$--0.95 (average of $z=0.53\pm0.25$; median of 0.48) and robust \OIIIa\ upper
limits for \Nrel\ galaxies at $z=0.04$--0.96 (average of $z=0.52\pm0.23$; median of
0.48). We refer readers to \citet[hereafter Paper II]{MACTII}, which presents the
first scientific results on the evolution of the \MZ\ relation using the \Te\ method.
For wider use of our spectroscopic data set, Paper I also includes the one-dimensional
Keck and MMT spectra (within a tar file) for the sample of \Ntot\ galaxies where we
have measured the weak \OIIIa\ line. We will release the remaining spectra and
ancillary information (e.g., multi-band photometry, spectroscopic information) in
forthcoming \Sname\ publications. The outline of Paper I is as follows.
In Section~\ref{sec:SDF} we describe the imaging survey of the SDF, the selection of over
9000 emission-line galaxies, and the follow-up optical spectroscopy with Keck and MMT.
In Section~\ref{sec:sample} we discuss detections and measurements of nebular emission
lines, which yield a spectroscopic sample with \OIIIa\ detections and reliable
non-detections. Section~\ref{sec:Prop} describes how we determine the: (1) dust
attenuation, (2) electron temperature and gas-phase (O/H) metallicity, (3) dust-corrected
SFR, (4) stellar properties from spectral energy distribution (SED) fitting, and (5) gas
density for the samples of \OIIIa\ detections and non-detections. We summarize the \Sname\
survey and our \OIIIa-detected and \OIIIa-non-detected samples in Section~\ref{sec:End}.

Throughout this paper, we adopt a flat cosmology with $\Omega_{\Lambda}=0.7$, $\Omega_M=0.3$,
and $H_0=70$ km s$^{-1}$ Mpc$^{-1}$. Magnitudes are reported on the AB system \citep{oke74}.
For reference, we adopt \OH$_{\sun}$ = 8.69 \citep{prieto01} as solar metallicity, \zsun.
Unless otherwise indicated, we report 68\% confidence measurement uncertainties, and
``\OIII'' alone refers to the 5007 \AA\ emission line.


\section{THE SUBARU DEEP FIELD}
\label{sec:SDF}


\input tab1.final.tex

The SDF has the most sensitive optical imaging in several narrowband and intermediate-band
filters in the sky, and is further complemented with ultra-deep multi-band imaging
between 1500 \AA\ and 4.5 \mm.
A summary of the ancillary imaging is available in \cite{ly11b} and later in
Section~\ref{sec:SED}.
The emission-line galaxies in this paper are selected from imaging in five narrowband
filters (NB704, NB711, NB816, NB921, and NB973), and two intermediate-band filters
(IA598 and IA679). A summary of the filter properties is in Table~\ref{tab:SDF_sample}
and \citet[hereafter Ly14]{ly14}.

These SDF data were acquired with Suprime-Cam \citep{miy02}, the optical imager mounted
at the prime focus of the Subaru telescope, between 2001 March and 2007 May. The
acquisition and reduction of these data have been discussed extensively in
\cite{kas04,kas06}, \citet[hereafter Ly07]{ly07}, as well as in \cite{OIIpop} for the
narrowband data, and in \cite{nag08} for the intermediate-band data. In brief, data were
obtained mostly under photometric conditions with average seeing of 0\farcs9--1\farcs0
for all five narrowband and two intermediate-band filters. These data were reduced
following standard reduction procedures using \textsc{sdfred} \citep{yagi02,ouc04}, a
software package designed especially for Suprime-Cam data.

The most prominent emission lines entering these filters are \Ha, \OIII, \Hb, and \OII,
at well-defined redshift windows between $z=\za$ and $z=\zb$. This results in probing
64\% of the redshift space and 67\% of the comoving volume at $z\leq1.03$.
Compared to the previous narrowband-selected metallicity study at $z=0.24$--0.85
\citep{hu09}, our survey probes 4.7 (3.8) times more redshift (volume) space, and is
deeper by $\approx$1.5 mag in the narrowband imaging.


\subsection{Selection of Emission-line Galaxies}\label{sec:NBem}
\begin{figure*}
  \epsscale{0.56}
  \plotone{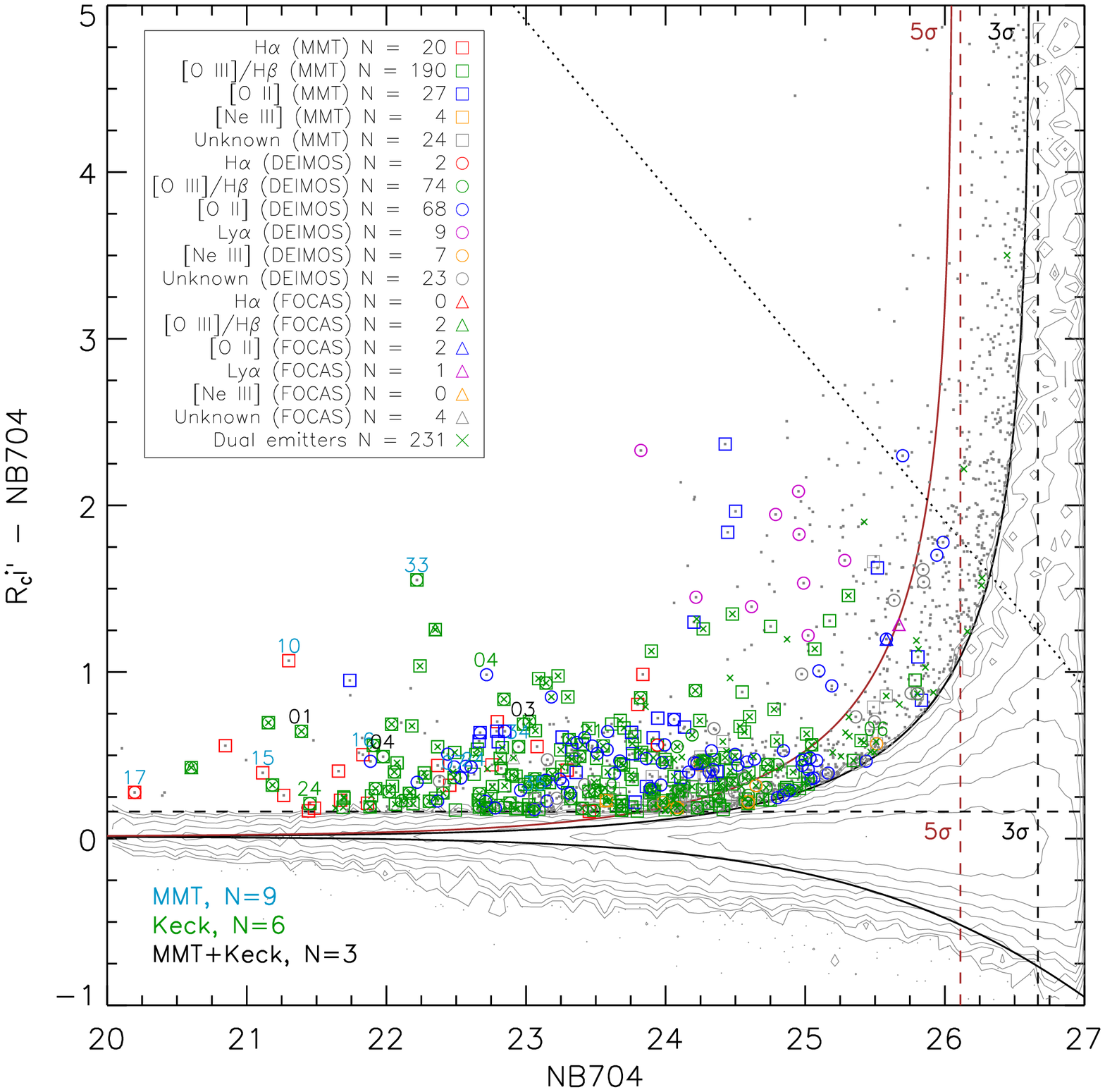}
  \plotone{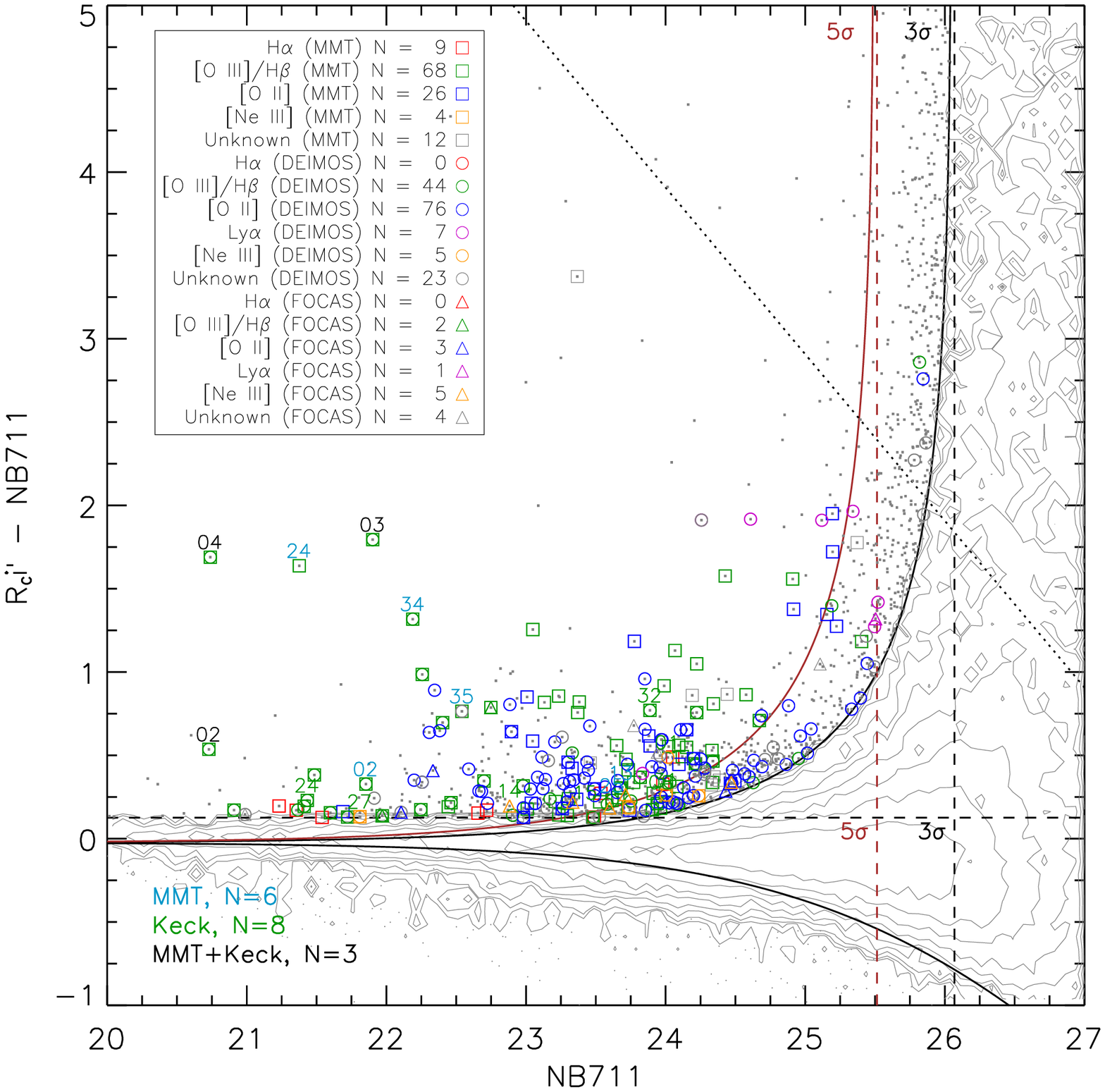}
  \caption{Narrowband excess plots for NB704 (left) and NB711 (right) excess emitters.
    Solid curves show detections that are significant at 5.0$\sigma$ (brown) and
    3.0$\sigma$ (black). Contours of source densities are shown following logarithmic
    intervals. Spectroscopically targeted sources are overlaid with symbols representing
    different instruments: Keck/DEIMOS (circles), MMT/Hectospec (squares), and
    Subaru/FOCAS (triangles). \Ha, \OIII, \OII, Ly$\alpha$, and unidentified sources are
    distinguished by red, green, blue, purple, and gray colors, respectively. Sources
    above the diagonal line are undetected in the \Rc\ and $i$\arcmin-band at 3$\sigma$.
    Vertical lines refer to 3$\sigma$ (black) and 5$\sigma$ (brown) NB704 or NB711
    magnitude limits. Galaxies in the \OIIIa-detected sample are labeled with their
    two-digit identification number in blue (MMT), green (Keck), and black (MMT+Keck).}
  \label{fig:Ex_NB1}
\end{figure*}

\begin{figure}
  \epsscale{1.1}
  \plotone{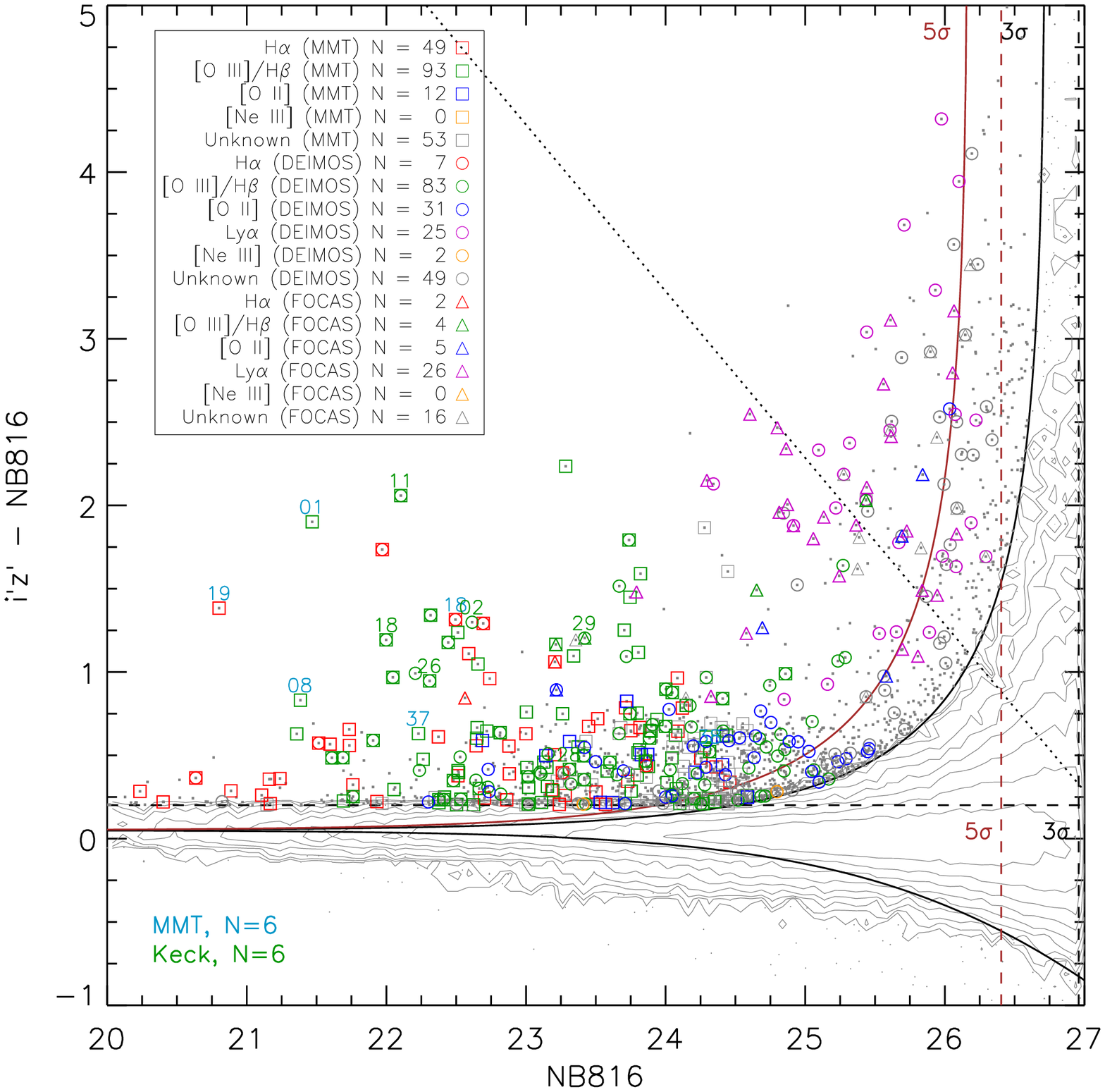}
  \caption{Same as Figure~\ref{fig:Ex_NB1} but for NB816 excess emitters.}
  \label{fig:Ex_NB2}
\end{figure}

\begin{figure*}
  \epsscale{0.56}
  \plotone{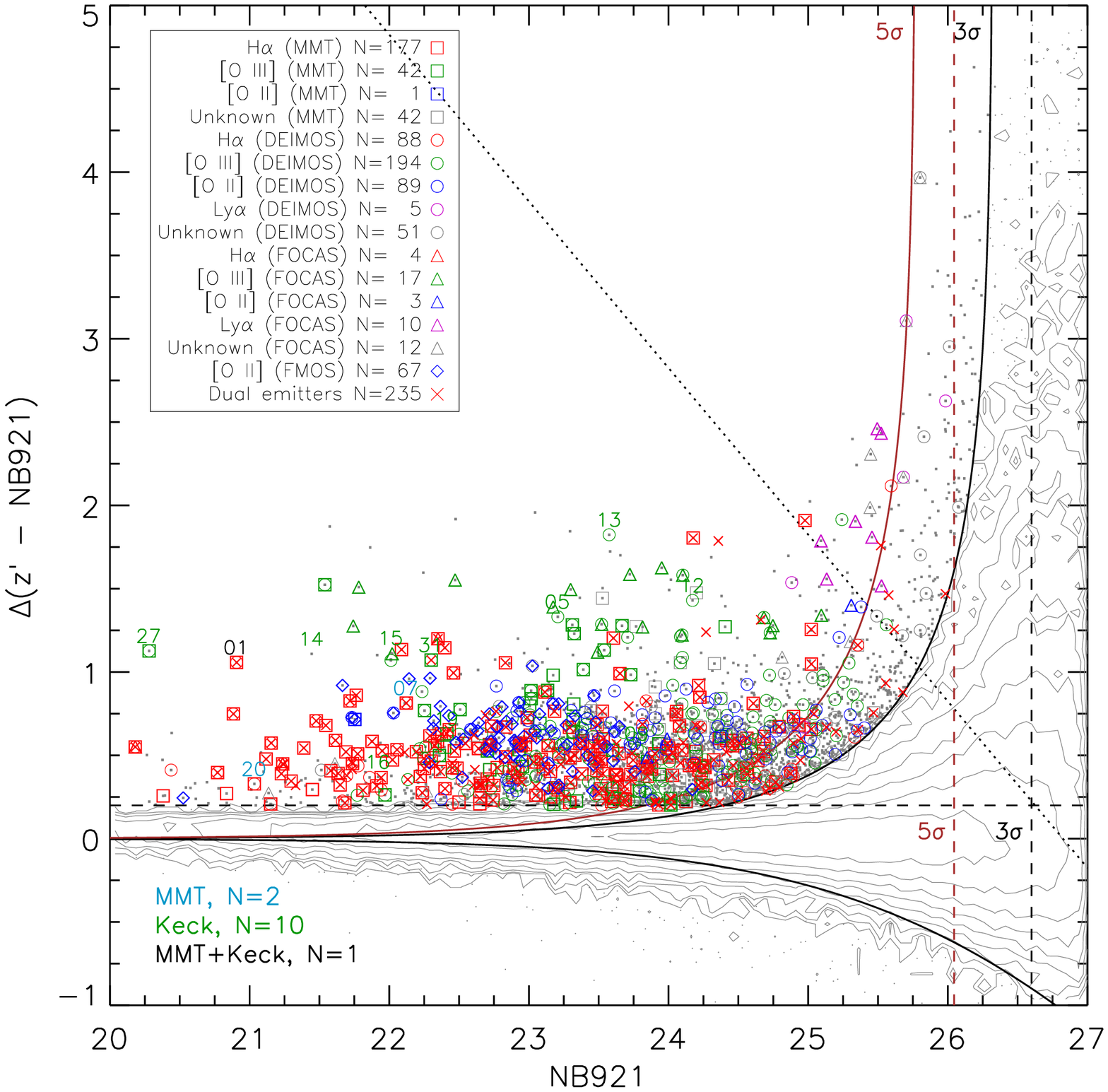}
  \plotone{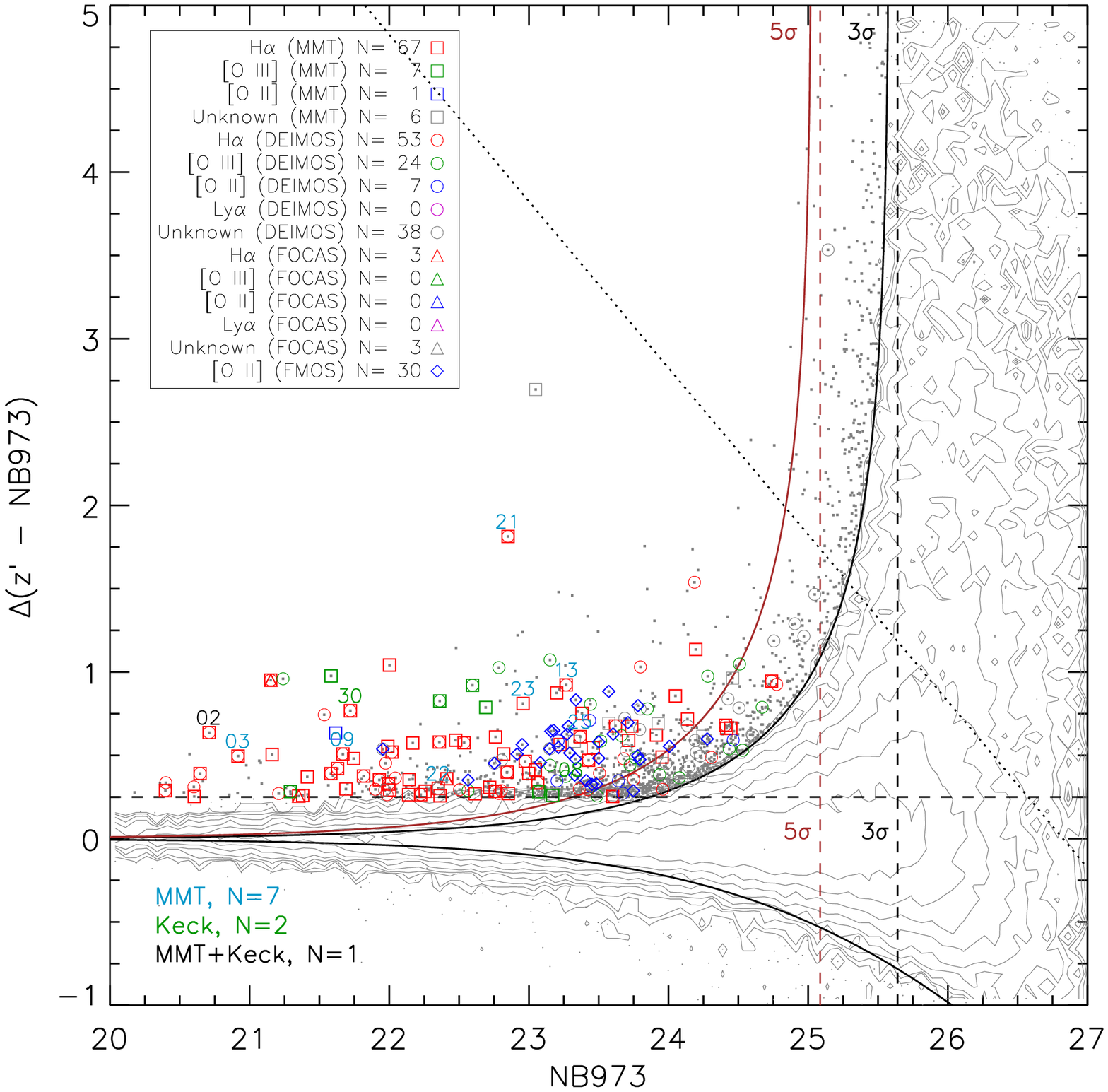}
  \caption{Same as Figures~\ref{fig:Ex_NB1}--\ref{fig:Ex_NB2} but for NB921 (left) and
    NB973 (right) excess emitters. \OII\ emitters that are spectroscopically confirmed
    with Subaru/FMOS \citep{hay15} are shown as blue diamonds.}
  \label{fig:Ex_NB3}
\end{figure*}

\begin{figure*}
  \epsscale{0.56}
  \plotone{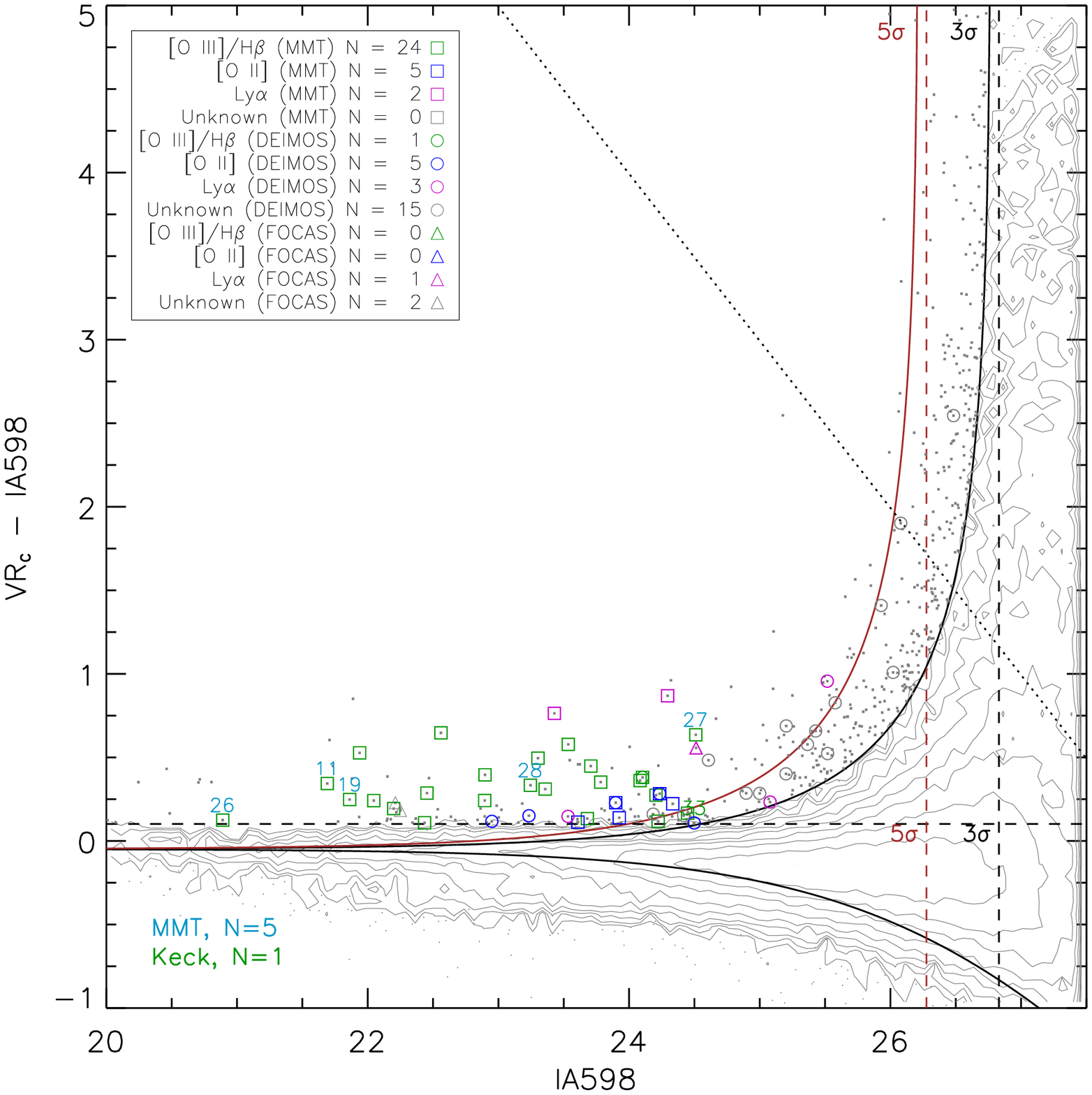}
  \plotone{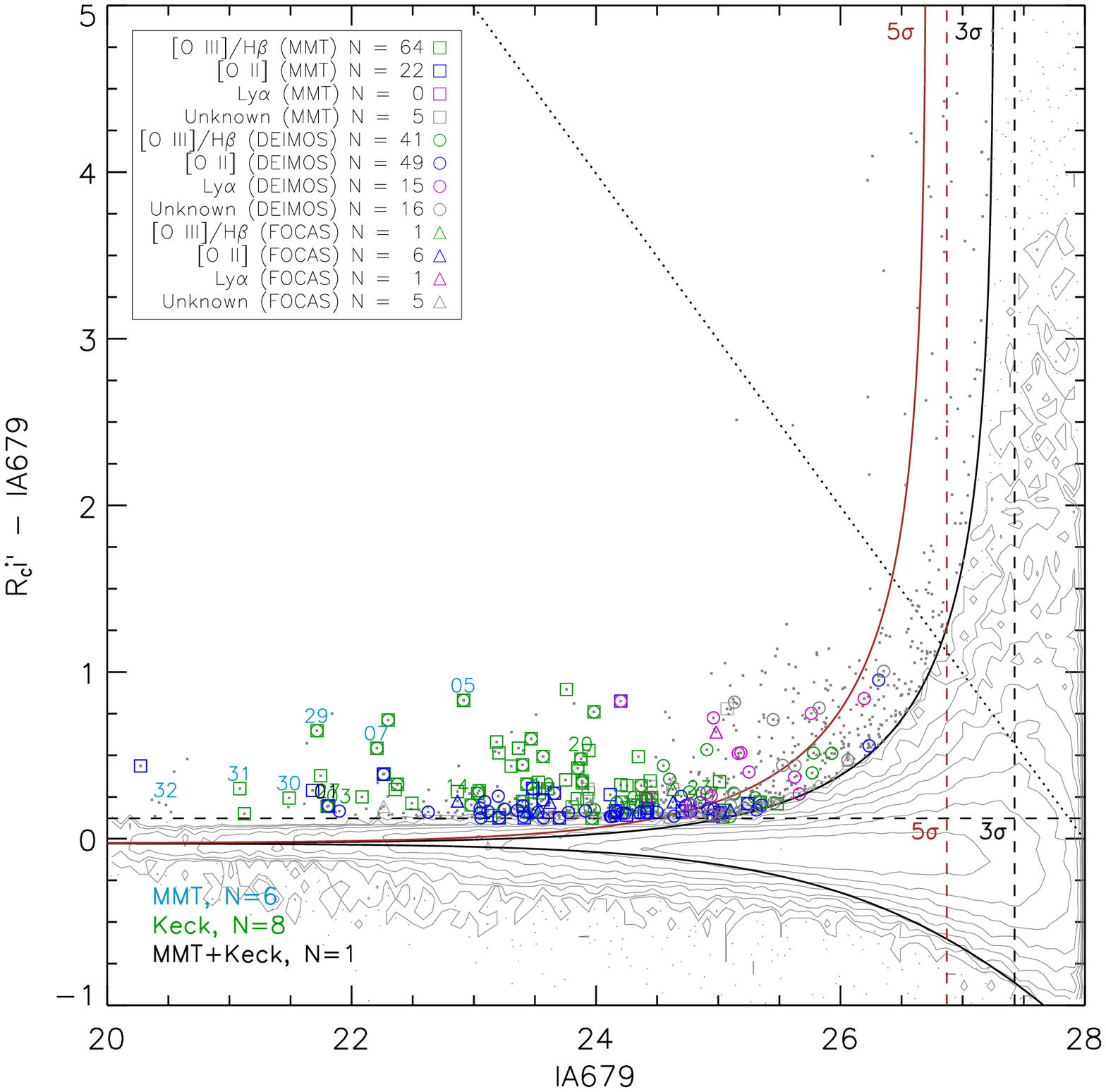}
  \caption{Same as Figures~\ref{fig:Ex_NB1}--\ref{fig:Ex_NB3} but for IA598 (left) and
    IA679 (right) excess emitters.}
  \label{fig:Ex_NB4}
\end{figure*}

To select narrowband and intermediate-band excess emitters due to the presence of
nebular emission line(s), we use the standard color excess selection, where photometry in
these filters is compared against magnitudes for adjacent broadband filters that sample
the continuum. Since the technique has been extensively used, we briefly summarize it
below, and refer readers to \cite{fuj03}, \cite{newha}, and \cite{lee12}.
Table~\ref{tab:SDF_sample} provides an overview of the SDF emission-line survey, and
we illustrate the selection of NB704 and NB711 excess emitters in Figure~\ref{fig:Ex_NB1},
NB816 excess emitters in Figure~\ref{fig:Ex_NB2}, NB921 and NB973 excess emitters in
Figure~\ref{fig:Ex_NB3}, and IA598 and IA679 excess emitters in Figure~\ref{fig:Ex_NB4}.

The measured continuum adjacent to the line is determined by the two broadband filters
closest to the narrow bandpass. For NB921 and NB973, we start with the $z$\arcmin-band.
Since the central wavelengths of these filters are longer than that of the $z$\arcmin\
filter, we correct for the differences using the $i$\arcmin--$z$\arcmin\ color
\citep{HaSFR,OIIpop}.
The remaining filters use a flux-weighted combination of either the $V$- and \Rc-band
(IA598), the \Rc- and $i$\arcmin-band (NB704, NB711, and IA679), or the $i$\arcmin- and
$z$\arcmin-band (NB816):
\begin{equation}
  f_{\rm cont} = \epsilon f_{\rm blue} + (1-\epsilon) f_{\rm red},
  \label{eqn:f_cont}
\end{equation}
where $f_{\rm blue}$ and $f_{\rm red}$ are the flux density in erg s$^{-1}$ cm$^{-2}$ Hz$^{-1}$
for the bluer and redder broadband filters, respectively (e.g., $i$\arcmin\ and $z$\arcmin\
for NB816), and $\epsilon=0.45$ (IA598), 0.5 (NB704, NB711), 0.75 (IA679), and 0.6 (NB816).

Photometric measurements for broadband data are obtained by running SExtractor
\citep[version 2.5.0;][]{ber96} in ``dual-image'' mode, where the respective narrowband
or intermediate-band image is used as the ``detection'' image.
This works well because all broadband, intermediate-band, and narrowband mosaicked
images have very similar seeing, so that excess colors are determined within the same
physical scale on the galaxies. For the extraction of fluxes and selection of sources,
we use a 2\arcsec-diameter circular apertures.
The 3$\sigma$ sensitivities for the $V$, \Rc, $i$\arcmin, and $z$\arcmin\ images are between
26.27 and 27.53 mag, which are generally $\approx$1 mag deeper than the narrowband
and intermediate-band imaging.

We also masked out regions affected by poor coverage and contamination by bright foreground
stars. The unmasked regions cover \areaa\ arcmin$^2$ for all filters with the exception of
NB973, which covers \areab\ arcmin$^2$. The latter is smaller due to higher systematic noise
in one of the ten CCDs, which we mask to avoid a large number of spurious detections. In
total, we detect between $\approx$85,000 and $\approx$140,000 sources in the unmasked
regions of the narrowband or intermediate-band mosaics (see Table~\ref{tab:SDF_sample}).
We select narrowband and intermediate-band excess emitters with a minimum Cont--NB or
Cont--IA color of 0.15 mag for IA598, IA679, NB704, NB711, and NB816, 0.20 mag for NB921,
and 0.25 mag for NB973. These minimum color excesses are shown in
Figures~\ref{fig:Ex_NB1}--\ref{fig:Ex_NB4} as dashed horizontal lines. In addition, we
require that the Cont--NB or Cont--IA color exceeds
\begin{equation}
  -2.5\log\left(1 - \frac{\sqrt{f_{\nu,3\sigma{\rm NB/IA}}^2 + f_{\nu,3\sigma{\rm Cont}}^2}}{f_{\nu,\rm NB/IA}}\right),
\end{equation}
where $f_{\nu,3\sigma{\rm Cont}}$ and $f_{\nu,3\sigma{\rm NB/IA}}$ are the 3$\sigma$ flux density
limits for the continuum band(s) and the narrowband or intermediate-band, respectively, and
$f_{\nu,{\rm NB/IA}}$ is the flux density for a given narrowband or intermediate-band
magnitude.
This selection yielded between $\approx$640 and $\approx$2360 excess emitters in each
narrowband or intermediate-band filter.

Some of our sources are independently selected by more than one filter. This is due to some
fortuitous redshift overlap of our narrowband/intermediate-band filters such that
different emission lines (e.g., \Ha\ and \OIII) are detectable at the same redshift.
Accounting for duplicate galaxies, the complete SDF emission-line galaxy sample consists
of \Nem\ galaxies mostly at $z=\za$ to $z=\zb$ with some at higher redshift due to
Ly$\alpha$ emission. We note that the sample presented in this paper supersedes our
earlier multi-narrowband study \citepalias{ly07}.
Specifically, our previous study computed the continuum for narrowband/intermediate
excess selection using magnitudes instead of flux densities (see
Equation~(\ref{eqn:f_cont})). Also, a fifth narrowband filter (NB973) was later
included \citep{HaSFR,OIIpop}.


\subsection{Optical Spectroscopy}\label{sec:spectra}
The primary results of this paper are based on optical spectroscopy conducted with Keck's
Deep Imaging Multi-Object Spectrograph \citep[DEIMOS;][]{fab03} and MMT's Hectospec
\citep{fab05}. These spectrographs are complementary in sensitivity, spectral coverage,
and multiplexing capabilities. Specifically, DEIMOS has greater sensitivity than Hectospec
(especially above 8000 \AA), however, it can only target $\approx$100 galaxies within a
5\arcmin$\times$17\arcmin\ slitmask field of view, and lacks spectral coverage below
$\approx$6000 \AA. Hectospec, with optical fibers, can target up to $\approx$270 galaxies
over a much larger 1\arcdeg\ diameter field of view, and has spectral coverage
extending down to $\approx$3700 \AA.
To maximize the scientific productivity of observing time on both spectrographs, we use
Hectospec to primarily target our lower redshift sample ($z\lesssim0.5$) and DEIMOS for
the higher redshift sample ($z\gtrsim0.6$).  We include some overlap at $z\sim0.4$--0.65
to improve the spectral coverage (i.e., observe redder emission lines, e.g., \Ha) and for
consistency checks on emission-line measurements (see Section~\ref{sec:f_calib} for
further discussion).
In total, we obtain \Nspec\ optical spectra for \Nspecg\ narrowband/intermediate-band
excess emitters (roughly 20\% of our narrowband/intermediate-band excess samples),
and successfully detect emission lines to determine redshift for \Nspecgz\ galaxies or
78\% of the targeted sample. A summary of the spectroscopically confirmed excess emitters
is provided in Table~\ref{tab:SDF_sample}.

As illustrated in Figures~\ref{fig:Ex_NB1}--\ref{fig:Ex_NB4}, our optical spectroscopy
spans six magnitudes in luminosity and three magnitudes in narrowband or intermediate-band
excess flux. Observations prior to 2014 targeted emission-line galaxies in a more uniform
manner, while more recent observations utilized available information
(narrowband/intermediate-band excess flux, previous spectra) to target galaxies where a
high success rate of obtaining gas metallicities would be achieved. Specifically, there
was a moderate bias toward galaxies with stronger nebular emission lines.


\begin{figure*}
  \epsscale{1.1}
  \plottwo{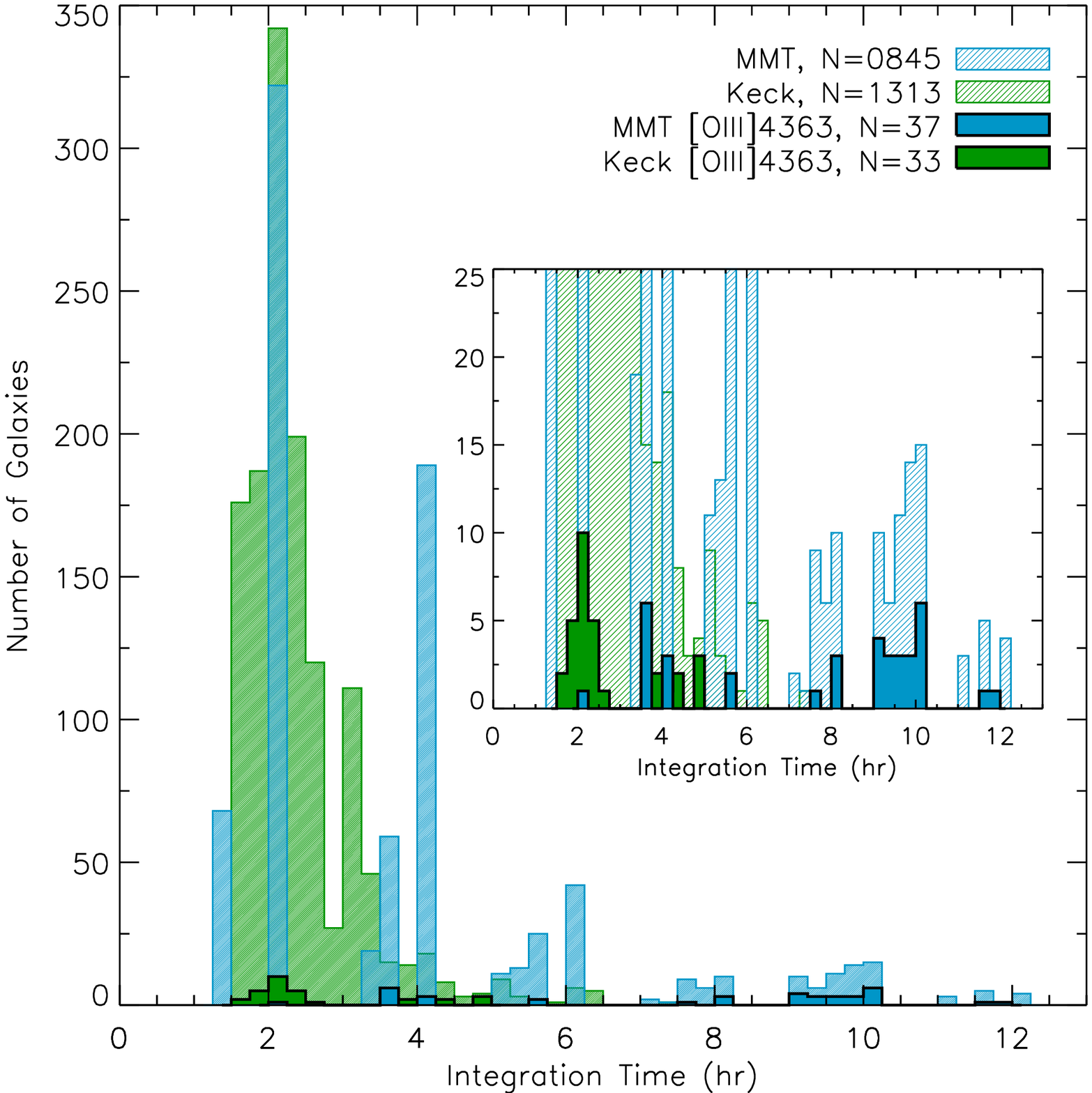}{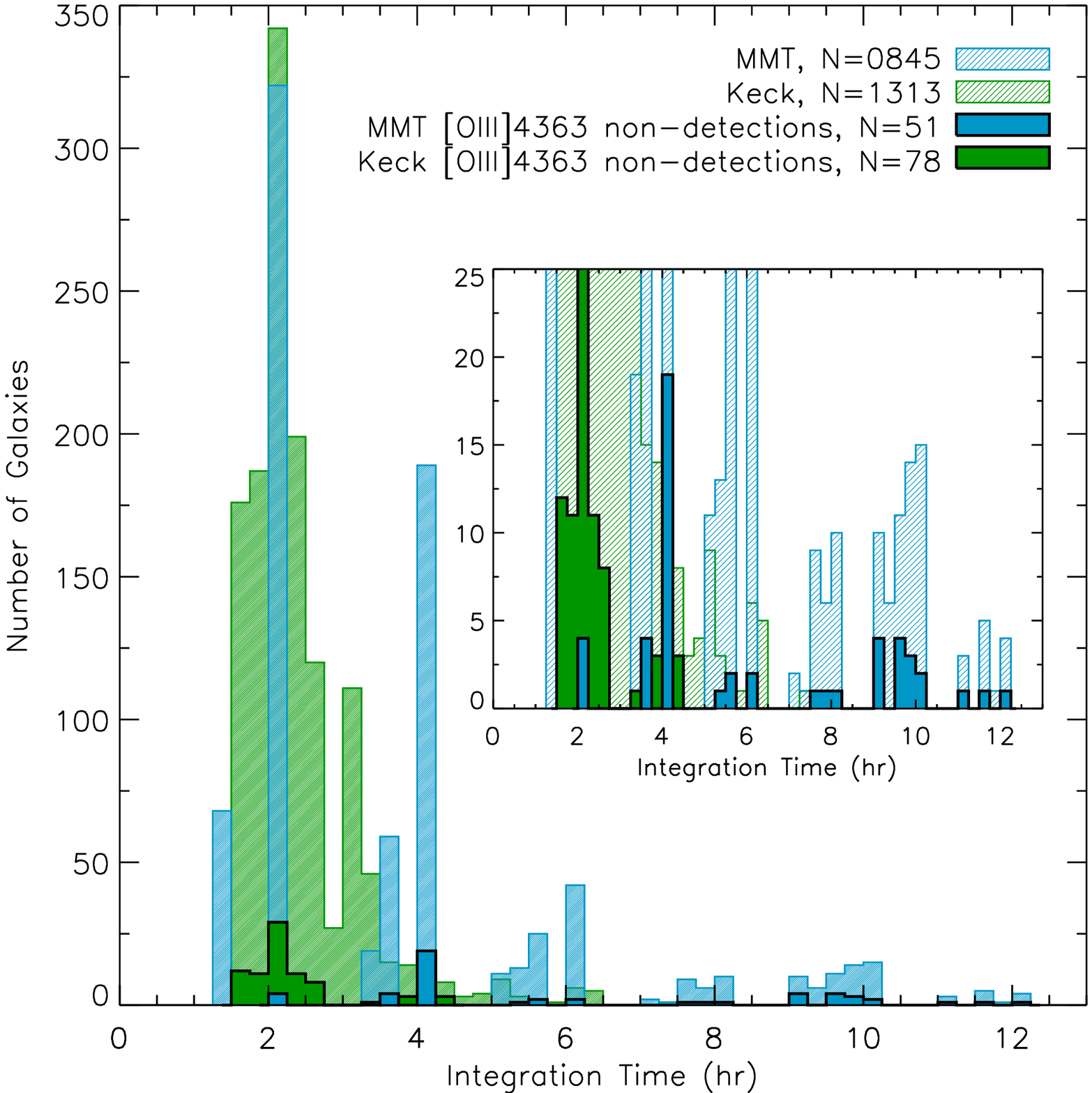}
  \caption{On-source integration time for MMT (blue shades) and Keck (green shades)
    spectroscopy. The full samples are shown by the line-filled distributions. The
      solid-filled distributions with black outlines show the \OIIIa-detected and
      \OIIIa-non-detected samples in the left and right panels, respectively.}
  \label{fig:int_time}
\end{figure*}


\subsubsection{MMT/Hectospec Observing Program}
The MMT observations were conducted on 2008 March 13, 2008 April 10--11, 2008 April 14,
2014 February 27--28, 2014 March 25, and 2014 March 28--31, and correspond to the
equivalent of three full nights. The sky was either clear or had thin cirrus clouds, with
seeing of 0\farcs7--1\farcs4. We utilized the 270 mm$^{-1}$ grating blazed at 5200 \AA, to
yield spectral coverage of 3650--9200 \AA\, with a spectral resolution of $\approx$5 \AA,
at a dispersion of 1.2 \AA\ pixel$^{-1}$.
Because of the queue observing mode for Hectospec, observations consisted of 60--120
minutes of on-source integration (each night), with 20--25 minutes for each individual
exposure. A total of \NMMTspec\ spectra that targeted \NMMTspecg\
narrowband/intermediate-band excess emitters were obtained. These MMT spectra provided
redshifts for 708 galaxies (84\% successful confirmation); the majority of the
unsuccessful spectra targeted fainter galaxies.  These observations were reduced using
External SPECROAD,\footnote{\url{http://astronomy.mnstate.edu/cabanela/research/ESPECROAD/}}
an \textsc{iraf}-based reduction pipeline developed by the Smithsonian Astrophysical
Observatory, but with improvements by Juan Cabanela \citep{hum11}. Since spectra for
any given source were obtained over multiple nights, we stacked the data, with weights
that are inversely proportional to the variance ($\sigma^2$) in the spectrum. In general,
the rms noise in the spectrum improves as the square root of the integration time,
$t_{\rm int}$. The integration times vary between 80 and 725 minutes with an average
(median) of 228 (220) minutes. The distribution of on-source integration times is
illustrated in Figure \ref{fig:int_time}.


\subsubsection{Keck/DEIMOS Observing}
The Keck observations were conducted on 2004 April 23--24, 2008 May 01--02, 2009 April
25--28, 2014 May 2, and 2015 March 17/19/26. The majority of the observations were
obtained in 2014--2015. The 2004 spectroscopic observations have been discussed in
\cite{kas06} and \citetalias{ly07}, and the 2008--2009 data have been discussed in
\cite{kas11}. In brief, we constructed 22 slitmasks with 1\arcsec\ slit widths, and
used the 830 line mm$^{-1}$ grating and GG495 order-cut filter. This configuration
yielded a spectral resolution of $R\sim3600$ at 8500 \AA, at a dispersion of 0.47
\AA\ pixel$^{-1}$.
A total of \NKeckspec\ spectra that targeted \NKeckspecg\ narrowband/intermediate-band
excess emitters were obtained. These Keck spectra provided redshifts for 1031 galaxies
(79\% successful confirmation). The data were reduced using the Keck DEIMOS spec2d
pipeline \citep{coo12},\footnote{\url{http://deep.ps.uci.edu/spec2d/primer.html}}
with improvements made by Peter
Capak.\footnote{\url{http://www.astro.caltech.edu/~capak/software/deimos.html}}
A subset of our galaxy sample was observed twice on different dates with Keck. We
stacked those data, following the same approach as described for MMT data. The
on-source integration times, which vary between 91.0 and 447.8 minutes with an average
(median) of 142 (120) minutes, are illustrated in Figure~\ref{fig:int_time}.

\subsection{Accurate Flux Calibration of MMT and Keck Spectra}
\label{sec:f_calib}

To obtain \OIIIa-based metallicities, which use emission-line diagnostics from
rest-frame $\approx$3700 to $\approx$5010 \AA, proper flux calibration of our spectra
is important. Since the spectroscopic data are obtained from two different instruments,
MMT/Hectospec and Keck/DEIMOS, using different observational configurations (fibers
versus slits), and under various observational conditions (seeing: 0\farcs5--1\farcs5),
we use the approach developed by \citetalias{ly14} to flux calibrate and unify the
spectroscopic data. The approach is as follows.
 
First, spectro-photometric standard stars are observed. These observations are generally
obtained on the same night; however, in some cases, these calibration data are taken a
few days apart.\footnote{For MMT, the Hectospec team provides all observers during a
  semester with reduced spectra for a few spectrophotometric standards to use.}
For DEIMOS, the sensitivity is not identical for all eight CCDs (four detectors for the
blue side, four for the red side). To address this issue, we dithered the telescope so
that the same standard star is available in all detectors.

Reducing the calibration data in identical or similar ways to our science spectra, we
determine the sensitivity function(s) using standard \textsc{iraf} processing techniques
(i.e., the \textit{onedspec} package). We then apply these sensitivity functions to our
data to yield ``first-pass'' flux-calibrated spectroscopic products.

Second, we examine the reliability of our flux calibration for each slitmask or fiber
configuration by comparing them to broadband photometric data. Here we only consider the
brightest galaxies such that the continuum is well-detected. For the MMT and DEIMOS data,
we restrict the calibration sample to $R_{\rm C}\leq22.0$ mag and $i\arcmin\leq23.0$ or
23.5 mag, respectively. This typically has $\approx$10--20 galaxies per set-up. To avoid
the effects of OH skylines, we generate smoothed spectra with a boxcar median.
From there we convolved our smoothed spectra with the filter responses and determine
fluxes. These fluxes are compared against the photometric data to examine slit/fiber
losses and if any wavelength-dependent corrections are needed. In most cases, specifically
MMT/Hectospec observations, the comparison against photometric data only shows a systematic
offset due to fiber losses. For Keck/DEIMOS, we found that the photometric--spectroscopic
comparison required an additional first-order wavelength-dependent correction.
  
\begin{figure*}
  \epsscale{1.1}
  \plottwo{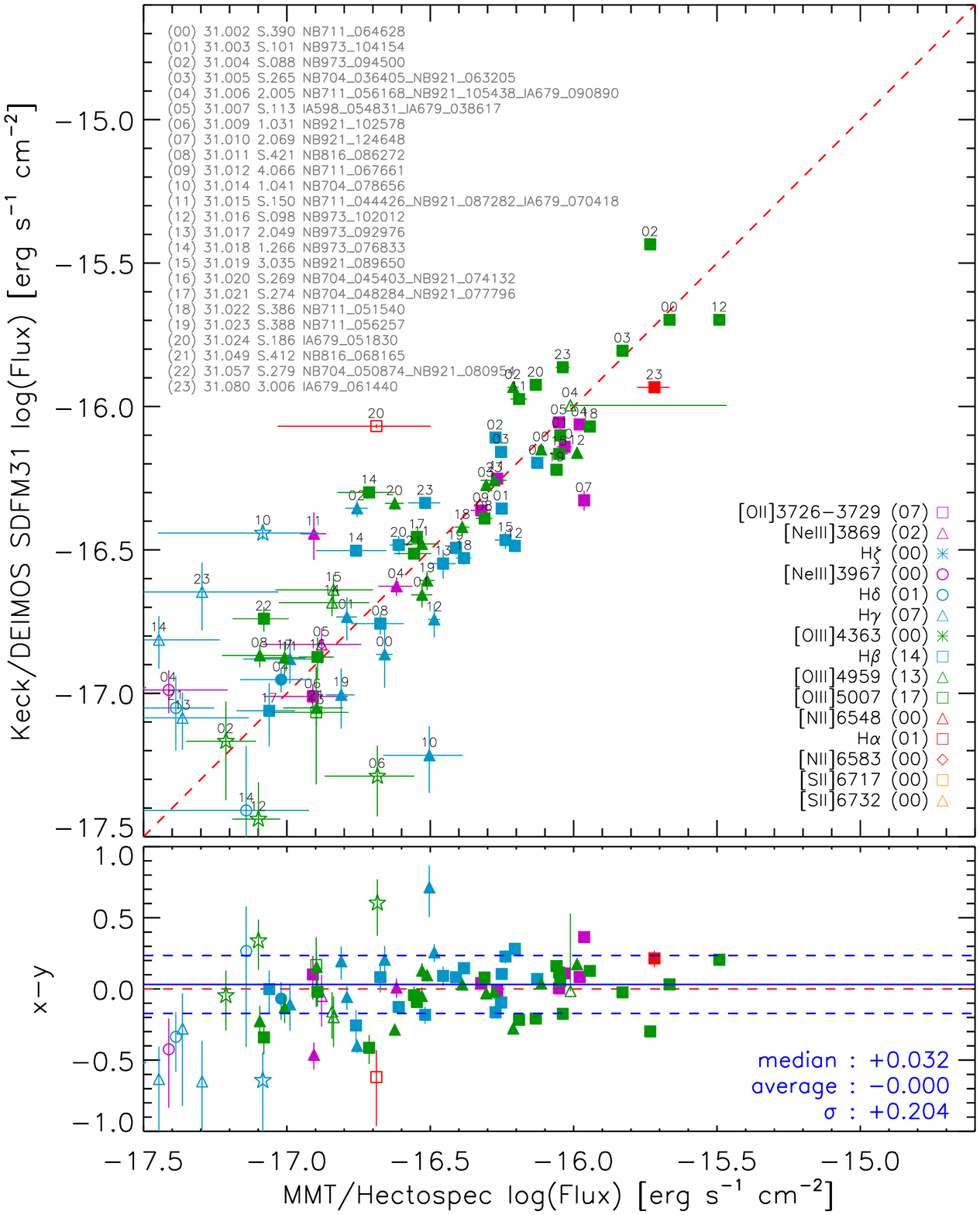}{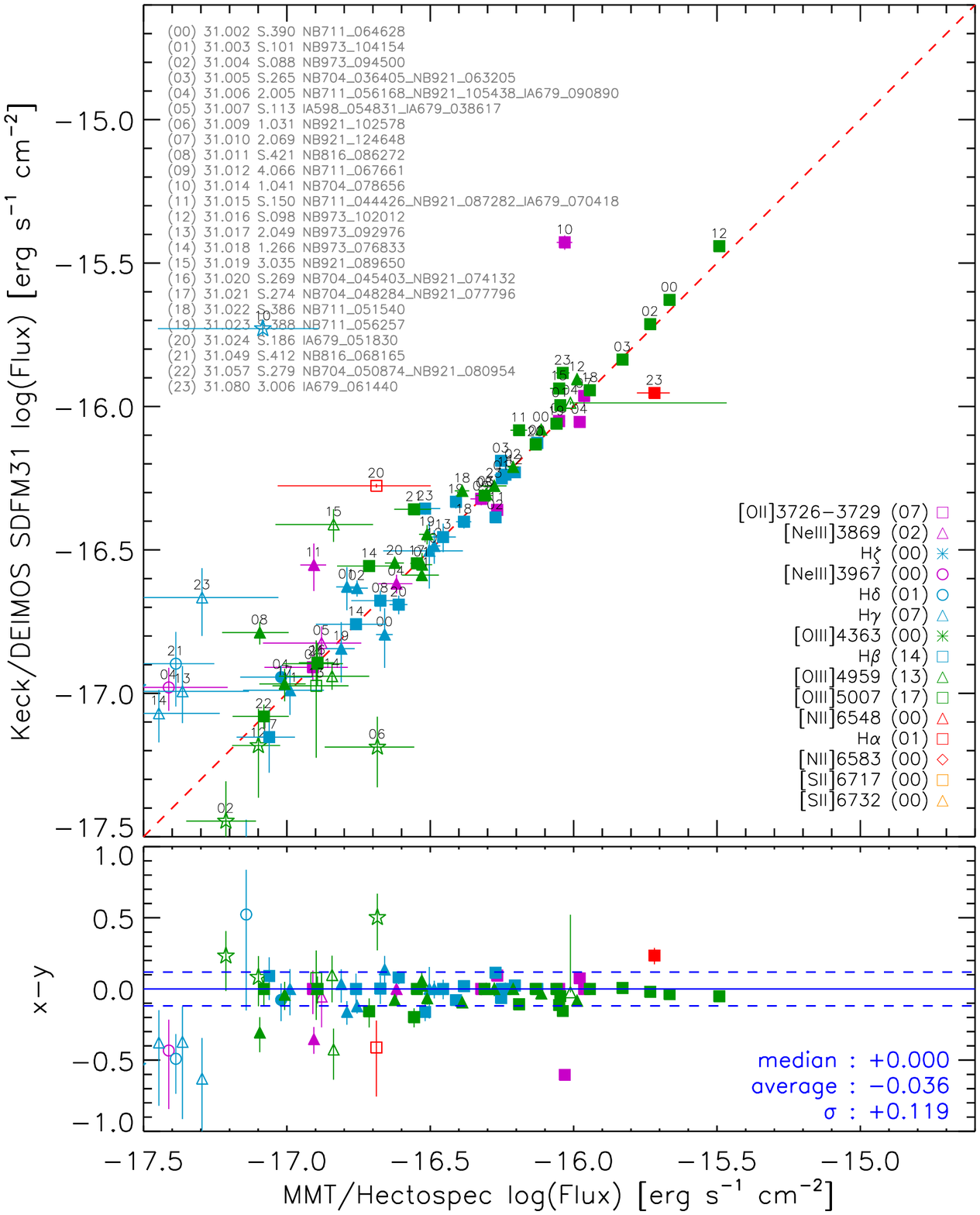}
  \caption{Illustration of the reliable emission-line fluxes obtained from Keck
    ($y$-axes) and MMT ($x$-axes). Here we compare measurements for one DEIMOS slitmask
    (SDFM31), which has 24 galaxies observed with both telescopes. Different colors and
    symbols indicate nebular emission lines (see the legend in the lower-right corner).
    Each data point is indicated with a two-digit number to specify the galaxy (see
    upper-left legend). The median, average, and dispersion are determined when emission
    lines are detected above S/N = 3 by both MMT and Keck (filled points); unfilled data
    points are those with low S/N. The red dashed lines show one-to-one correspondence.
    The left panel illustrates the comparison in flux after applying (1) the sensitivity
    function of the instrument and (2) slit-loss correction based on photometric data
    (see Section~\ref{sec:f_calib}). The right panel illustrates the improvement,
    $\approx$2 times lower in dispersion, after a median-based correction is applied for
    each individual galaxy. The bottom panels illustrate the difference with the median
    and 1$\sigma$ dispersion indicated by the blue solid and blue dashed lines,
    respectively. The results of the MMT--Keck comparisons are provided for all DEIMOS
    slitmasks in Table~\ref{tab:MMTKeck_flux}.}
  \label{fig:MMTKeck}
\end{figure*}

Next, we compare the emission-line fluxes of the MMT and Keck spectra when galaxies
are observed with both telescopes. In this analysis, we compare Keck measurements of
several emission lines against those from MMT. We chose MMT as the reference because (1)
Hectospec has an atmospheric dispersion compensator to correct for wavelength-dependent
fiber losses; (2) the spectral coverage is consistently the same unlike Keck slit
spectroscopy, which depends on the position on the slitmask; and (3) independent analyses
from the Hectospec instrument team have demonstrated that the MMT flux calibration is
accurate and precise \citep{fab08}. For the MMT--Keck analyses, we perform them on each
slitmask to ensure that the observing conditions did not affect measurements. To account
for differences between MMT and Keck, we apply corrections on an individual source basis
by determining the median offset based on several emission-line measurements. We
illustrate the comparison before and after the median-based correction is applied for one
Keck/DEIMOS slitmask in Figure~\ref{fig:MMTKeck}. Table~\ref{tab:MMTKeck_flux} summarizes
the MMT--Keck comparison and the improvements with the median-based correction applied on
an individual basis. We find that the steps taken to flux calibrate the spectra yield
consistent fluxes between MMT and Keck, and the dispersion is further reduced when the
median-based corrections are applied.


\input tab2.final.tex

\begin{figure*}
  \epsscale{1.1}
  \plottwo{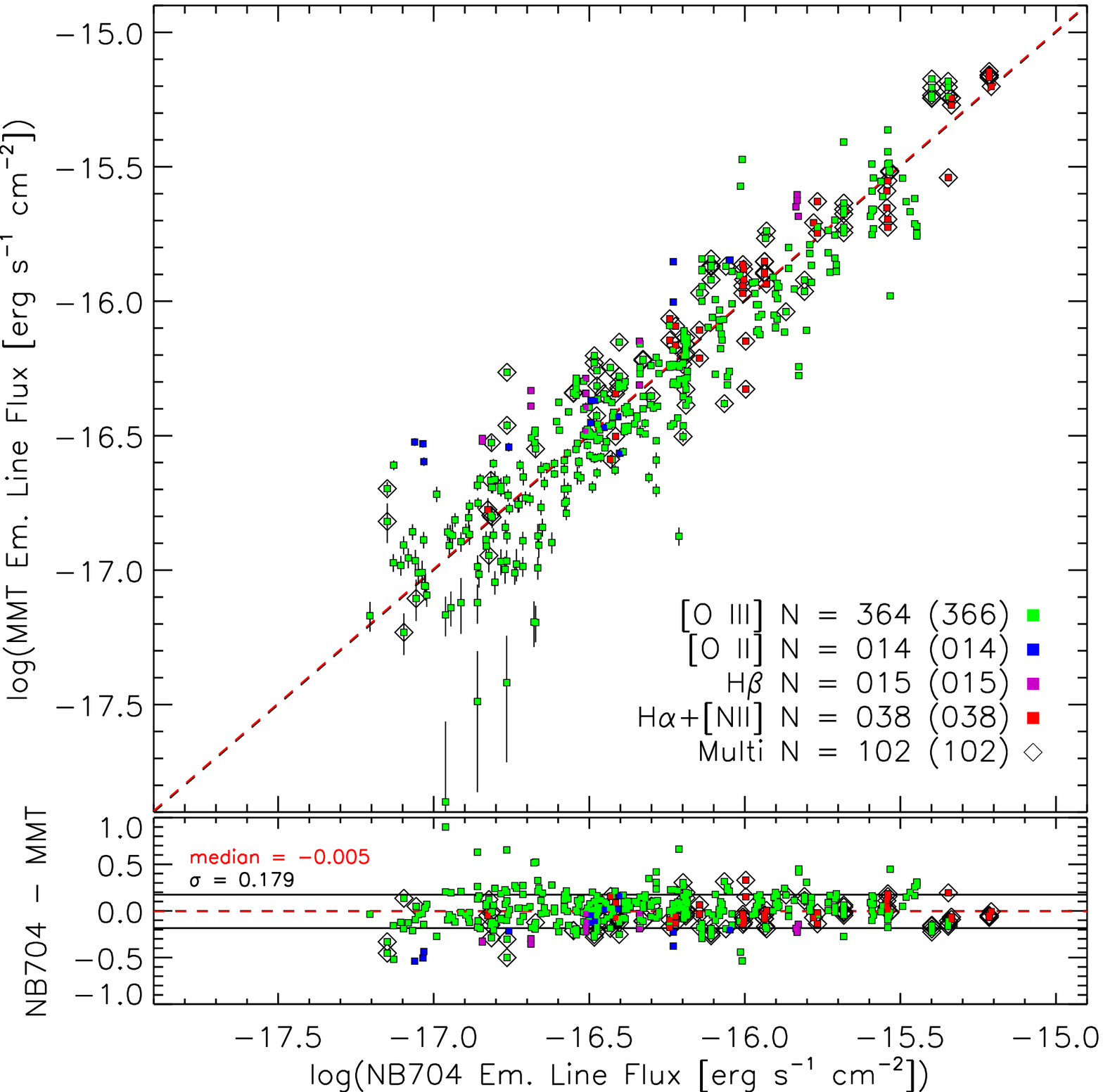}{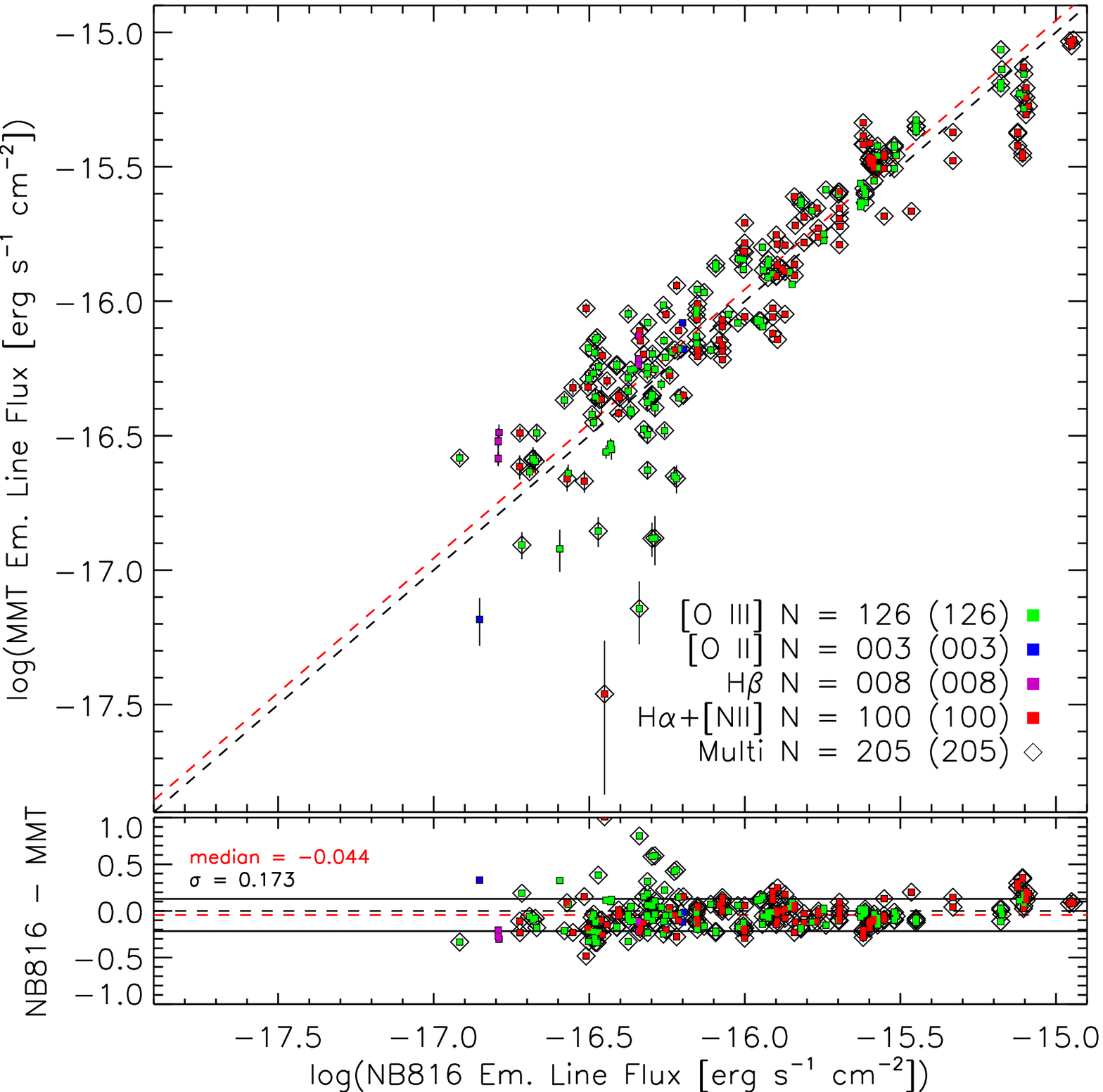}
  \caption{Comparisons between MMT/Hectospec emission-line flux measurements ($y$-axes)
    against NB704 (left) and NB816 (right) excess fluxes. The narrowband excess fluxes are
    derived from Equation~(\ref{eqn:NBF}). The top panels show the comparison while the
    bottom panels show the difference (narrowband excess flux -- MMT flux) on the
    $y$-axes. Green, blue, purple, and red squares indicate \OIII$\lambda\lambda$4959,\,5007,
    \OII, \Hb, and \Ha+\NII\ excess emitters, respectively. Black diamonds indicate
    measurements with more than one emission line in the narrowband filter (e.g., the
    \OIII\ doublet, \Ha\ and \NII). The red dashed lines in the top and bottom panels
    show the median difference between narrowband excess fluxes and MMT emission-line
    fluxes, while the black solid lines in the bottom panels show the $1\sigma$ dispersion.
    These comparisons illustrate a good agreement between photometric and spectroscopic
    measurements (black dashed lines indicate one-to-one correspondence), and demonstrate
    that the flux calibration is performed at a reliable level over 3 dex in emission-line
    flux. The results of the narrowband--spectroscopy comparison are provided in
    Table~\ref{tab:f_calib}.}
  \label{fig:calib1}
\end{figure*}

\begin{figure*}
  \epsscale{1.1}
  \plottwo{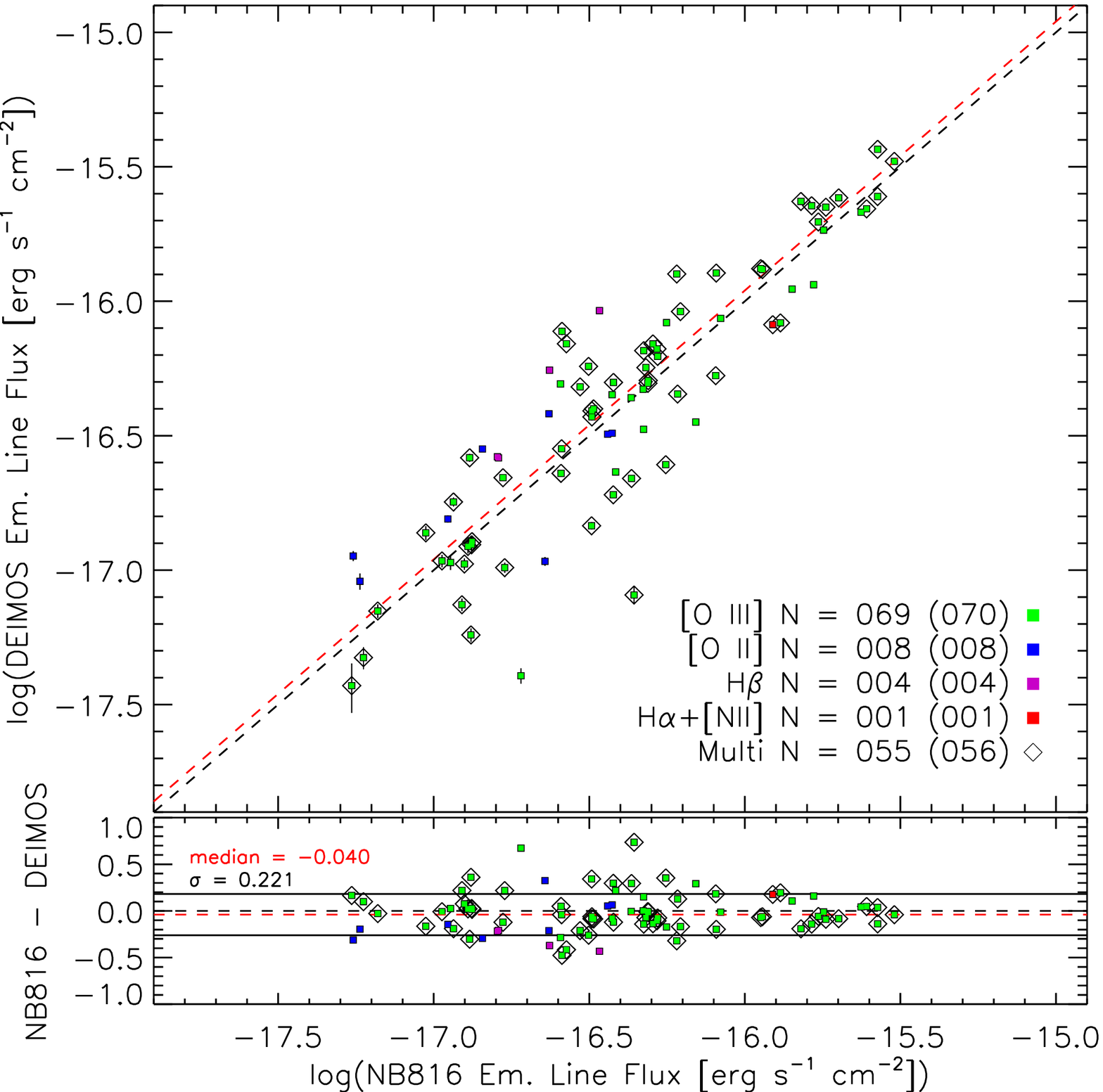}{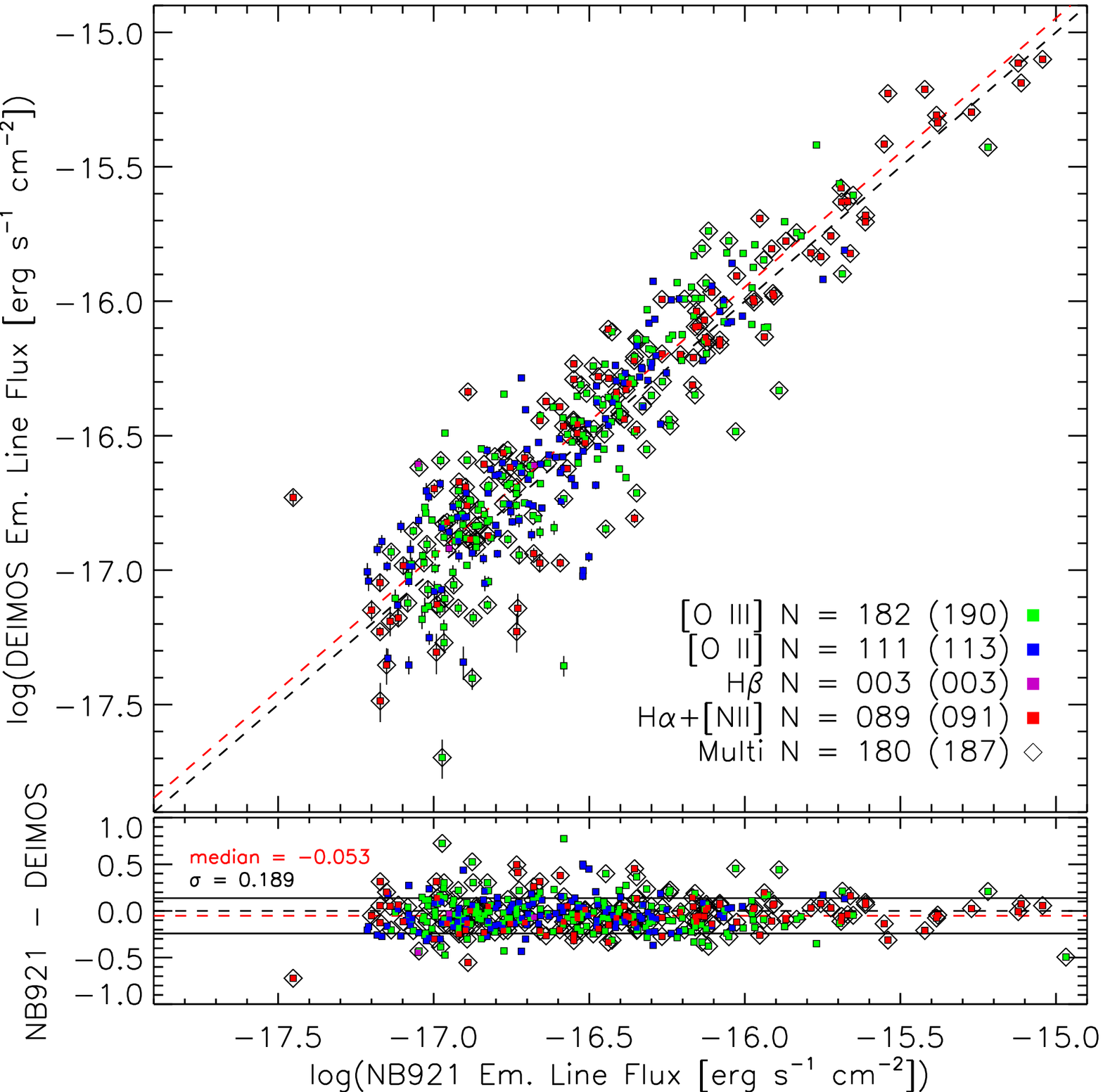}
  \caption{Same as Figure~\ref{fig:calib1} but for Keck/DEIMOS measurements and
    comparisons are illustrated against NB816 (left) and NB921 (right) excess fluxes.
    The results of the narrowband--spectroscopy comparison are provided in
    Table~\ref{tab:f_calib}.}
  \label{fig:calib2}
\end{figure*}

Finally, to ensure that the flux calibration is reliable, we use our measurements of
nebular emission lines from narrowband imaging data to compare against the spectroscopic
measurements. The narrowband excess fluxes are derived from a combination of narrowband
and broadband fluxes:
\begin{equation}
  F_{\rm Line} = \Delta{\rm NB}\frac{f_{\rm NB}-f_{\rm BB}}{1-\epsilon(\Delta{\rm NB}/\Delta{\rm BB})},
  \label{eqn:NBF}
\end{equation}
where $f_{\rm NB}$ and $f_{\rm BB}$ are the flux densities for the narrowband and broadband
filters, respectively, in units of erg s$^{-1}$ cm$^{-2}$ Hz$^{-1}$, and $\Delta{\rm NB}$
and $\Delta{\rm BB}$ are the FWHM of the filters. We use the spectroscopic redshift to
correct the narrowband flux for when the emission line(s) falls in the filter's wing.
This comparison is illustrated in Figure~\ref{fig:calib1} for the MMT spectra and
Figure~\ref{fig:calib2} for Keck. For brevity, we present the comparison against MMT
spectroscopic measurements for NB704 and NB816 excess emitters, and NB816 and NB921
excess emitters for Keck measurements. Table~\ref{tab:f_calib} provides the accuracy and
precision of the flux calibration based on comparison with narrowband excess fluxes.
Typically, the accuracy is no more than $\pm0.05$ dex with a precision of $\lesssim$0.2
dex. These flux calibration comparisons against narrowband excess fluxes are conducted
on individual spectra, as well as stacked spectra. Overall, our analyses suggest that we
have reliable flux calibration across the full spectral coverage for both MMT and Keck.


\input tab3.final.tex


\section{SAMPLE SELECTION FOR \citetalias{MACTII}}
\label{sec:sample}


\subsection{The \OIIIa-detected Sample}

To extract fluxes of emission lines in these spectra, we fit each emission line with a
Gaussian profile\footnote{Since the \OII\,$\lambda\lambda$3726,\,3729 doublet is
  resolved with Keck, we fit both emission lines with a double Gaussian profile.}
using the IDL routine \textsc{mpfit} \citep{mar09}. The expected location of emission
lines was based on a priori redshift determined by either the \OIII\ or \Ha\ (for lower
redshift) lines. A local spectral median, $\left<f\right>$, is computed within a 200 \AA\
wide band, excluding regions affected by OH skylines and nebular emission lines. In
addition, the standard deviation $\sigma(f)$ is measured locally.
To determine the significance of emission lines, we integrate the spectrum between
$l_{\rm C}-2.5\sigma_{\rm G}$ and $l_{\rm C}+2.5\sigma_{\rm G}$, where $l_{\rm C}$ is the
central wavelength and $\sigma_{\rm G}$ is the Gaussian width:
\begin{equation}
  {\rm Flux} \equiv \sum_{-2.5\sigma_{\rm G}}^{+2.5\sigma_{\rm G}} \left[f(\lambda-l_{\rm C})-\left<f\right>\right] \times l\arcmin.
\end{equation}
Here, $l$\arcmin\ is the spectral dispersion (1.21 \AA\ pixel$^{-1}$ for MMT and
$\approx$0.47 \AA\ pixel$^{-1}$ for Keck).
We then compute the signal-to-noise ratio (S/N) of the line by dividing the integrated
flux by:
\begin{equation}
  {\rm Noise} \equiv \sigma(f) \times l\arcmin \times \sqrt{N_{\rm pixel}},
\end{equation}
where $N_{\rm pixel}=5\sigma_G/l$\arcmin.

\begin{figure*}
  \epsscale{1.15}
  \plottwo{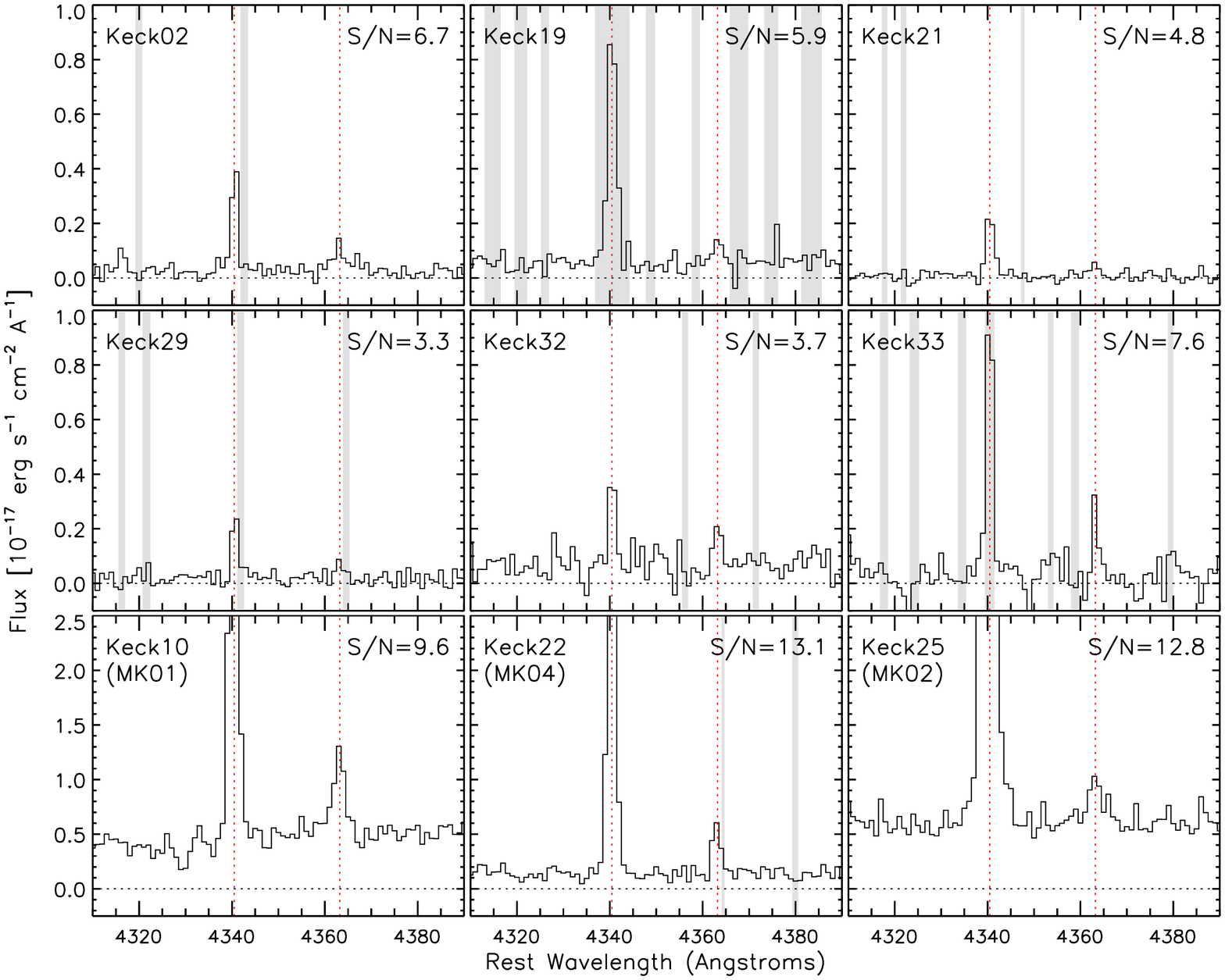}{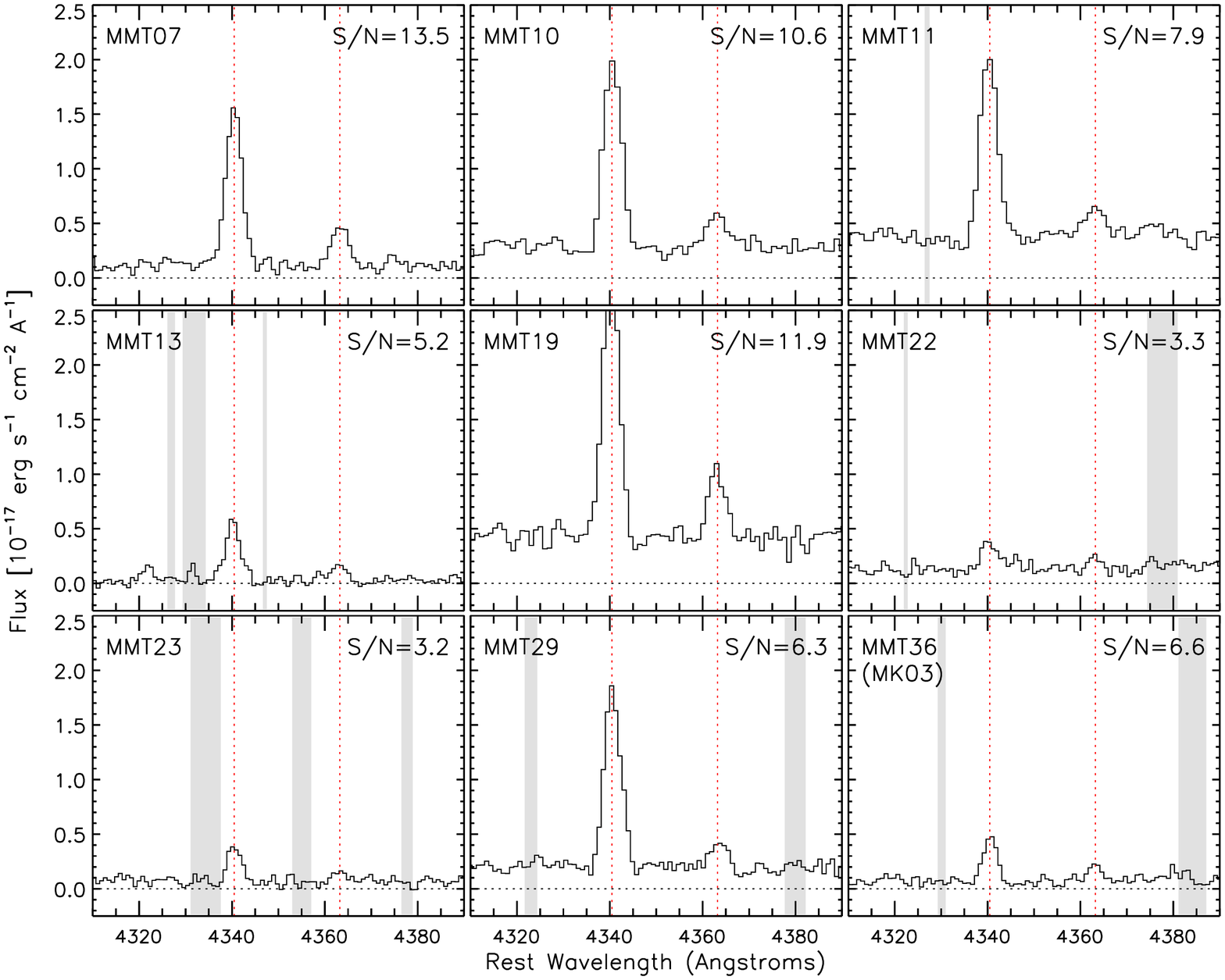}
  \caption{Detections of \OIIIa\ in 18 galaxies, nine with Keck (left) and nine with MMT
    (right). These examples illustrate the full range of S/N (indicated in the upper right)
    present in our \OIIIa-detected sample. The rest-frame spectra are shown in black, with
    vertical red dashed lines indicating the locations of \Hg$\lambda$4340 and \OIIIa. The
    wavelengths contaminated by OH skylines are indicated by light gray vertical bands.
    The galaxy identifying names are provided in the upper left.}
  \label{fig:oiii_spec}
\end{figure*}

\begin{figure*}
  \epsscale{1.1}
  \plotone{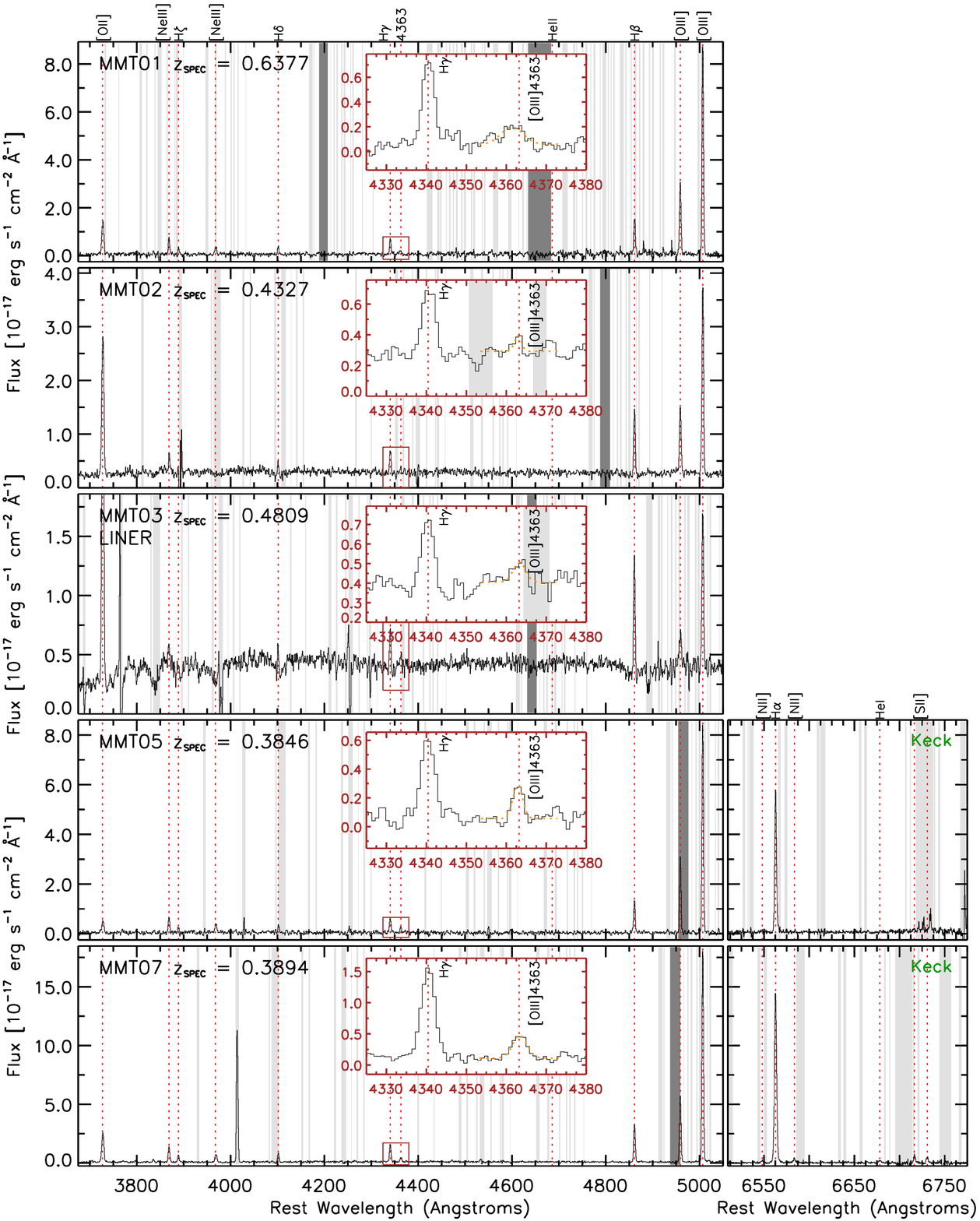}
  \caption{MMT spectra for a subset of \OIIIa-detected galaxies from the SDF. The blue-ends
    of the spectra, spanning \OII$\lambda$3727 to \Hb+\OIII$\lambda$5007, are shown on
    the left. Where available, we also illustrate (on the right) the red-ends of the spectra
    that span \Ha+\NII$\lambda\lambda$6548,\,6583 and \SII$\lambda\lambda$6716,\,6731. The
    insets provide a closer view of \Hg\ and \OIIIa\ (rectangular region outlined in brown)
    with the best Gaussian fits to the \OIIIa\ lines shown by the dotted orange lines.
    For the right panels, Keck spectra are shown where available or if the Keck measurements
    are more reliable than MMT. The vertical light gray bands indicate spectral regions
    contaminated by night-sky lines while the darker gray bands indicate the telluric A-
    and B-bands. Nebular emission lines are indicated by the vertical dashed red lines.}
  \label{fig:MMT_spec1}
\end{figure*}

\begin{figure*}
  \epsscale{1.1}
  \plotone{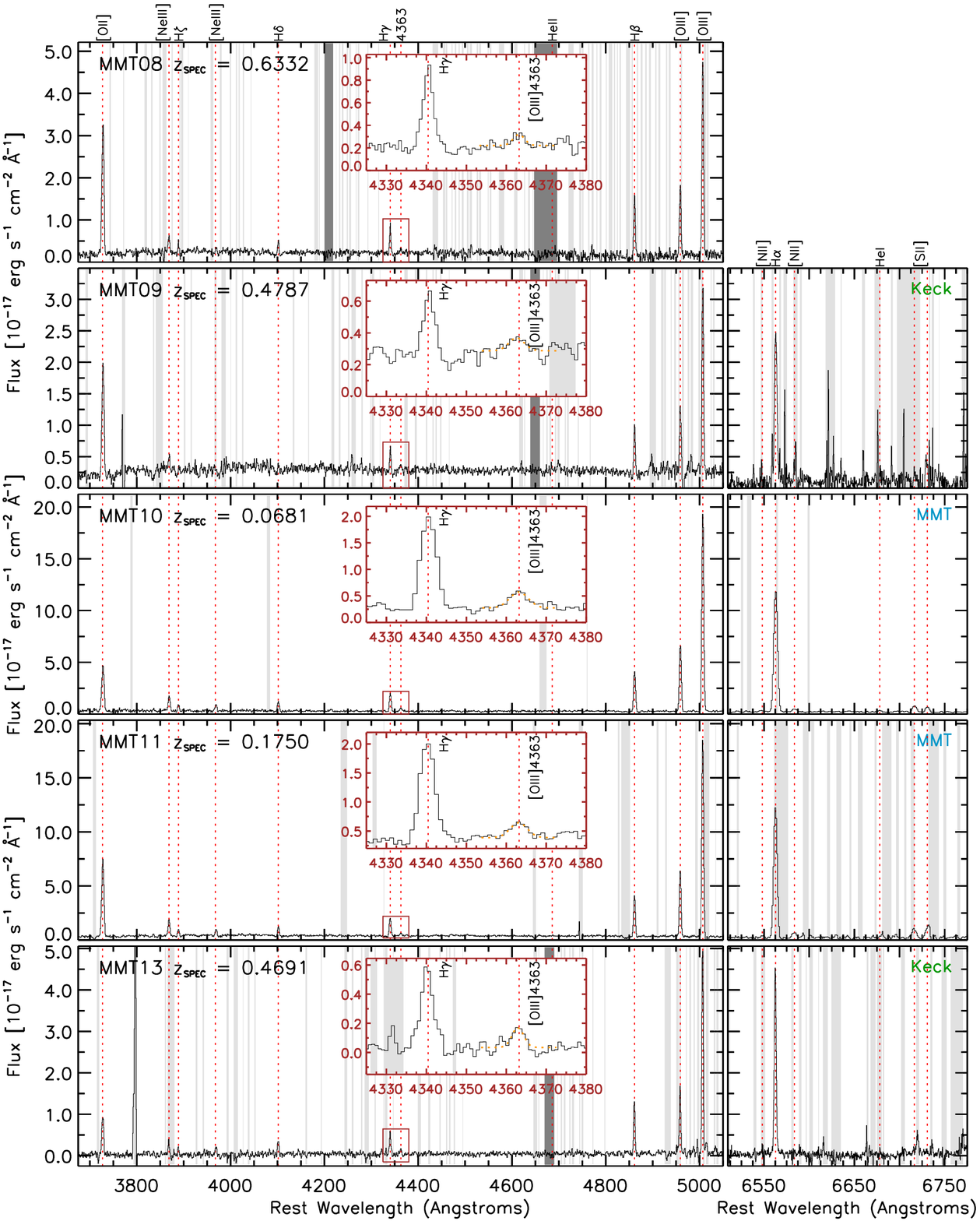}
  \caption{Same as Figure~\ref{fig:MMT_spec1}.}
  \label{fig:MMT_spec2}
\end{figure*}

\begin{figure*}
  \epsscale{1.1}
  \plotone{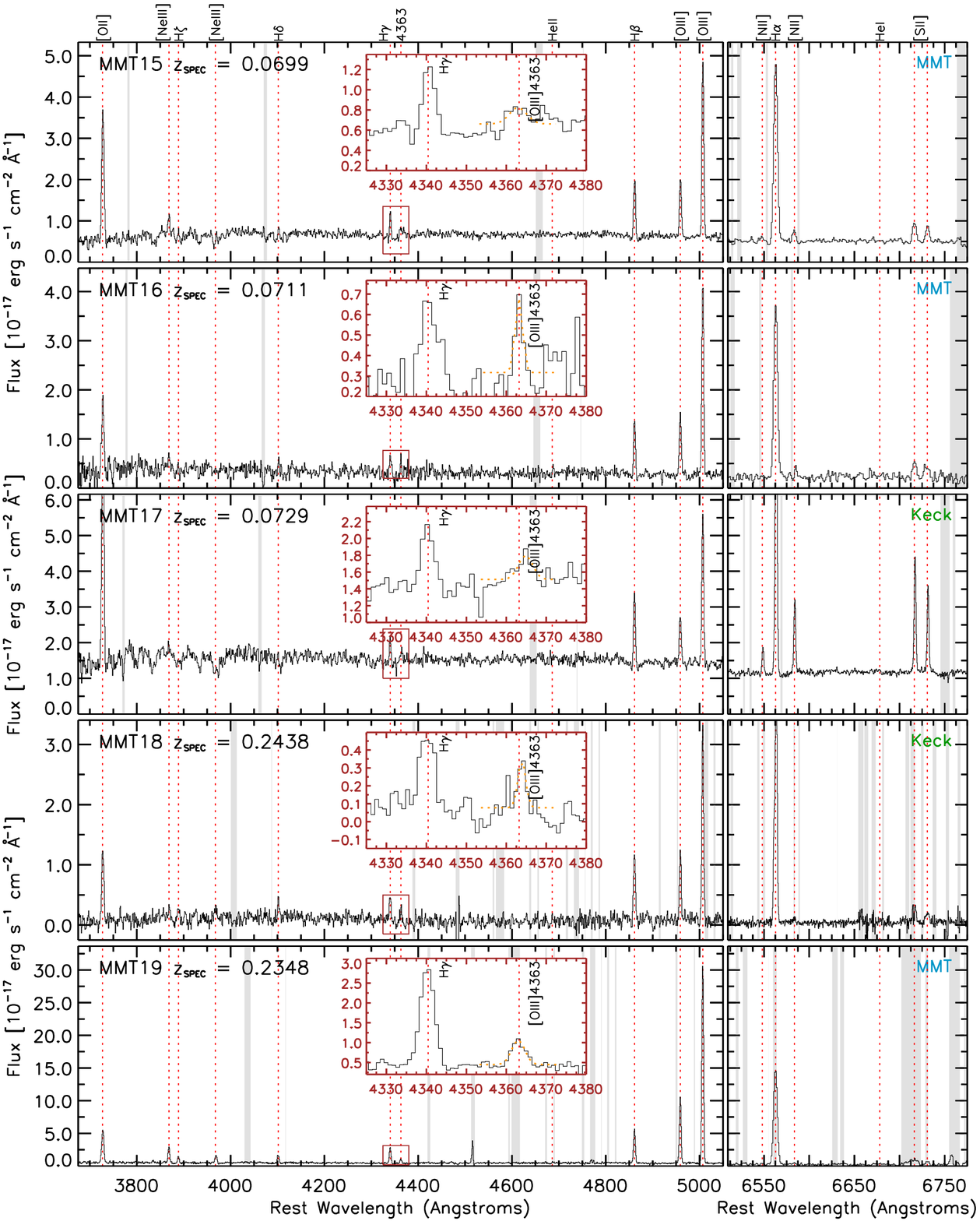}
  \caption{Same as Figures~\ref{fig:MMT_spec1}--\ref{fig:MMT_spec2}.}
  \label{fig:MMT_spec3}
\end{figure*}

\begin{figure*}
  \epsscale{1.1}
  \plotone{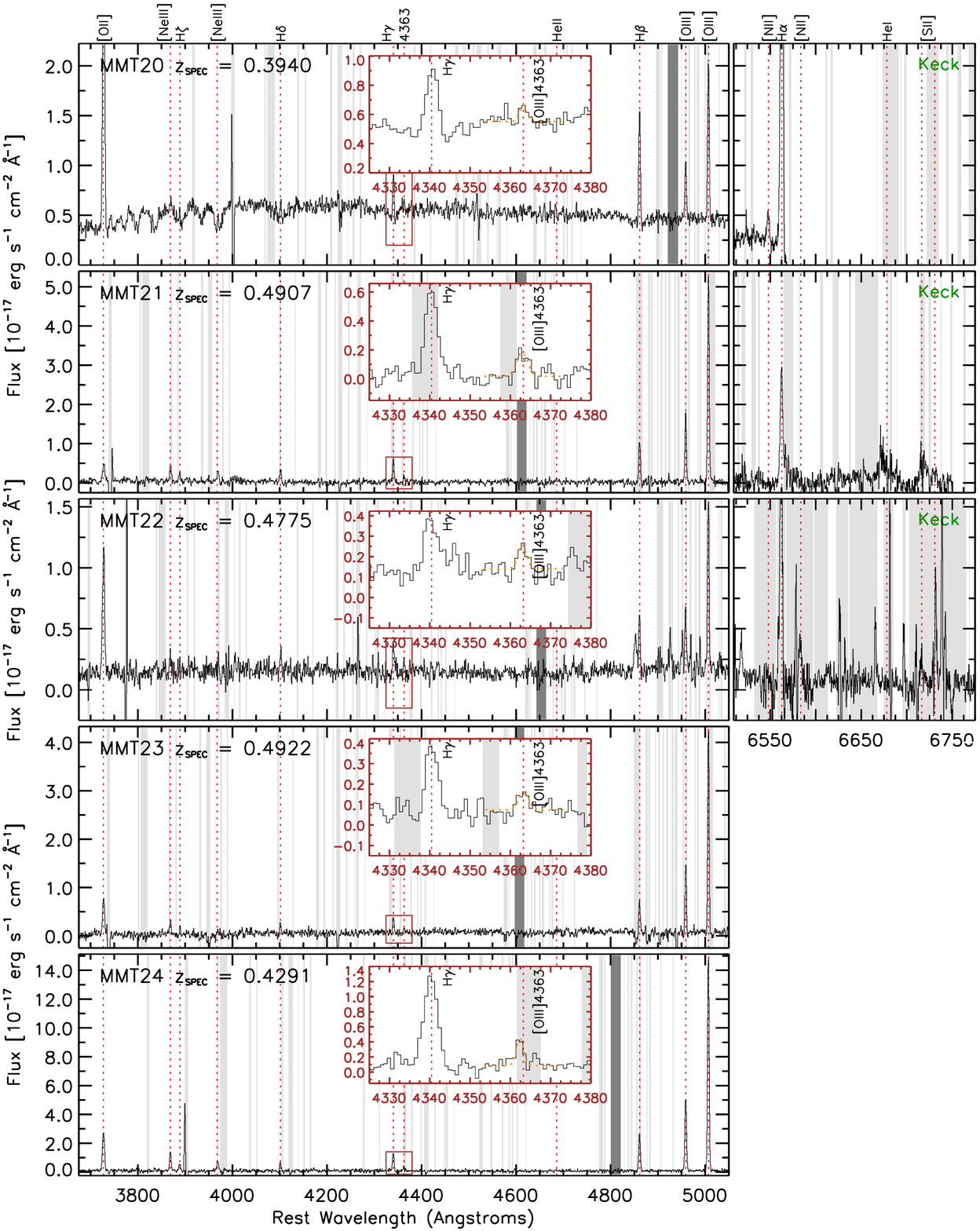}
  \caption{Same as Figures~\ref{fig:MMT_spec1}--\ref{fig:MMT_spec3}.}
  \label{fig:MMT_spec4}
\end{figure*}

\begin{figure*}
  \epsscale{1.1}
  \plotone{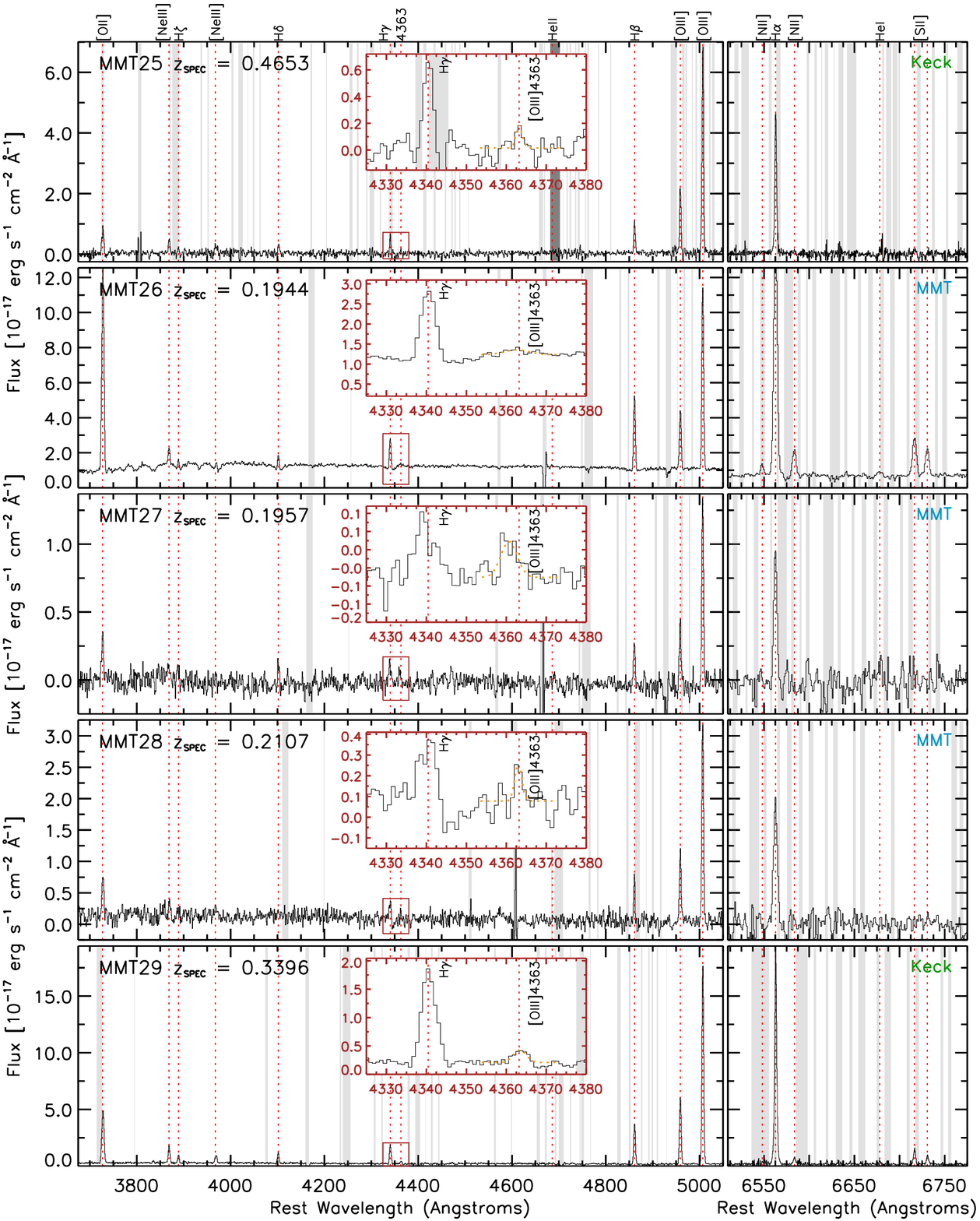}
  \caption{Same as Figures~\ref{fig:MMT_spec1}--\ref{fig:MMT_spec4}.}
  \label{fig:MMT_spec5}
\end{figure*}

\begin{figure*}
  \epsscale{1.1}
  \plotone{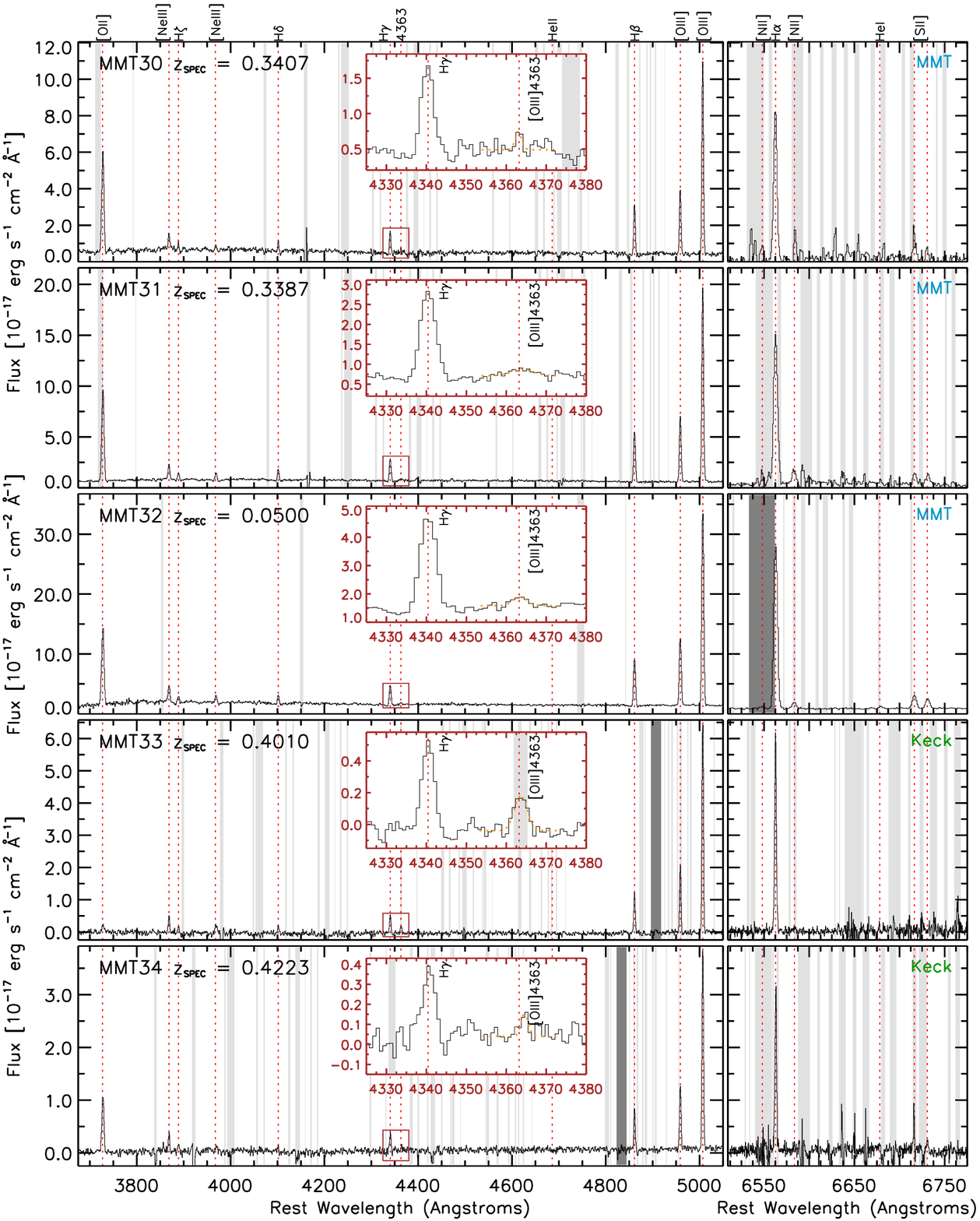}
  \caption{Same as Figures~\ref{fig:MMT_spec1}--\ref{fig:MMT_spec5}.}
  \label{fig:MMT_spec6}
\end{figure*}

\begin{figure*}
  \epsscale{0.819}
  \plotone{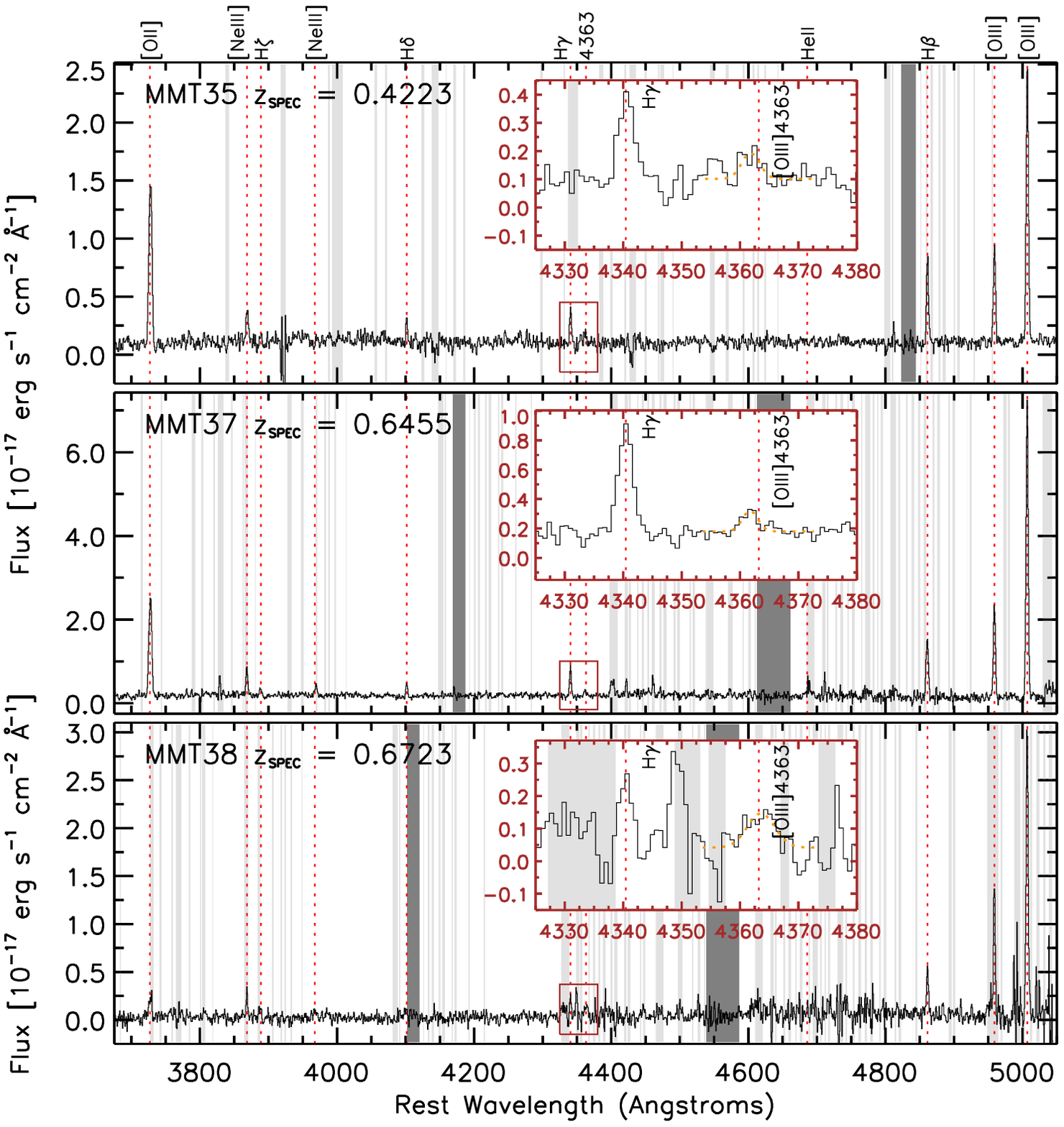}
  \caption{Same as Figures~\ref{fig:MMT_spec1}--\ref{fig:MMT_spec6}.}
  \label{fig:MMT_spec7}
\end{figure*}

\begin{figure*}
  \epsscale{0.819}
  \plotone{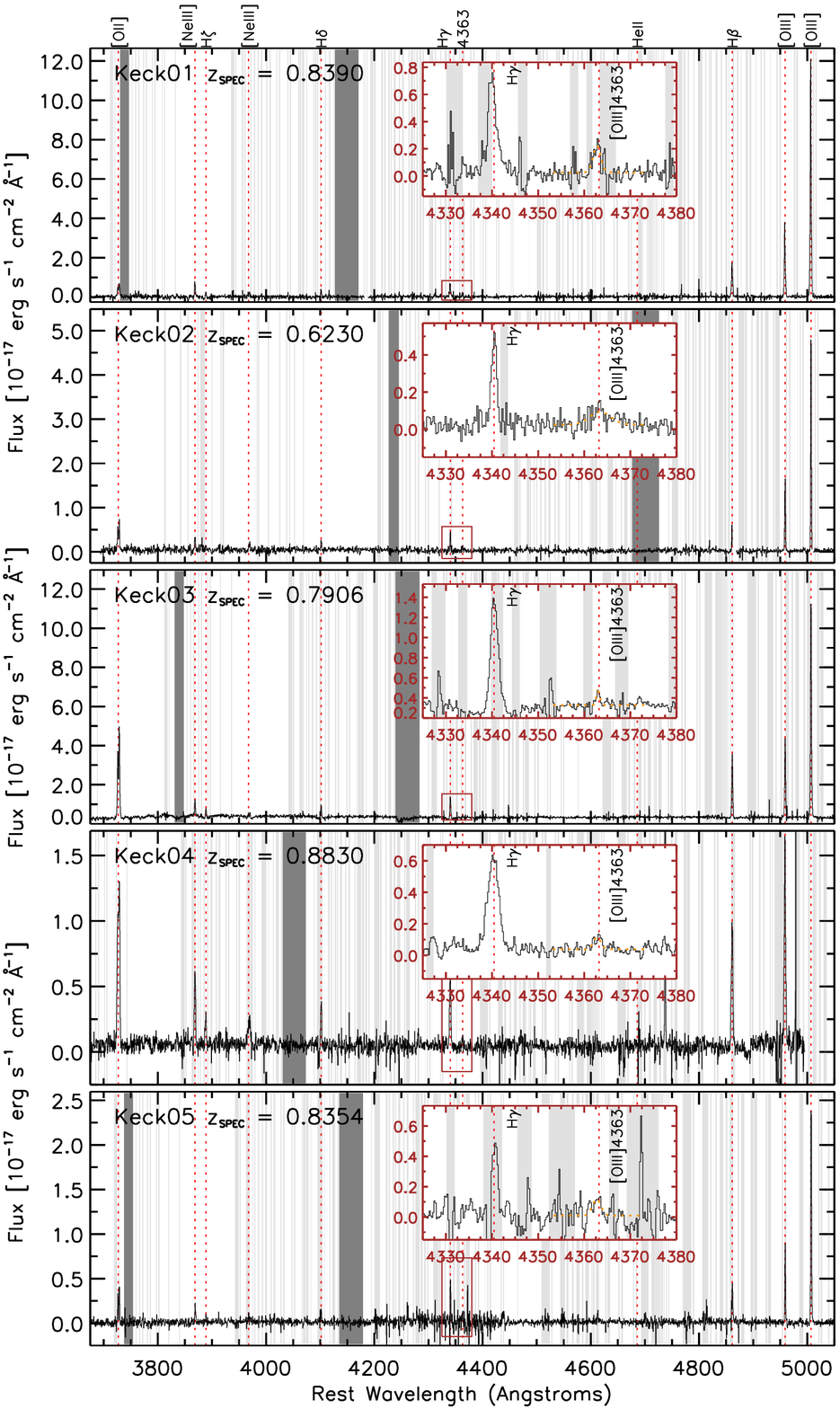}
  \caption{Keck spectra for a subset of \OIIIa-detected galaxies from the SDF. The
    blue-ends of the spectra, spanning \OII$\lambda$3727 to \Hb+\OIII$\lambda$5007, are
    shown on the left. Where available, we also illustrate (on the right) the red-ends
    of the spectra that span \Ha+\NII$\lambda\lambda$6548,\,6583 and
    \SII$\lambda\lambda$6716,\,6731. The insets provide a closer view of \Hg\ and \OIIIa\
    (rectangular region outlined in brown) with the best Gaussian fits to the \OIIIa\
    lines shown by the dotted orange lines. For the left panels, we also overlay the MMT
    spectra (where available) in light blue when the Keck spectra lacks the bluest
    spectral coverage. The vertical light gray bands indicate spectral regions
    contaminated by night-sky lines while the darker gray bands indicate the telluric
    A- and B-bands. Nebular emission lines are indicated by the vertical dashed red lines.}
  \label{fig:Keck_spec1}
\end{figure*}

\begin{figure*}
  \epsscale{0.819}
  \plotone{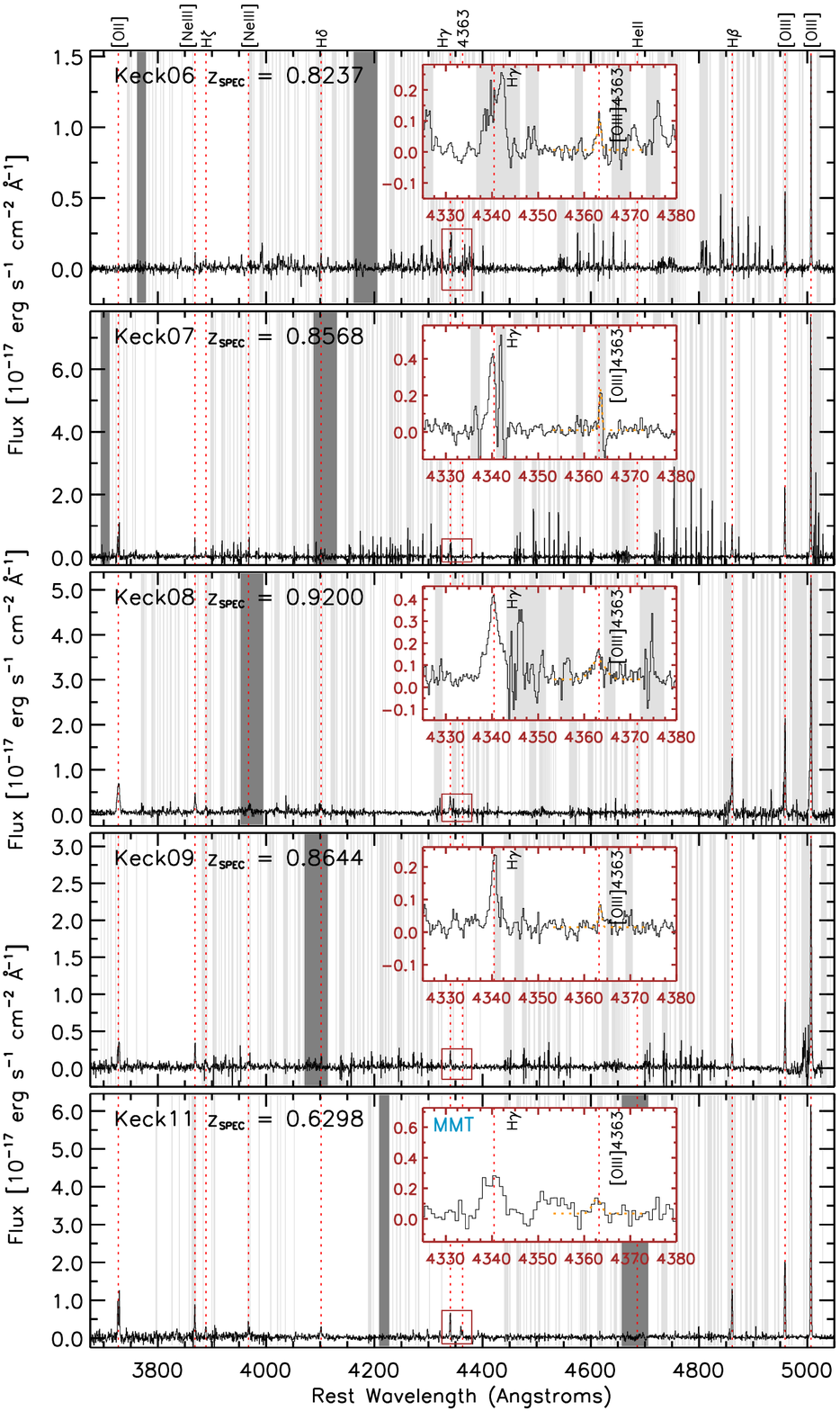}
  \caption{Same as Figure~\ref{fig:Keck_spec1}. The MMT spectrum is illustrated for
    the inset of Keck11 since \OIIIa\ is contaminated by a cosmic ray in the Keck
    spectrum.}
    \label{fig:Keck_spec2}
\end{figure*}

\begin{figure*}
  \epsscale{0.819}
  \plotone{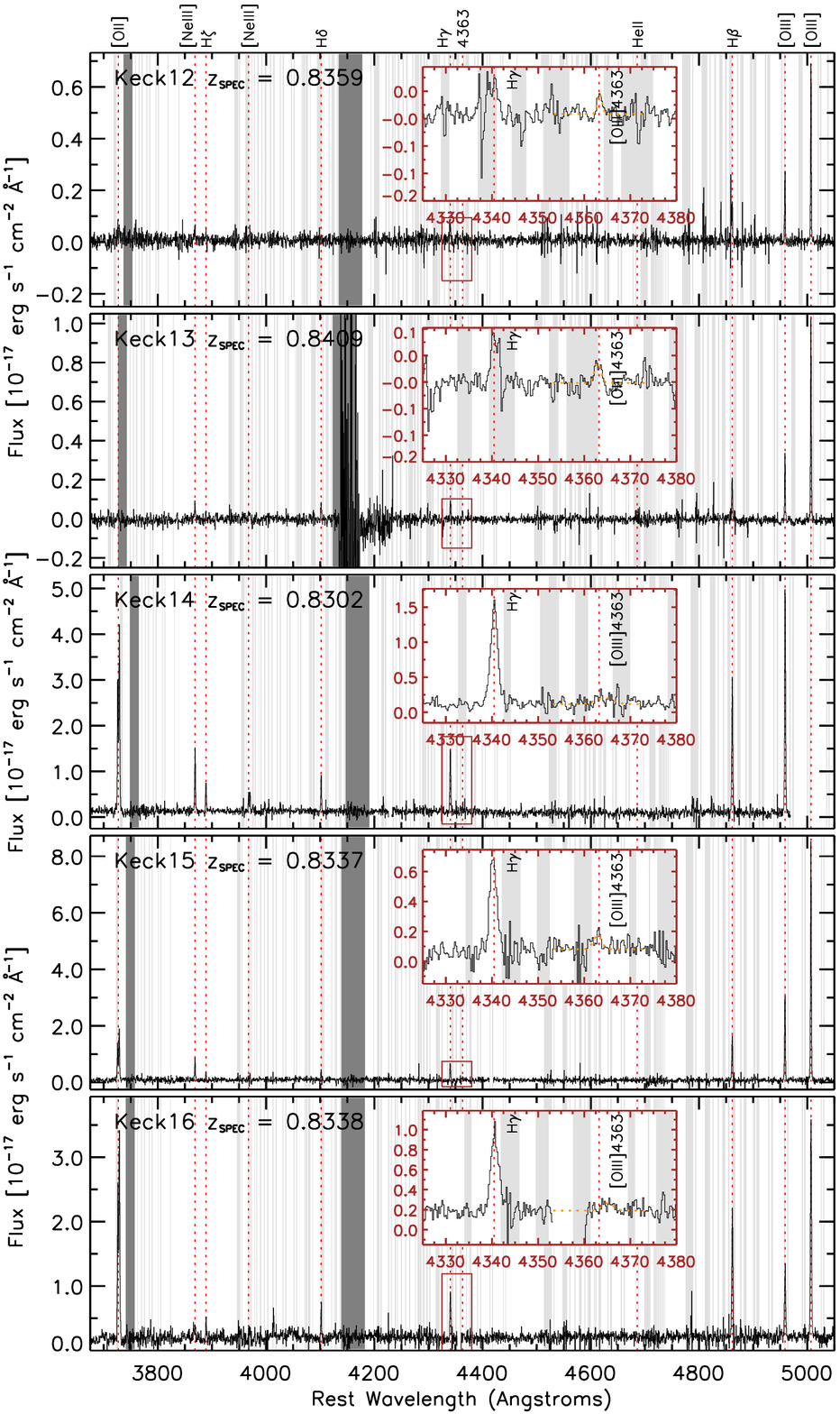}
  \caption{Same as Figures~\ref{fig:Keck_spec1}--\ref{fig:Keck_spec2}.}
  \label{fig:Keck_spec3}
\end{figure*}

\begin{figure*}
  \epsscale{1.1}
  \plotone{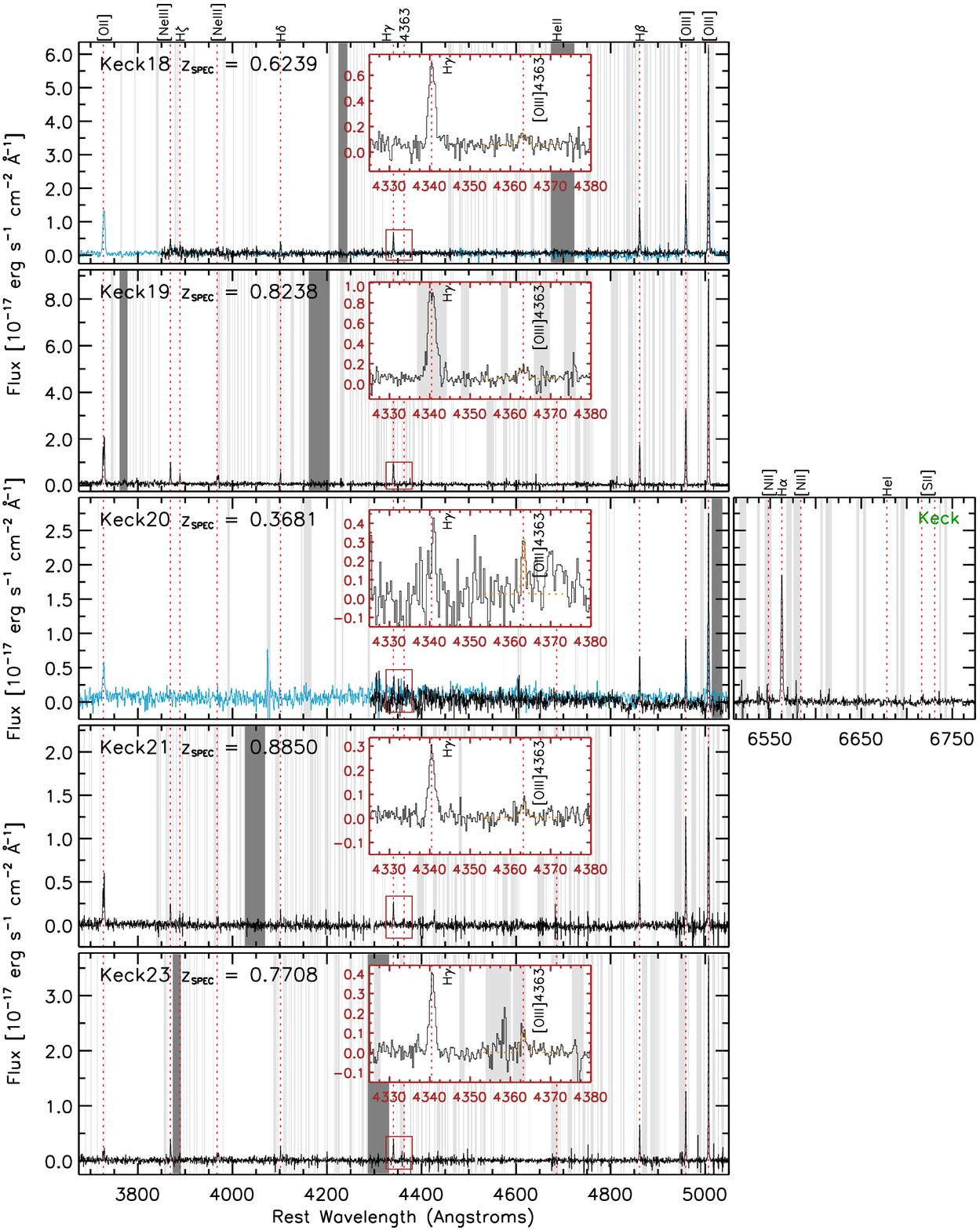}
  \caption{Same as Figures~\ref{fig:Keck_spec1}--\ref{fig:Keck_spec3}.}
  \label{fig:Keck_spec4}
\end{figure*}

\begin{figure*}
  \epsscale{1.1}
  \plotone{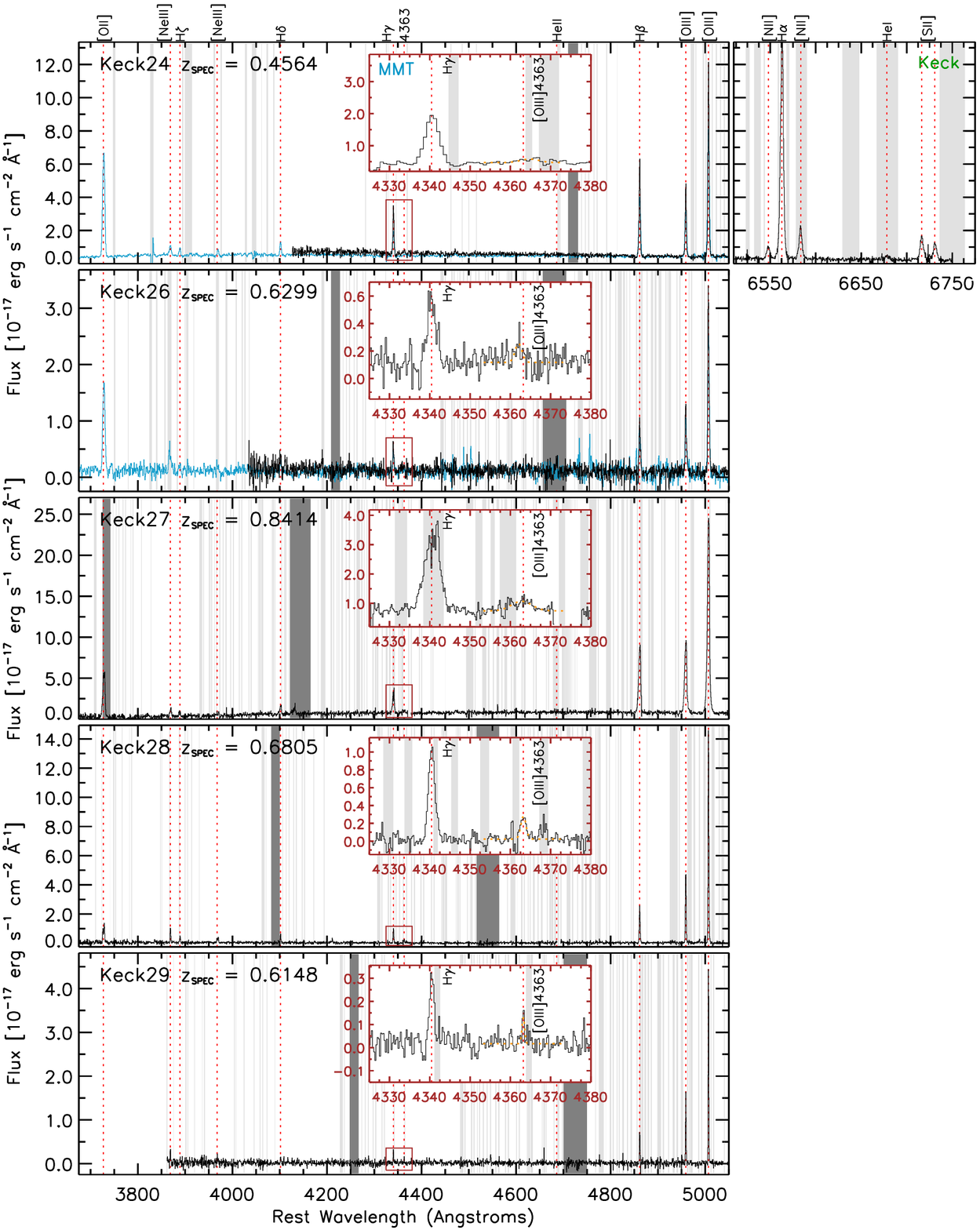}
  \caption{Same as Figures~\ref{fig:Keck_spec1}--\ref{fig:Keck_spec4}.}
  \label{fig:Keck_spec5}
\end{figure*}

\begin{figure*}
  \epsscale{1.1}
  \plotone{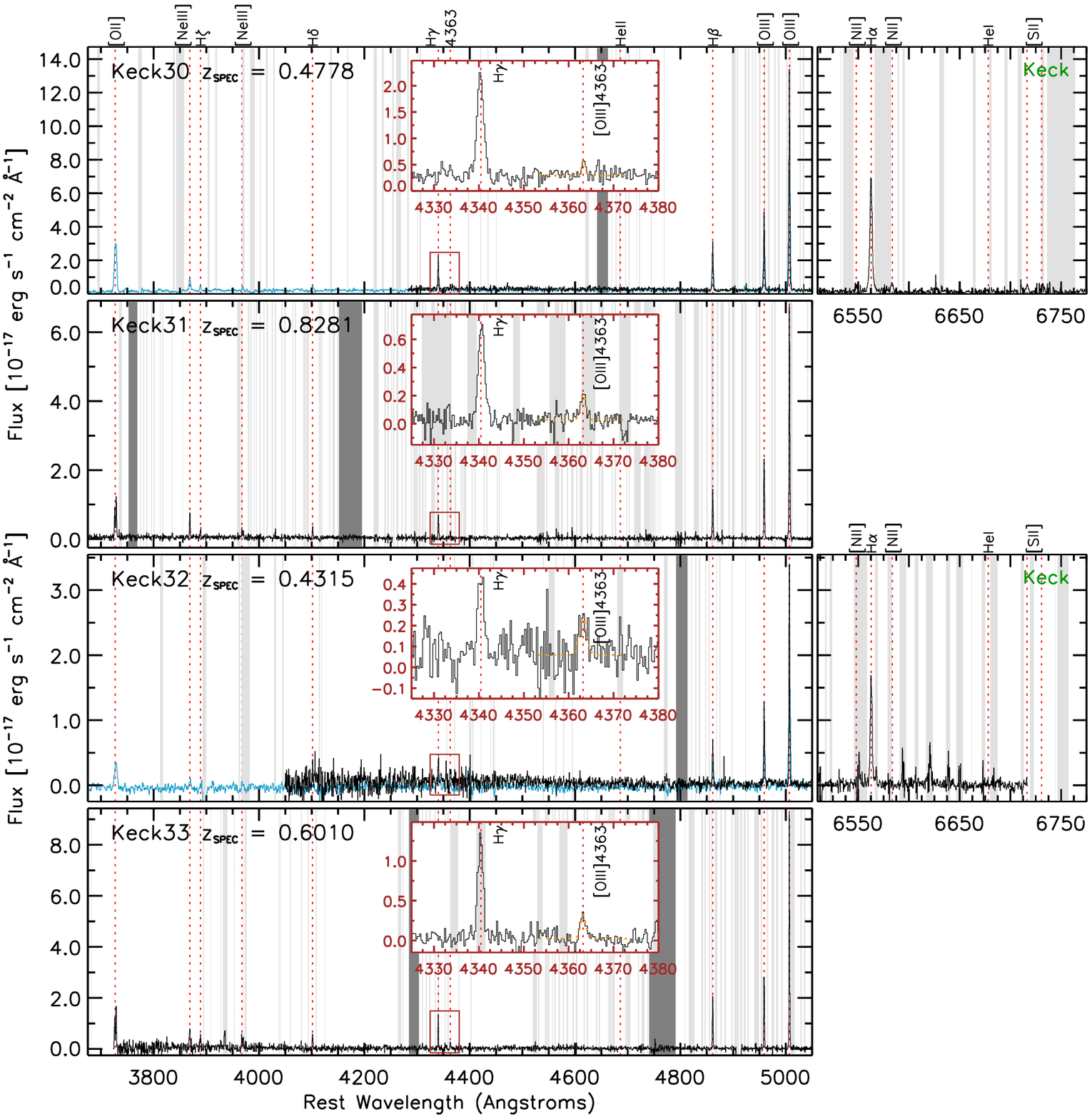}
  \caption{Same as Figures~\ref{fig:Keck_spec1}--\ref{fig:Keck_spec5}.}
  \label{fig:Keck_spec6}
\end{figure*}

\begin{figure*}
  \epsscale{1.1}
  \plotone{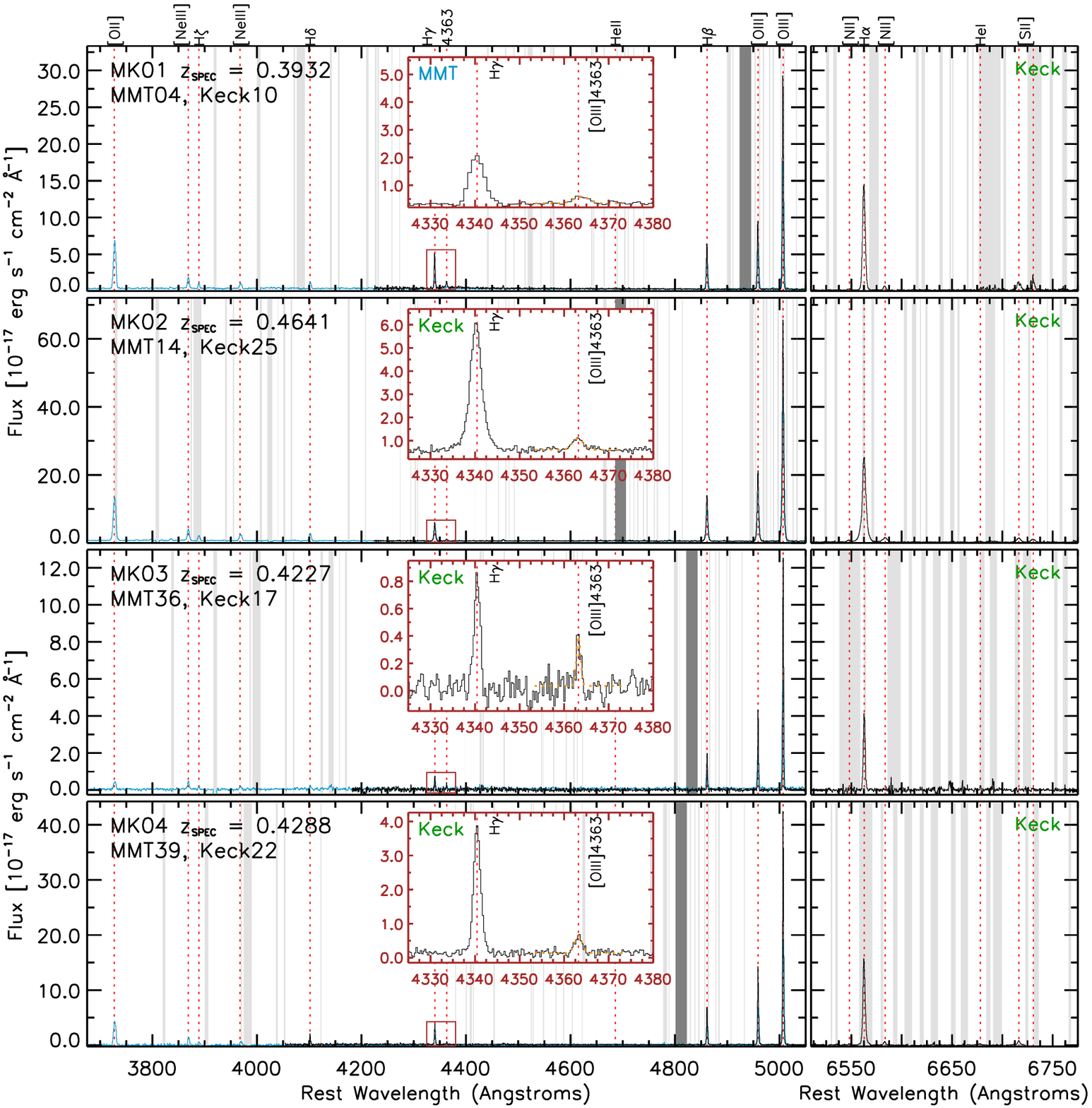}
  \caption{Same as Figures~\ref{fig:MMT_spec1}--\ref{fig:Keck_spec6} but for galaxies
    where \OIIIa\ is detected in both MMT and Keck spectra. The Keck spectra are
    illustrated in black (except for the inset panel for MK01), while MMT spectra are
    shown in light blue.}
  \label{fig:merged_spec}
\end{figure*}

Adopting a minimum significance threshold of 3$\sigma$, we identify \NOIIIMMT\ and
\NOIIIKeck\ \OIIIa\ detections with MMT and Keck, respectively.
We visually inspected each \OIIIa\ detection. For MMT, we find that OH skylines
contaminated \OIIIa\ in 14 cases and \Hb\ in one other galaxy. For two galaxies, we
lack spectral coverage of \OIII, and for another 13 galaxies, the \OIIIa\ detections
are marginal or are likely affected by cosmic rays.
Cosmic rays are easily identified when narrow peaks (less than a spectral resolution
element) are seen. We classify detections as marginal when a visual comparison of the
strength of the \OIIIa\ line against the local rms in the spectra indicated that such
detections may be misjudged as being above S/N = 3.
This results in a final MMT sample of \NMMTf\ \OIIIa\ detections. This sample includes
12 of the 14 MMT detections from our previous study \citepalias{ly14}, with two galaxies
(MMT06 and MMT12) not meeting the \OIIIa\ S/N = 3 cut.\footnote{MMT12 is in the
  \OIIIa-non-detected sample (see Section~\ref{sec:reliable}).}
For Keck, OH skylines contaminated \OIIIa\ in 55 cases and \OII\ measurements in one
other case, while 20 galaxies lack full spectral coverage (missing \OII, \Hb, and/or
\OIII\,$\lambda\lambda$4959,\,5007),\footnote{There are four galaxies (Keck02, 04, 14, 29)
  that we include in our sample for various reasons, discussed in \citetalias{ly14} and
  Table~\ref{tab:em_lines_strong}.} and 10 spectra were affected by cosmic rays on
\OIIIa\ or exhibited AGN properties (e.g., broad emission lines).
This reduced the \OIIIa\ Keck sample to \NKeckf\ galaxies. This includes all six Keck
detections from our previous study \citepalias{ly14}.
A subset of the MMT and Keck \OIIIa\ detections are shown in Figure~\ref{fig:oiii_spec}.
The full spectra are provided in Figures~\ref{fig:MMT_spec1}--\ref{fig:MMT_spec7}
for MMT and Figures~\ref{fig:Keck_spec1}--\ref{fig:Keck_spec6} for Keck.
Four of our \OIIIa-detected galaxies have both MMT and Keck spectra, providing independent
confirmations. Thus, our final sample of \OIIIa\ detections consists of \Ndet\ galaxies.
We refer to the overlapping cases, from MMT and Keck, as the merged sample with IDs
of MK01 to MK04. The MMT and Keck spectra for these galaxies are shown in
Figure~\ref{fig:merged_spec}. A summary of our \OIIIa-detected sample is provided in
Tables~\ref{tab:MMT_source_summary} and \ref{tab:Keck_source_summary}, and we provide
the emission-line fluxes for strong and weak emission lines in
Tables~\ref{tab:em_lines_strong} and \ref{tab:em_lines_weak}.

\begin{figure*}
  \epsscale{1.15}
  \plotone{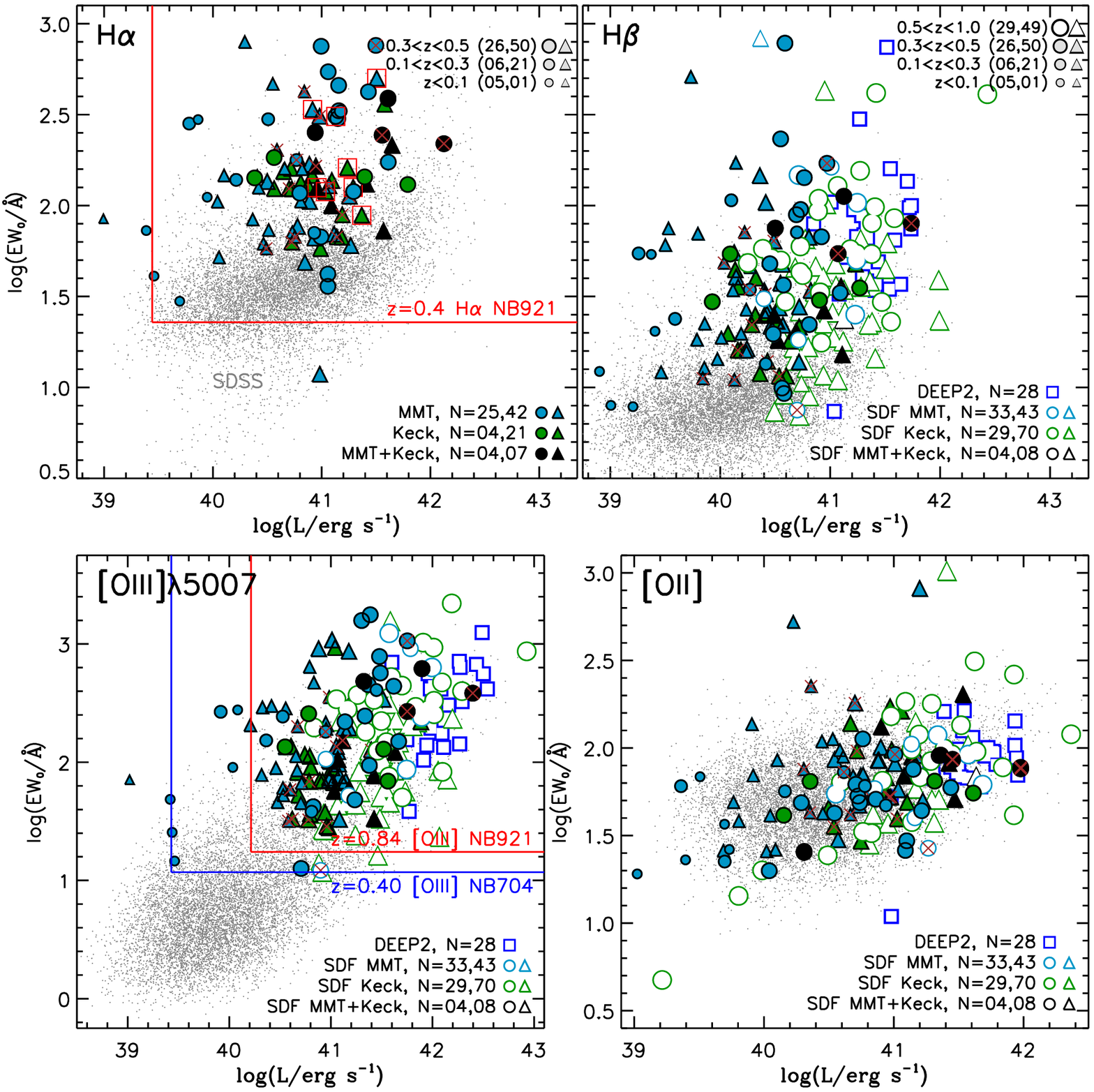}
  \caption{Emission-line luminosities and rest-frame EWs for our \OIIIa-detected (circles)
    and \OIIIa-non-detected (triangles) samples from MMT (light blue), Keck (green),
    and both (black). The triangles show galaxies that have a lower limit on \Te-based
    metallicity. Filled points are galaxies with \Ha\ measurements. The red squares in
    the top left panel identify those with \Ha\ fluxes from NB921 or NB973 excess fluxes.
    All fluxes are observed before correction for dust attenuation. The size of the data
    points indicates the redshift of the galaxies. Gray points illustrate the SDSS DR7
    sample with at least 5$\sigma$ detections in all four emission  lines. In addition,
    we overlay the $z\sim0.8$ \cite{ly15} DEEP2 \OIIIa-detected sample as dark blue
    squares. 
    Compared to local galaxies, the \OIIIa-detected and \OIIIa-non-detected samples
    consist of galaxies with higher emission-line EWs and luminosities. While this bias
    toward stronger nebular emission exists, it can be seen that the samples span 1.5 dex
    in EWs. Also, the \OIIIa-detected and \OIIIa-non-detected samples probe lower
    luminosities than \cite{ly15}. The wide range in EW and luminosity is due to deep
    spectroscopy.
    For comparison, we show the EW and luminosity limits of our NB921 imaging (solid red
    lines) for $z=0.4$ \Ha\ (top left) and $z=0.84$ \OIII\ (bottom left) and NB704 imaging
    (solid blue lines) for $z=0.4$ \OIII. AGNs and LINERs are indicated by red crosses.}
  \label{fig:EW_Lum}
\end{figure*}

Emission-line luminosities and rest-frame equivalent widths (EW$_0$) are illustrated in
Figure~\ref{fig:EW_Lum}.
The emission-line luminosities are primarily determined from spectroscopy, with the
exception of a subset of galaxies where \Ha\ measurements are unavailable (indicated
with red squares in Figure~\ref{fig:EW_Lum}), so we use either the NB921 or NB973 excess
measurements.
The line luminosity is $L = 4\pi d_L^2 F_{\rm Line}$, where $d_L$ is the luminosity distance,
and $F_{\rm Line}$ is the emission-line flux. The EWs are determined from the ratio of
the measured emission-line flux and the continuum. Since the continuum is not always
well-measured from spectra, we measure it from the broadband SEDs, with corrections for
emission-line contamination (see Section~\ref{sec:SED}).
In Figure~\ref{fig:EW_Lum}, we compare our \OIIIa-detected sample against local galaxies
from SDSS (gray points). Since \citetalias{MACTII} will also incorporate metal-poor
galaxies with \OIIIa-based metallicities from the DEEP2 Survey \citep{ly15}, we also
overlay these DEEP2 galaxies as blue squares. Compared to local galaxies, the
\OIIIa-detected sample consists of galaxies with higher emission-line EWs and luminosities.
While this bias exists, it can be seen that the \OIIIa-detected sample spans 1.5 dex in
EWs. Additionally, the \OIIIa-detected sample probes lower luminosities than \cite{ly15}.
The wide range in EW and luminosity is due to the deep spectroscopy of \Sname.


\subsection{The \OIIIa-non-detected Sample}
\label{sec:reliable}
An advantage of our study is the significant number of deep optical spectra (see Figure
\ref{fig:int_time}). They supplement our primary sample of \OIIIa\ detections, by
providing reliable {\it upper limits} on \Te.
The construction of a sample of reliable non-detections reduces the selection bias of
targeting more strongly star-forming and/or metal-poor galaxies in our primary sample,
which we discuss in Section~\bias\ of \citetalias{MACTII}. We construct our reliable
\OIIIa-non-detected samples as follows.

First, we estimate the sensitivity for each spectrum based on the rms in the continuum
near \OIII$\lambda$5007. The sensitivity varies by a factor of 2--3 between the deepest
and shallowest observations (see Figure \ref{fig:int_time}).
Due to the inhomogeneity in the observations, we have chosen to adopt a \OIII$\lambda$5007
line flux limit that, to first order, corresponds to a minimum S/N of 100. By adopting a
flux limit, rather than a S/N limit, it is more straightforward to model the selection
function.
Adopting a flux limit also includes galaxies with very high \OIII$\lambda$5007 fluxes,
but have shallow spectra. This results in a less biased selection than a S/N cut.

The flux limit is determined by considering the distribution of
100 $\times$ $\sigma$(\OIII$\lambda$5007) in several redshift bins. To minimize selection
bias, we adopt a less restrictive \OIII$\lambda$5007 flux limit that encompasses 85\% of
galaxies in each redshift bin. These flux limits for MMT and Keck are provided in
Table~\ref{tab:nondet_sel}.

\begin{figure*}
  \epsscale{1.1}
  \plottwo{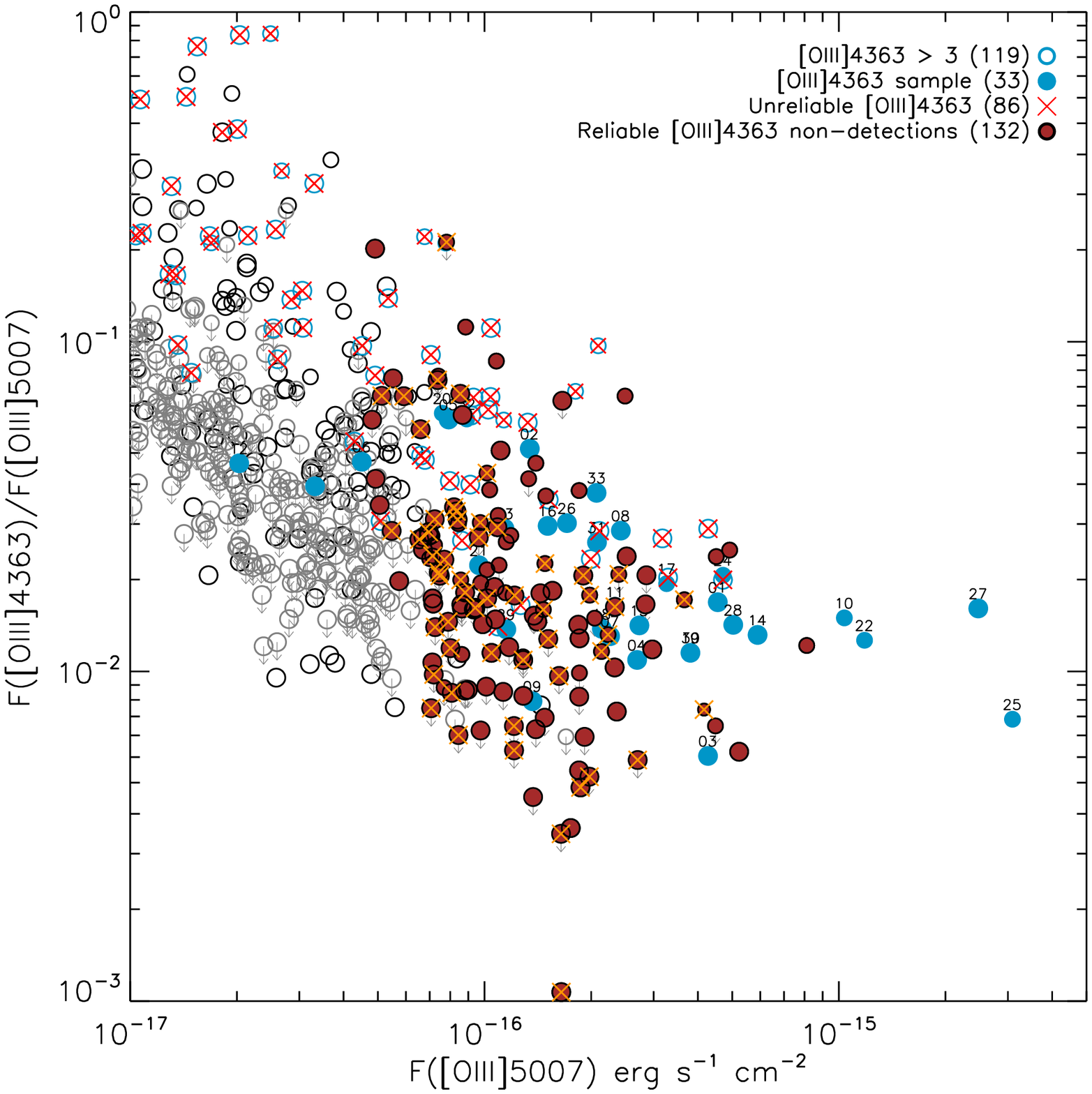}{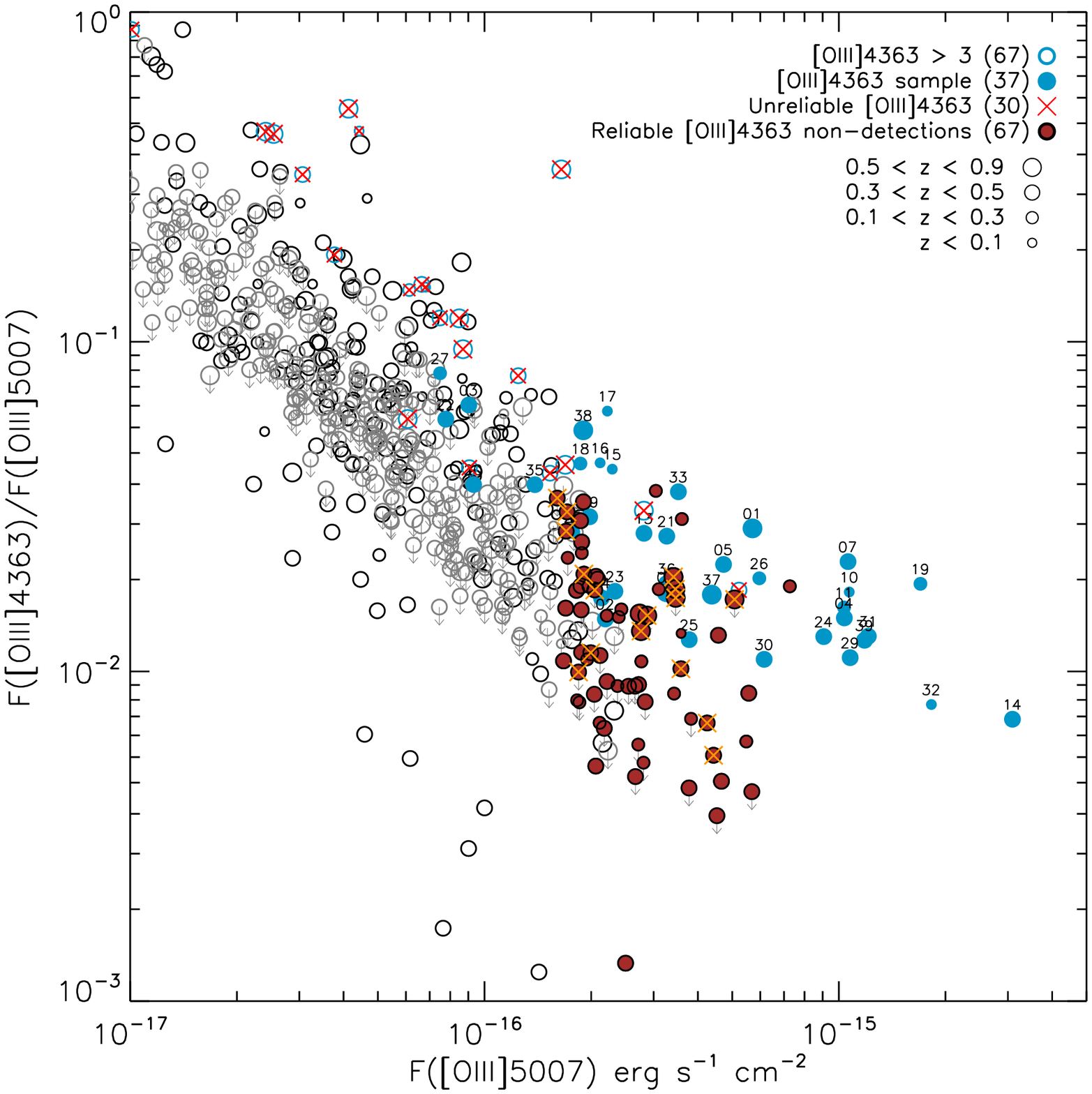}
  \caption{Comparison of the \OIIIa/\OIII\,$\lambda$5007 flux ratio against
    \OIII\,$\lambda$5007 line flux for Keck (left) and MMT (right) spectra. The samples
    with \OIIIa\ detections at $\geq3\sigma$ (i.e., the \OIIIa-detected sample) are
    illustrated in filled blue circles. Those excluded by visual examination of the
    spectra are indicated by red crosses on unfilled blue circles. The \OIIIa-non-detected
    samples are shown by the brown circles with yellow crosses indicating those excluded
    by visual examination of spectra. Arrows indicate \OIIIa\ 1$\sigma$ limits, and
    symbol sizes indicate redshift. This figure illustrates that by including reliable
    \OIIIa\ non-detections, high-EW galaxies that are more metal-rich than the
    \OIIIa-detected galaxies are incorporated in our study.}
  \label{fig:nondet}
\end{figure*}

These selections initially give us 67 galaxies from MMT and 132 from Keck. We then visually
inspected the spectra for these galaxies and excluded those where \OIIIa\ is compromised by
nearby night skylines or telluric bands. This reduced the sample to 51 MMT and 78 Keck
galaxies, eight of which are in both \OIIIa-non-detected samples. We refer to these eight
galaxies as MK05 to MK12.  We illustrate the \OIIIa-non-detected samples in
Figure~\ref{fig:nondet}. Compared to galaxies with \OIIIa\ detections, these plots
show that our non-detection limits extend to lower \OIIIa/\OIII$\lambda$5007 flux ratio,
and hence lower \Te\ (higher metallicity).
A summary of our \OIIIa-non-detected sample is provided in
Tables~\ref{tab:MMT_source_summary_nondet} and \ref{tab:Keck_source_summary_nondet}.

The above \OIII\ flux limits are adopted to define the \OIIIa-non-detected samples.
For analyses involving the \MZ\ relation and its dependence on SFR, which are presented
in \citetalias{MACTII}, we adopt stricter flux limits\footnote{Encompass 75\% of
  the distribution in 100 $\times$ $\sigma$(\OIII$\lambda$5007).}
to avoid galaxies with weaker emission lines. These stricter flux limits and the sample
sizes are provided in Table~\ref{tab:nondet_sel} in parentheses. The \OIIIa-non-detected
samples with the stricter flux limits consist of 40 galaxies from MMT and 63 from Keck,
with five galaxies in both \OIIIa-non-detected samples. For clarity, we indicate those
that are excluded with the stricter flux limit in
Tables~\ref{tab:MMT_source_summary_nondet} and \ref{tab:Keck_source_summary_nondet}.


\section{DERIVED PROPERTIES}
\label{sec:Prop}


\subsection{Dust Attenuation Correction from Balmer Decrements}
\label{sec:dust}

To correct the emission-line fluxes for interstellar reddening, we use Balmer decrement
measurements (\Ha/\Hb\ and \Hg/\Hb) obtained from a combination of our spectroscopy and
narrowband imaging. Since the lines in our \OIIIa-detected emission-line galaxies tend
to have high EWs, 42, 59, and \Ndet\ galaxies have \Hd, \Hg, and \Hb\ detected at
$\gtrsim$10$\sigma$, respectively.
In addition, \Ha\ measurements are available for \NHa\ galaxies (i.e., half of the sample).
For MMT23, the \Ha\ flux is available from NB973 excess measurement.
We note that for MMT03, the NB921 flux is not reliable for Balmer decrement
determinations; MMT03 may suffer from significant \NII\ contamination in the NB921
filter as a LINER candidate.

\begin{figure*}
  \epsscale{0.55}
  \plotone{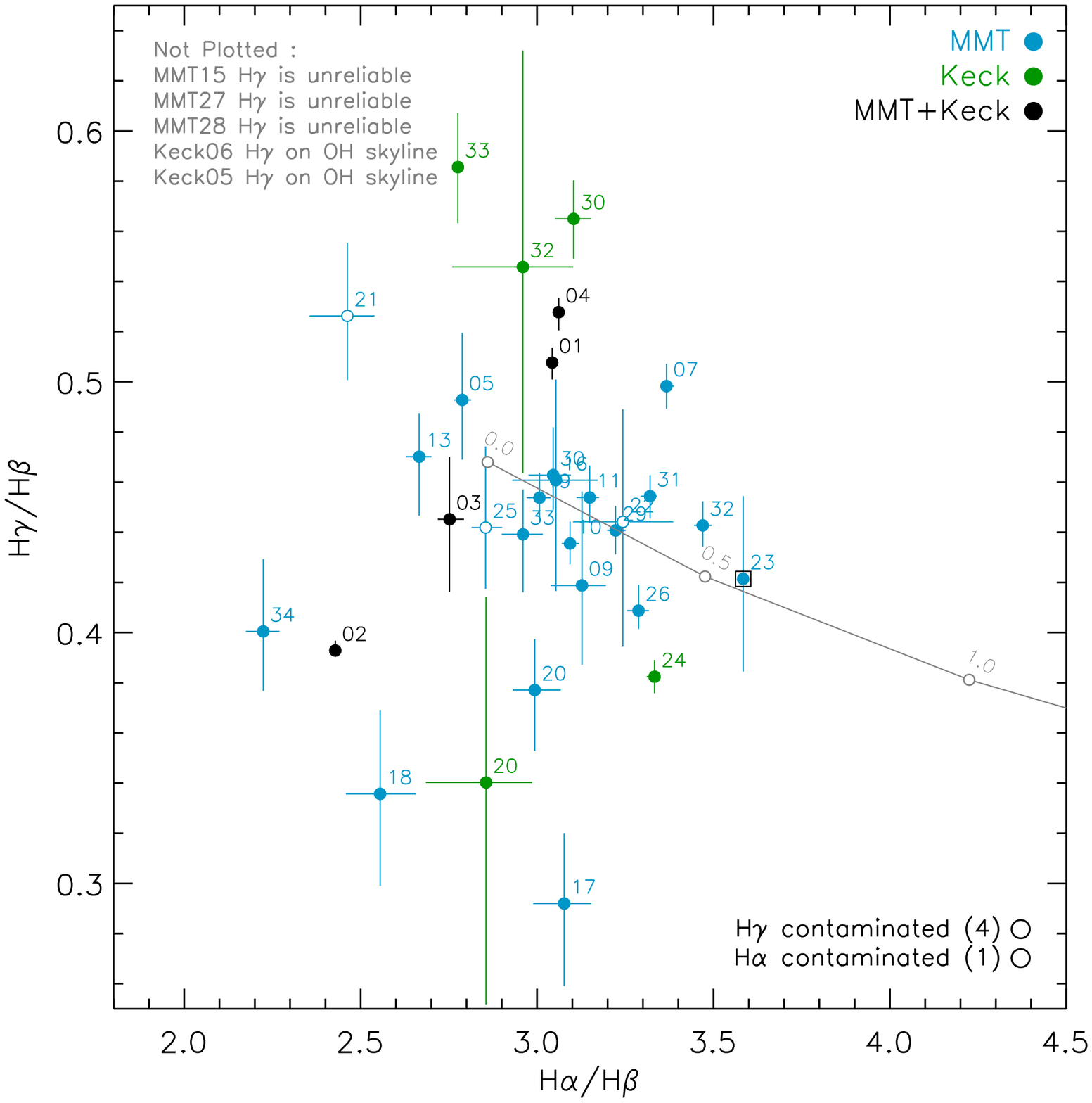}
  \plotone{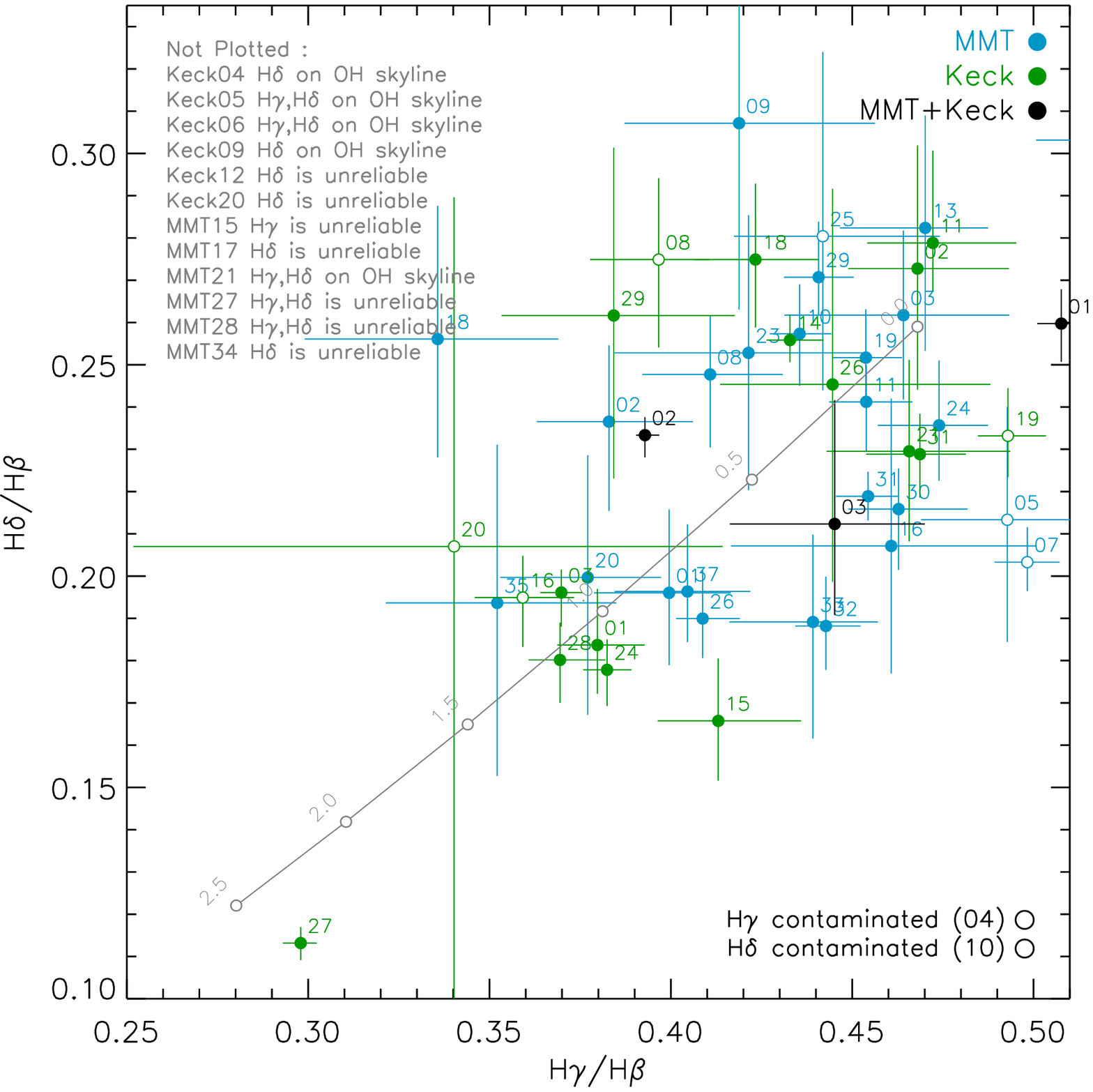}
  \caption{Reddening estimation by using Balmer emission-line ratios from the MMT
    (light blue), Keck (green), and merged (black) \OIIIa-detected samples. The left 
    panel shows those galaxies at low enough redshifts to have observed \Ha\ fluxes (\Ha\
    measurement from NB973 imaging is indicated by the black square). The \Ha/\Hb\ ratio
    is plotted against the second most reliable reddening indicator, \Hg/\Hb. The right
    panel instead compares \Hg/\Hb\ with \Hd/\Hb.
    The error bars show Balmer decrements for individually labeled galaxies.  The curves
    show the expected values assuming an intrinsic Case B recombination ratio and a
    reddening law of \cite{car89}.  Tick marks indicate increments of 0.5 mag in A(\Ha).
    Galaxies without reliable Balmer decrement measurements due to contamination from
    OH night skylines are denoted with open circles (see Table~\ref{tab:balmer} for
    further details).}
  \label{fig:Balmer}
\end{figure*}

A significant problem encountered when using Balmer decrements to determine dust attenuation
is correcting for the underlying stellar absorption. For a subset of our \OIIIa-detected
sample, the spectra have sufficient S/N on the continuum to fit and remove the stellar
absorption, or use \textsc{iraf}'s splot command to re-measure the continuum from the
absorption trough. For the remaining galaxies, the S/N of our spectra are insufficient
to model the stellar absorption. To correct these galaxies, we adopt
EW$_{\rm abs}$(\Hd) = 2 \AA\ and EW$_{\rm abs}$(\Hg) = 1 \AA. These values correspond to
thresholds where the amount of stellar absorption is difficult to measure from visual
examination.
We assume the stellar absorption under \Hb\ and \Ha\ is negligible. This is reasonable,
since the measured rest-frame emission-line EWs are very large (\Hb: median of 54\AA,
average of 82 \AA; see Figure~\ref{fig:EW_Lum}). The stellar absorption corrections
are provided in Table~\ref{tab:balmer}, and the Balmer decrements are illustrated in
Figure~\ref{fig:Balmer}.

Under the assumption of Case B recombination, the intrinsic Balmer flux ratios are:
(\Ha/\Hb)$_0$ = \HaHbi, (\Hg/\Hb)$_0$ = 0.468, and (\Hd/\Hb)$_0$ = 0.259 for
\Te\ = 10$^4$ K \citep{ost06}.
If dust attenuates these measurements, then the observed ratios are:
\begin{equation}
  \frac{(\Hyd n/\Hbe)_{\rm obs}}{(\Hyd n/\Hbe)_0} = 10^{-0.4\EBV[k(\Hyd n)-k(\Hbe)]},
  \label{eqn:balmer}
\end{equation}
where \EBVa\ is the {\it nebular} color excess, and $k(\lambda)\equiv A(\lambda)$/\EBVa\
is the reddening curve at $\lambda$. We adopt the reddening curve of \cite{car89},
which gives $k(\Hae)=2.535$, $k(\Hbe)=3.61$, $k(\Hge)=4.17$, and $k(\Hde)=4.44$,
illustrated in Figure~\ref{fig:Balmer} along with the observed Balmer decrements.
Adopting a \cite{cal00} reddening curve would not alter the metallicity by more than
$\sim$0.01 dex; it would increase the dust-corrected SFR estimates by an average
(median) of 0.05 dex (0.03 dex).

Although the scatter in individual galaxies is substantial, the extinctions inferred
from different pairs of Balmer lines (e.g., \Ha/\Hb\ vs. \Hg/\Hb) are generally
consistent with each other. Specifically, the median (average) difference in \EBVa\ is
--0.02 mag (--0.04 mag) with a dispersion of 0.25 mag. The average and median \EBVa\ for
our \OIIIa-detected sample are 0.13 and 0.07 mag, respectively.

For the \OIIIa-non-detected sample, the galaxies have lower emission-line EWs and thus
lower S/N on the Balmer lines for reliable individual measurements. We instead stacked
the MMT and Keck spectra and fitted the Balmer line profiles with a double Gaussian
(emission and stellar absorption). The stacking was performed in bins of \Hb\
emission-line luminosity, the highest S/N hydrogen recombination line available for
all galaxies.
Three (four) bins of 17 (19--20) spectra are used for the MMT (Keck) samples. Where
available, we use Keck measurements since those are more sensitive. The average (median)
reddening for the \OIIIa-non-detected sample is \EBVa\ = 0.14 (0.11). We provide the
Balmer decrements from spectral stacking in Table~\ref{tab:balmer_nondet}.


\subsection{\Te\ and Metallicity Determinations}
\label{sec:Te}
To determine the gas-phase metallicity for our galaxies, we first estimate \Te(\OIII)
using the auroral-to-nebular \OIII\ flux ratio, $\mathcal{R}$:
\begin{eqnarray}
  \mathcal{R} &\equiv& \frac{F(\lambda4363)}{F(\lambda5007)+F(\lambda4959)},\\
  T_e &=& a \left(-\log(\mathcal{R})-b\right)^{-c},{\rm~where}\label{eqn:R}
\end{eqnarray}
$a$ = 13205, $b$ = 0.92506, and $c$ = 0.98062 \citep{nic14}. These coefficients
utilized the latest collision strength data for O$^{++}$ \citep{pal12}. In general,
the above relation yields $T_e$ that are lower than \cite{izo06b} by 5\% \citep{nic13}.
We correct the nebular-to-auroral \OIII\ flux ratio for dust attenuation (see
Section~\ref{sec:dust}), which increases our estimated \Te.

\begin{figure*}
  \epsscale{1.0}
  \plottwo{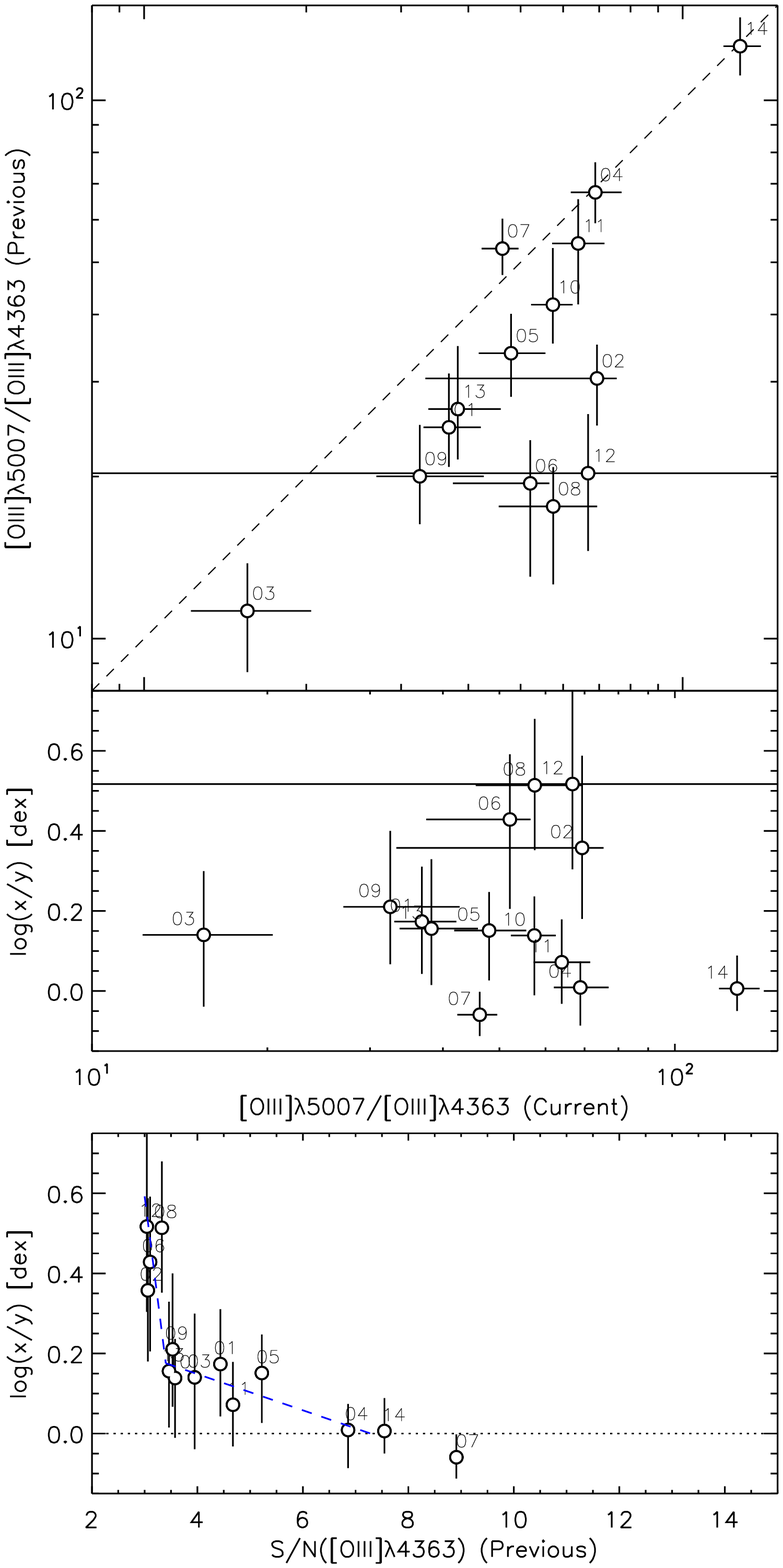}{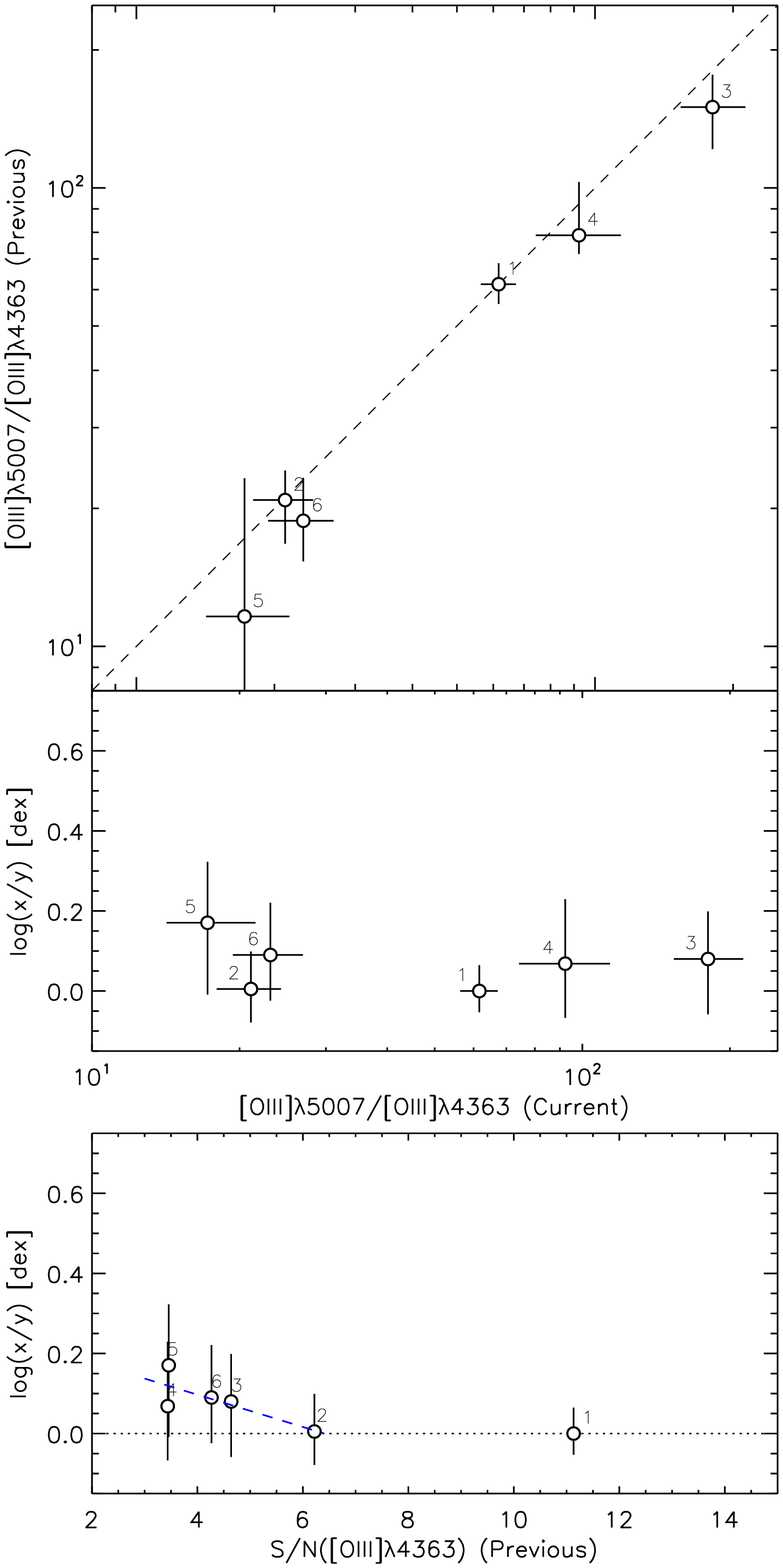}
  \caption{Comparisons between previous measurements \citepalias{ly14} and
    measurements from this paper of \OIII$\lambda$5007/\OIIIa\ (top panels) for MMT
    (left) and Keck (right) sources. The previous measurements underestimated the
    \OIII$\lambda$5007/\OIIIa\ ratio (i.e., overestimated \Te) due to \cite{edd13} bias
    by 0.3 and 0.1 dex for MMT and Keck, respectively. The middle panels illustrate
    this difference as a function of the current \OIII$\lambda$5007/\OIIIa\ measurements.
    The bottom panels illustrate this difference as a function of S/N(\OIIIa), and
    show the necessary statistical corrections to apply to the \OIII$\lambda$5007/\OIIIa\
    flux ratio, based on the measured S/N. The blue dashed lines are the least-squares
    fits, and are provided in Equations~(\ref{eqn9})--(\ref{eqn13}).}
    \label{fig:bias}
\end{figure*}

We also correct $\mathcal{R}$ for \cite{edd13} bias, which can boost the \OIIIa\ line
flux for galaxies near the adopted S/N = 3 limit. Since there is a direct relation
between $\mathcal{R}$ and \Te, this bias can result in higher \Te. We perform this
correction empirically, by comparing the measured $\mathcal{R}$ values of previous
observations of the \citetalias{ly14} sample of 20 \OIIIa-detected galaxies. These
comparisons are illustrated in Figure~\ref{fig:bias}. It can be seen that the
\OIIIa/\OIII\,$\lambda$5007 line ratio is elevated when the observations are shallow.
Using the measured \OIIIa\ S/N, we apply the following corrections to
$\log\left(\mathcal{R}^{-1}\right)$ for MMT measurements:
\begin{eqnarray}
  3.718 - 1.042x {\rm~dex},~(3.00\leq &x& < 3.4),\label{eqn9}\\
  0.329 - 0.045x {\rm~dex},~(3.40\leq &x& \leq7.28),\\
             0.0 {\rm~dex},~(7.28 < &x&),
\end{eqnarray}
where $x$ = S/N(\OIIIa). For Keck measurements, the corrections are:
\begin{eqnarray}
  0.258 - 0.040x {\rm~dex},~(3.00\leq &x& \leq 6.41),\\
             0.0 {\rm~dex},~(6.41 < &x&).\label{eqn13}
\end{eqnarray}

Our \OIIIa\ measurements have a large dynamic range: the strongest \OIIIa\ line is
as much as 0.065 of the \OIII\ flux (MMT03 and Keck05), while the weakest is 0.007
(Keck03 and MK02). We find that the average and median $\lambda$4363/$\lambda$5007
flux ratios for our sample are 0.023 and 0.018, respectively.
The derived electron temperatures for our galaxy sample span $1\times10^4$ K to
$3.4\times10^4$ K. \Te\ measurements for all \Ndet\ objects are tabulated in
Table~\ref{tab:metals}.

For the \OIIIa-non-detected sample, the lack of a detection yields an upper limit on
\Te. We find from visual examination that spectra with \OIIIa\ detections at a level of
S/N = 2--3 are marginal, while those below S/N = 2 are weaker or undetected. For this
reason, we determine upper limit $\mathcal{R}$ values by adopting \OIIIa\ fluxes that
correspond to S/N = 2.5 for those with $2 \leq {\rm S/N} < 3$, 1.5 for
$1 \leq {\rm S/N} < 2$, and 1.0 for ${\rm S/N} < 1$. For the \OIIIa-non-detected sample,
we do not apply the \cite{edd13} bias correction because the lack of a detection suggests
that the bias is weak. The derived upper limits on \Te\ span $0.8\times10^4$--$3\times10^4$
K, and are provided in Table~\ref{tab:metals_nondet}.

Throughout this paper, we generate multiple realizations of the emission-line fluxes
to construct probability distribution functions  for all observed and derived
measurements. This is critical, as the distributions are non-Gaussian in the domain of
low S/N ratio \citepalias[$\lesssim$5;][]{ly14}, and it allows us to propagate our
measurement uncertainties, including \EBVa.


To determine the ionic abundances of oxygen, we use two emission-line flux ratios,
\OII\,$\lambda\lambda$3726,3729/\Hb\ and \OIII\,$\lambda\lambda$4959,5007/\Hb\
\citep{izo06b}:
\begin{eqnarray}
  12+\log{\left(\frac{{\rm O}^+}{{\rm H}^+}\right)} = \log{\left(\frac{\OII}{{\rm H}\beta}\right)}
  + 5.961 + \frac{1.676}{t_2}\\
  \nonumber
  - 0.4\log{t_2} - 0.034t_2+\log{(1+1.35x)}\\
  12+\log{\left(\frac{{\rm O}^{++}}{{\rm H}^+}\right)} = \log{\left(\frac{\OIII}{{\rm H}\beta}\right)}
  + 6.200+ \frac{1.251}{t_3}\label{eqn:O++}\\
  - 0.55\log{t_3} - 0.014t_3.
  \nonumber
\end{eqnarray}
Here, $t_2\equiv T_e$(\OII)/10$^4$ K and $t_3\equiv T_e$(\OIII)/10$^4$ K. For our
metallicity estimation, we adopt a standard two-zone temperature model with
$t_2 = 0.7t_3 + 0.17$ for $t_3 < 2.0$ and $t_2 = 1.57$ for $t_3\geq2.0$ \citep{and13}.
In computing O$^+$/H$^+$, we also correct the \OII/\Hb\ ratio for dust attenuation.

\begin{figure}
  \epsscale{1.1}
  \plotone{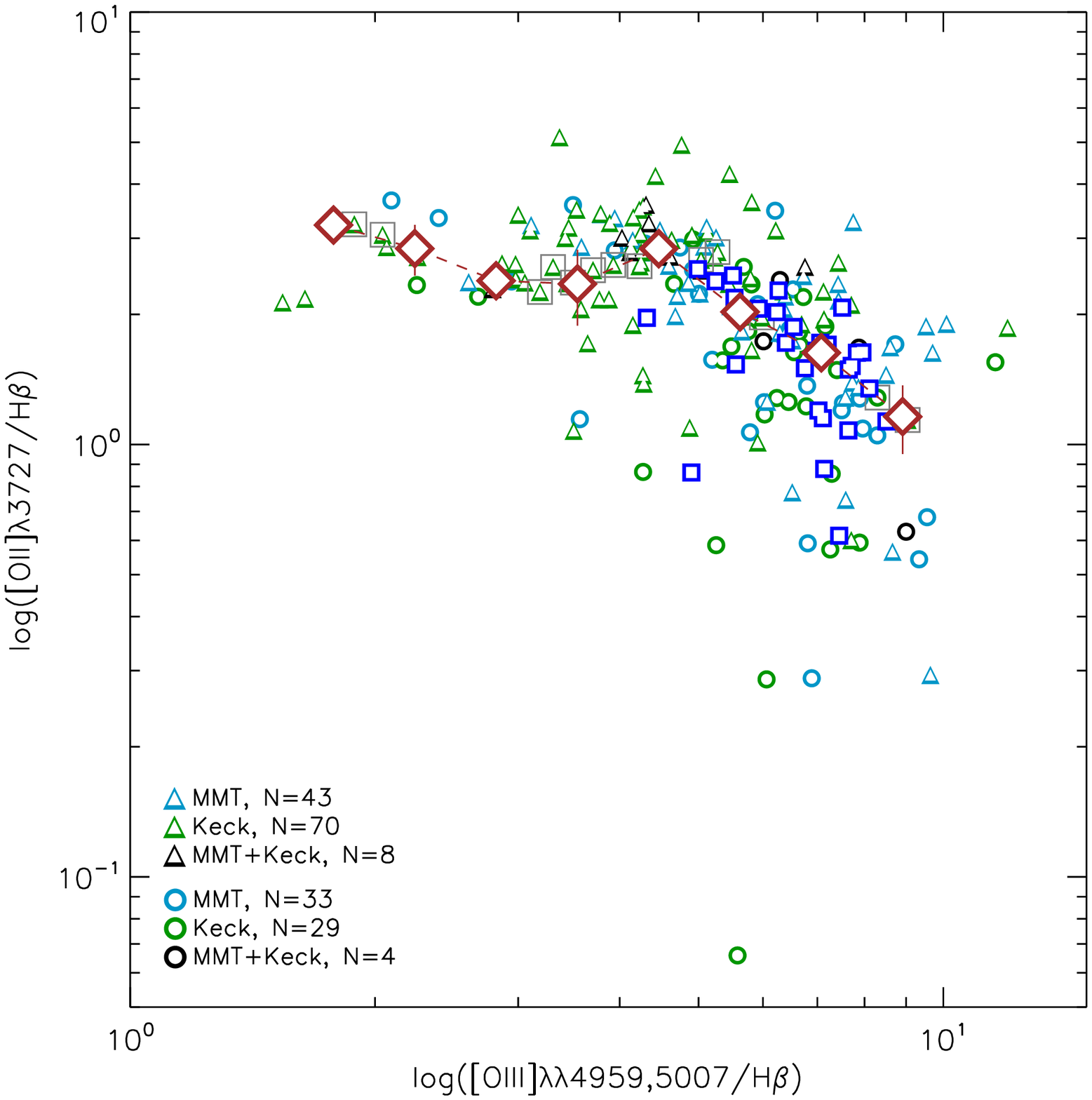}
  \caption{\OII$\lambda$3727/\Hb\ as a function of \OIII$\lambda\lambda$4959,\,5007/\Hb\
    for galaxies with \OIIIa\ detections (circles) and reliable \OIIIa\ non-detections
    (triangles). The MMT, Keck, and MMT+Keck samples are shown in light blue, green, and
    black, respectively. In addition, we overlay \OIIIa-detected galaxies from DEEP2 as
    blue squares \citep{ly15}. Median \OII/\Hb\ values are shown for each \OIII/\Hb\ bin
    as brown diamonds with uncertainties determined from bootstrapping. The dashed
    brown line shows a linear interpolation between the median values. We use these median
    emission-line ratios to estimate the \OII\ fluxes for 13 galaxies without \OII\
    spectral measurements (gray squares).}
  \label{fig:OIIOIII}
\end{figure}

Since the most abundant ions of oxygen in \textsc{H ii} regions are O$^+$ and O$^{++}$,
the oxygen abundances are given by: ${\rm O/H}=({\rm O}^++{\rm O}^{++})/{\rm H}^+$.
For 13 of our Keck galaxies (one in the \OIIIa-detected sample, 12 in the \OIIIa-non-detected
sample), our spectra do not include the \OII\,$\lambda\lambda$3726,3729 doublet emission
line as they are at lower redshifts. For these Keck galaxies (\#29, 43, 47, 53, 54, 58, 59,
73, 80, 91, 92, 93, 110), we estimate the \OII\ fluxes by assuming that they obey the
approximate inverse correlation between \OII/\Hb\ and \OIII/\Hb, shown in
Figure~\ref{fig:OIIOIII}, for the other SDF galaxies. The uncertainties on the assumed
\OII\ fluxes are based on the observed scatter in this correlation.


\begin{figure*}
  \epsscale{0.56}
  \plotone{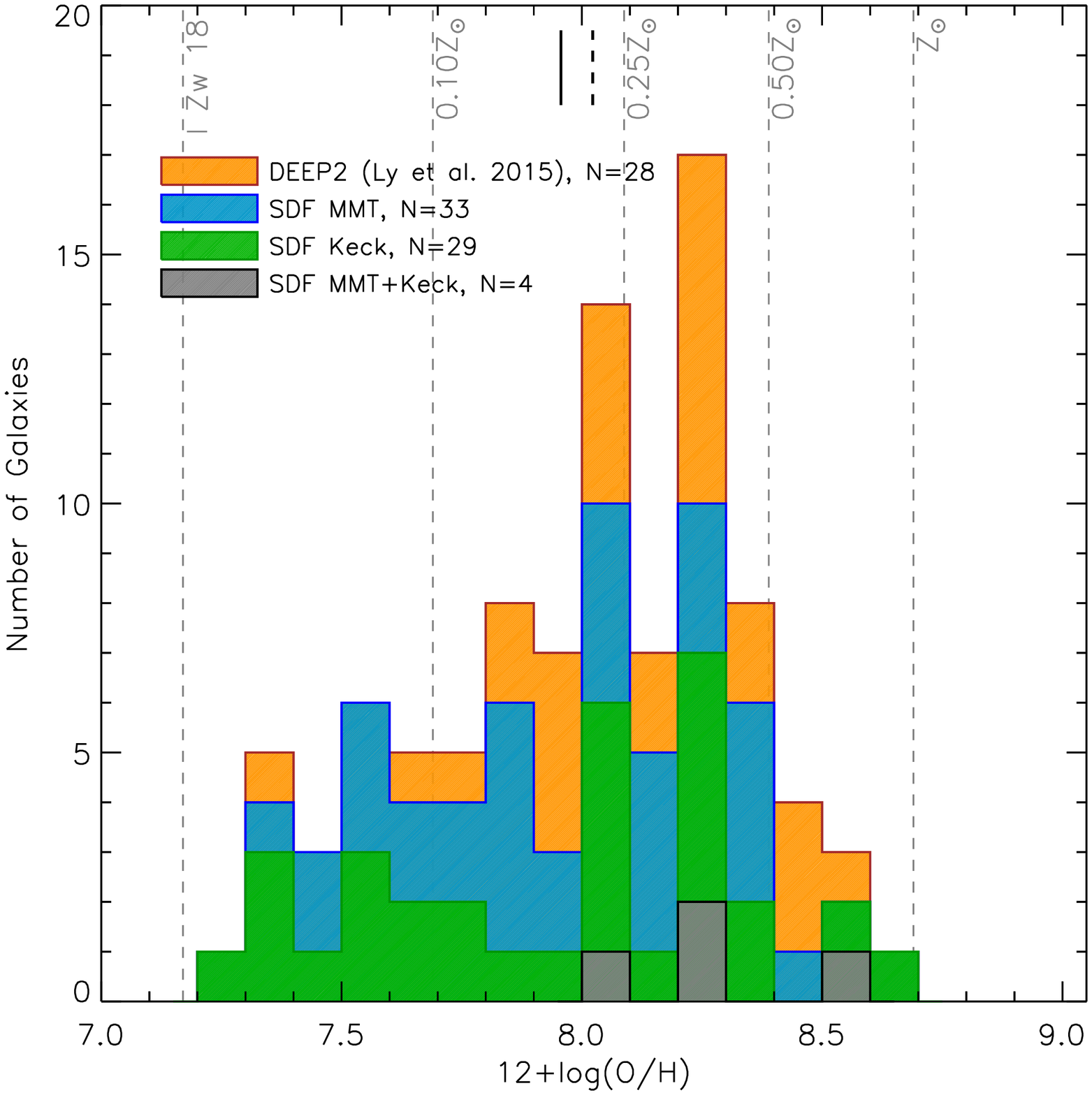}
  \plotone{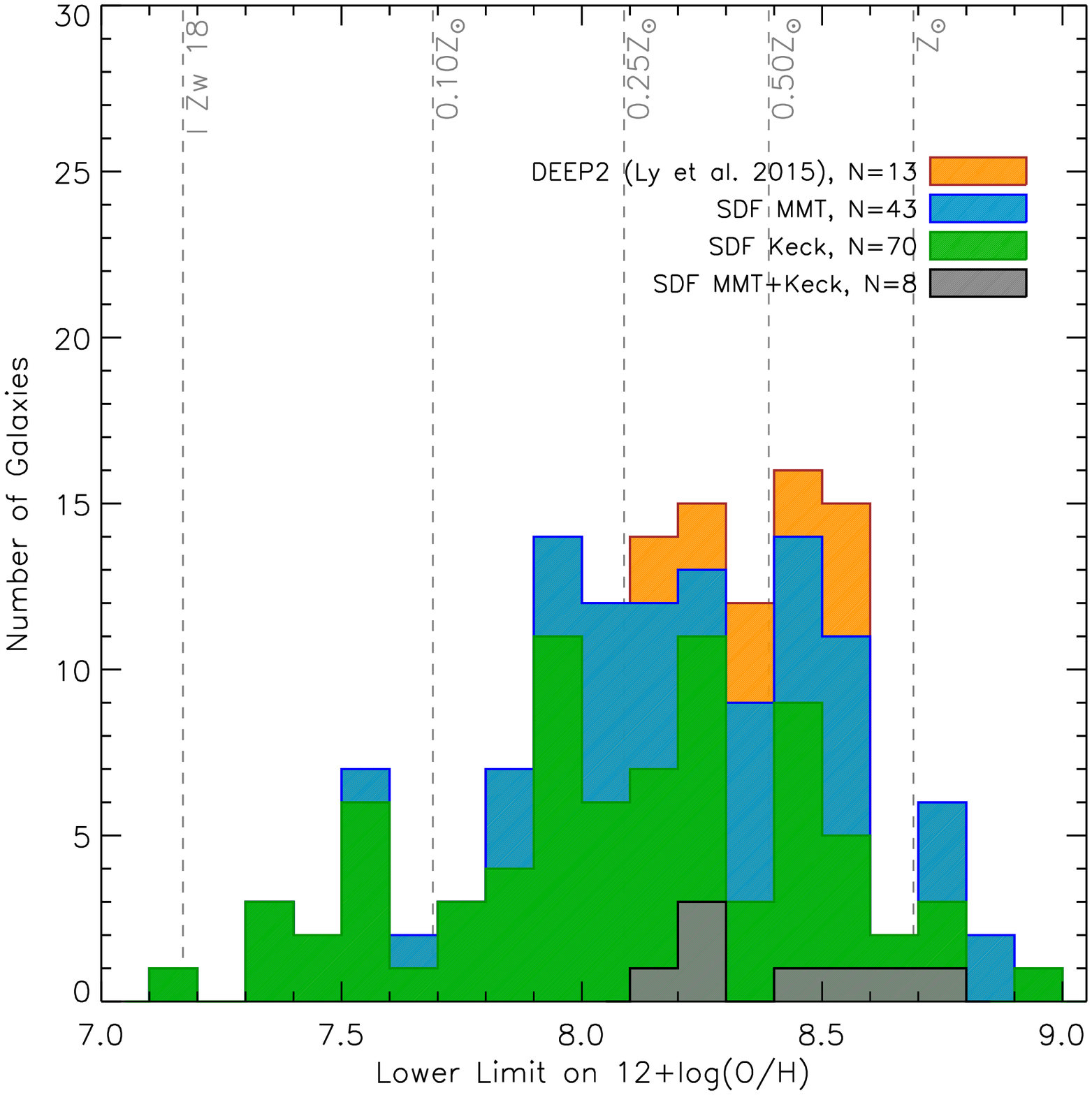}
  \caption{Distribution in \OH\ for the \OIIIa-detected sample (left) and lower limit on
    \OH\ for the \OIIIa-non-detected (right) sample. Our SDF samples are shown in light
    blue (MMT), green (Keck), and gray (MMT+Keck). The average (median) metallicity for
    the SDF samples is shown by the solid (dashed) black lines. Also overlaid in orange
    is the \cite{ly15} DEEP2 sample. Vertical dashed lines indicate abundances relative
    to solar. For comparison, the gas metallicity measurement for I Zw 18 \citep{izo06a}
    is shown.}
  \label{fig:metal_dist}
\end{figure*}

Table~\ref{tab:metals} provides observed and de-reddened flux ratios, estimates of
\Te(\OIII), $\log({\rm O^+/H^+})$, $\log({\rm O^{++}/H^+})$, and \OH\ for our
\OIIIa-detected sample.
Likewise, Table~\ref{tab:metals_nondet} provides these measurements for the
\OIIIa-non-detected sample. The distribution in \OH\ for the \OIIIa-detected sample
and lower limit on \OH\ for the \OIIIa-non-detected sample are illustrated in
Figure~\ref{fig:metal_dist}. The average and median metallicities for our \OIIIa-detected
sample are $\OHm = 7.97\pm0.34$ and 8.03, respectively. For the \OIIIa-non-detected
sample, the average and median lower limits on metallicity are $\OHm = 8.16\pm0.36$ and
8.17, respectively.


\subsection{Dust-corrected SFRs}
\label{sec:SFR}

We determine dust-corrected SFRs using the hydrogen recombination lines, which are
sensitive to the most recent star formation, $\lesssim$10 Myr.
Assuming a \cite{cha03} initial mass function (IMF) with minimum and maximum masses of
0.1 and 100 \Msun, and solar metallicity \citep{ken98}, the SFR can be determined from
the dust-corrected \Ha\ luminosity as:
\begin{equation}
  {\rm SFR}(M_{\sun}~{\rm yr}^{-1}) = 4.4\times10^{-42} L({\rm erg~s}^{-1}).
\end{equation}
This relation is calibrated at solar metallicity, and overestimates the SFR at lower
metallicities due to the greater escape of ionizing photons from more metal-poor
(less blanketed) O star atmospheres.
To account for the metallicity dependence, we use predictions of the \Ha\ luminosity
from Starburst99 spectral synthesis models \citep{lei99}. Here, we adopt the Padova
stellar tracks \citep{bre93,fag94a,fag94b}, a constant star formation history (SFH), a
\citet{kro01} IMF\footnote{Starburst99 does not allow for a \cite{cha03} IMF.}, and
metallicities of 0.02, 0.20, 0.40, 1.0, and 2.5\zsun. Since we are adopting
\cite{cha03} IMF, we apply an offset of $\approx$0.1 dex to the \Ha\ luminosities.
The best-fit to the metallicity-dependent $L$(\Ha)--SFR relation is:
\begin{equation}
  \log\left[\frac{{\rm SFR}}{L(\Hae)}\right] = -41.34 + 0.39 y + 0.127 y^2,
\end{equation}
where $y = \log({\rm O/H}) + 3.31$.\footnote{$y = 0$ corresponds to solar oxygen abundances.}
For galaxies without \Ha\ measurements, SFRs can be determined from  de-reddened \Hb,
by assuming the intrinsic Case B  ratio, (\Ha/\Hb)$_0 = \HaHbi$.

\begin{figure*}
  \epsscale{0.56}
  \plotone{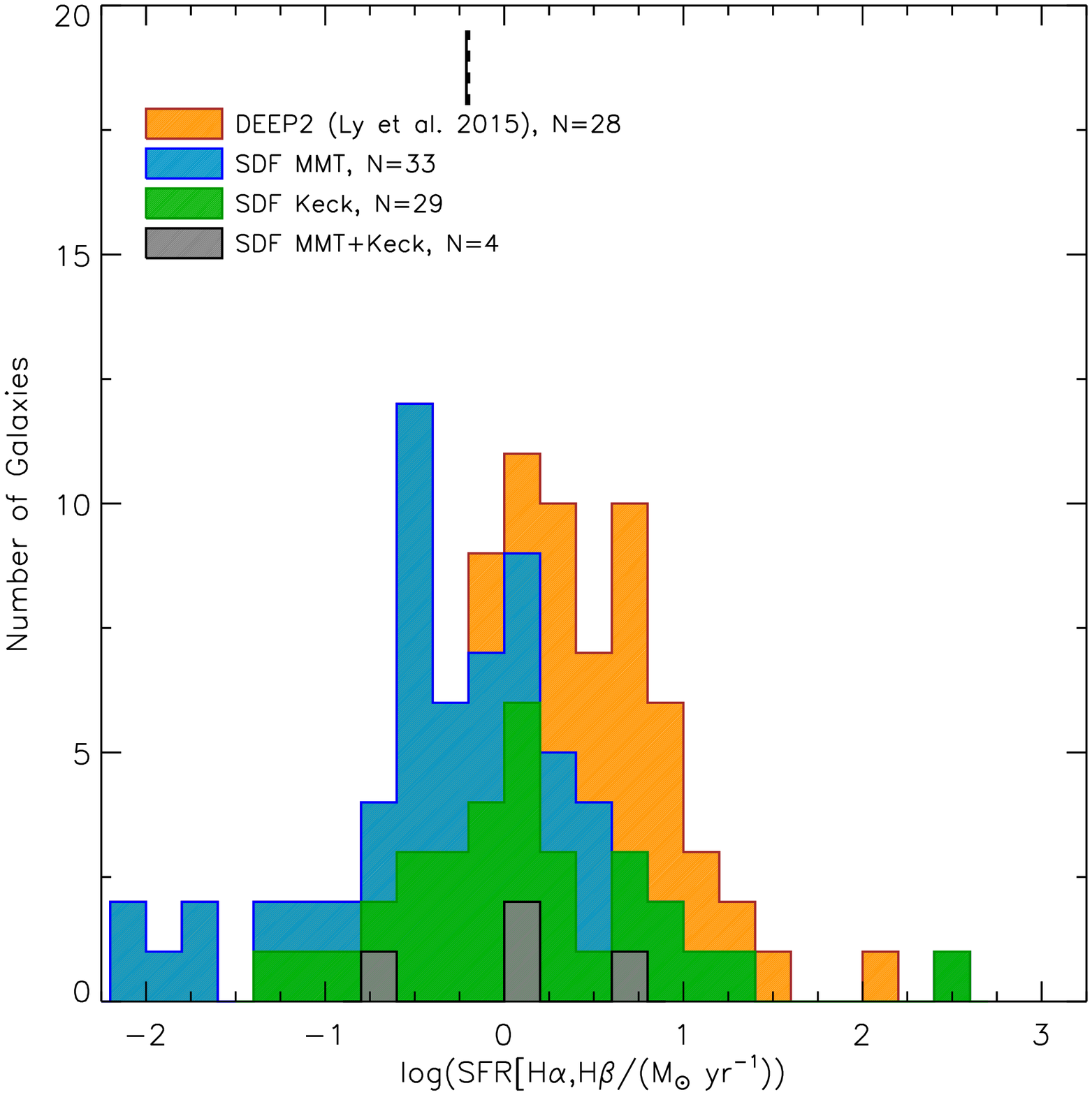}
  \plotone{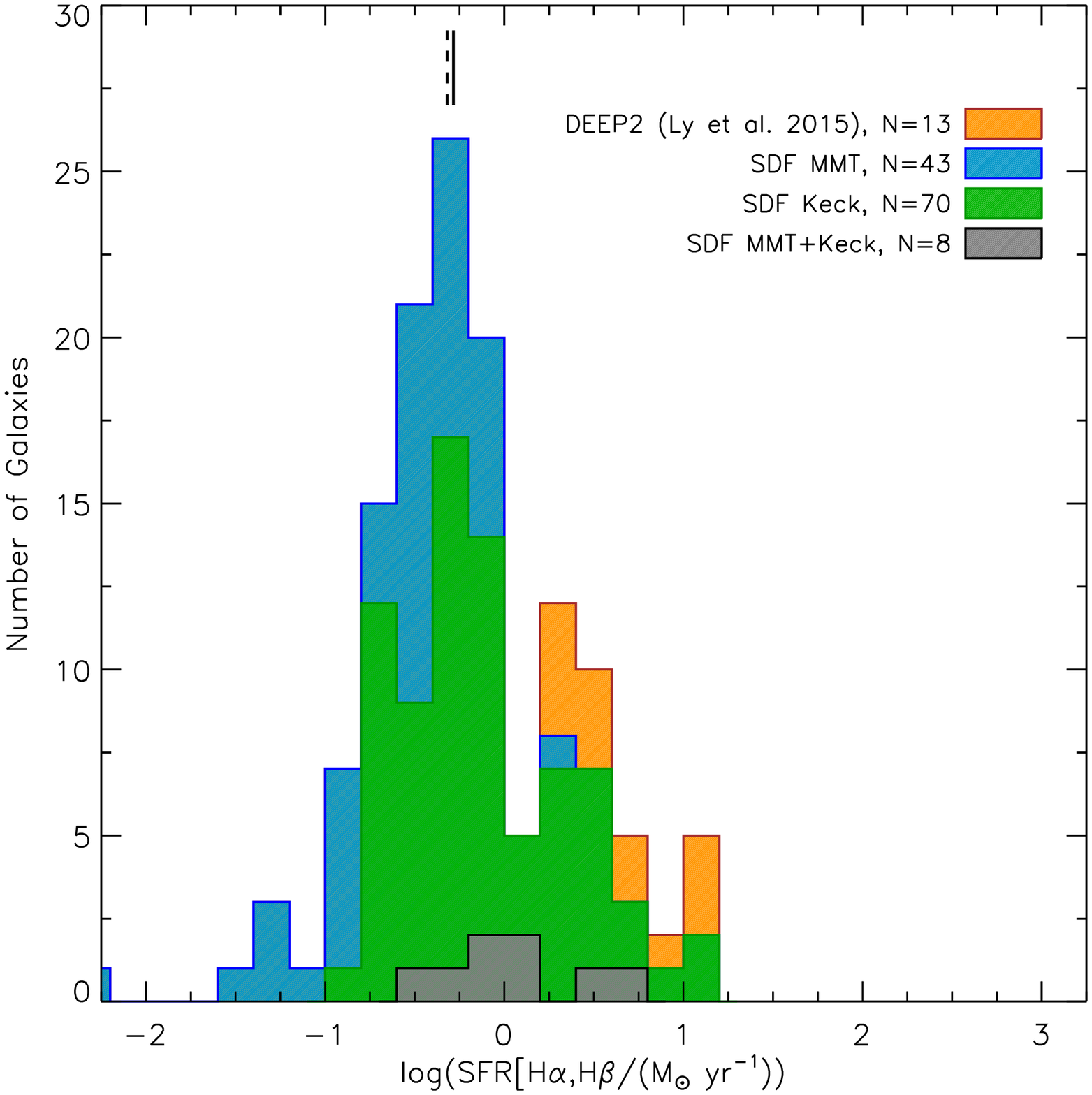}
  \caption{Distribution in SFR for the \OIIIa-detected (left) and \OIIIa-non-detected
    (right) samples determined from dust-corrected \Ha\ or \Hb\ luminosities (see
    Section~\ref{sec:SFR}). Our SDF samples are shown in light blue (MMT), green (Keck),
    and gray (MMT+Keck). The average (median) SFR for the SDF samples is shown by the
    solid (dashed) black lines. Also overlaid in orange is the \cite{ly15} DEEP2 sample.}
  \label{fig:SFR_dist}
\end{figure*}

Our SFR estimates are illustrated in Figure~\ref{fig:SFR_dist} and summarized in
Table~\ref{tab:SFR_Mass} for the \OIIIa-detected sample and
Table~\ref{tab:SFR_Mass_nondet} for the \OIIIa-non-detected sample. Our \OIIIa-detected
galaxies have dust-corrected $\log\left({\rm SFR}/M_{\sun}~{\rm yr}^{-1}\right) = $
--2.2 to 2.5 with an average (median) of \SFRA\ (\SFRM). The average and median for
the \OIIIa-non-detected sample are --0.26$\pm$0.51 and --0.29, respectively.


\subsection{SEDs and Stellar Masses}
\label{sec:SED}
One significant advantage of studying low-mass galaxies in the SDF is the ultra-deep
imaging in twenty-four bands from the UV to the IR, which allows us to characterize
their stellar properties.
The SDF has been imaged with: (1) \GALEX\ \citep{mar05} in both the \FUV\ and \NUV\
bands;\footnote{Details on the \GALEX\ imaging are available in \cite{ly09,ly11b}.}
(2) KPNO's Mayall telescope using MOSAIC in $U$; (3) Subaru telescope with Suprime-Cam
in 14 bands ($BV\Rcf i\arcmin z\arcmin z_b z_r$, and the five narrowband and two
intermediate-band filters as mentioned previously); (4) KPNO's Mayall telescope using
NEWFIRM \citep{pro08} in $H$; (5) UKIRT using WFCAM in $J$ and $K$; and (6)
{\it Spitzer} in the four IRAC bands (3.6, 4.5, 5.8, and 8.0 \mm).

Most of these imaging data have been discussed in \cite{ly11b}, except for the WFCAM
$J$-band data and most of the NEWFIRM $H$-band data. The more recent NEWFIRM imaging data
were acquired on 2012 March 06--07 and 2013 March 27--30 with photometric observing
conditions at KPNO, clear skies, and 0\farcs9 seeing. These conditions were significantly
better than in 2008 where the seeing was 1\farcs3 with transparency varying by as much
as $\approx$2 mag. The improved observing conditions yielded a mosaicked image that is
$\approx$1.8 mag deeper than our previous $H$-band observations \citep{ly11b}. These
NEWFIRM data were reduced following the steps outlined in \cite{newha} with the
\textsc{iraf} \textit{nfextern} package. The WFCAM data were obtained on 2005 April 14--15,
2010 March 15--20, and 2010 April 22--23 with photometric observing conditions and
$\approx$1\arcsec\ seeing. These $J$-band observations are $\approx$1.6 mag deeper than
previous NEWFIRM $J$-band observations obtained in poor observing conditions \citep{ly11b}.
The reduction of the WFCAM $J$-band data follows the procedure outlined in \cite{hay09}.

We obtained source catalogs for all narrowband/intermediate-band excess emitters by
running SExtractor in ``dual-image'' mode for all optical and near-IR data (i.e.,
imaging that spans $U$ to $K$). For \GALEX\ measurements, since almost all our galaxies
are virtually point sources,\footnote{MMT17 is resolved as an edge-on disk at
  $z\approx0.08$, so we use Kron aperture photometry.} we obtain more accurate photometry
in the \FUV\ and \NUV\ bands by point spread function (PSF) fitting with
\textsc{iraf}/\textsc{daophot} \citep[vers. 2.16;][]{ste87}.
Our examination of the residuals for each source suggests that the PSF-fitting is extremely
successful,\footnote{The rms in the residual image is consistent with Poisson noise from
  the background.} and that none of these galaxies is affected by contamination from
nearby sources. Among the Keck sample,  32 (21) galaxies are detected in the \NUV\ (\FUV).
Because the MMT \OIIIa\ galaxies are brighter, nearly all of them are detected in the
\NUV\ (36 of \NMMTf) and \FUV\ bands (35 of \NMMTf), allowing for robust UV SFR
determinations.\footnote{MMT05 is adjacent to a bright source, preventing accurate
  measurements in the UV images.} The photometric fluxes are provided in
Table~\ref{tab:SED}.

Since our galaxies have very high emission-line EWs, we correct the broadband photometry
for the contribution from nebular emission lines using emission-line measurements from our
spectroscopy and narrowband imaging.
We generate a spectrum for each galaxy with zero continuum and emission lines located at
the redshifted wavelengths for \OII, \NeIII, \Hb, \OIII, \Ha, higher order Balmer lines,
and other weaker emission lines.
These spectra are then convolved with the filter bandpasses to determine excess fluxes,
which are then are removed from the broadband photometry.
For \NHa\ galaxies, spectroscopy provided \Ha\ measurements. For MMT23, we use the NB973
excess as an estimate of the \Ha\ flux.
However, for our higher redshift galaxies, \Ha\ is redshifted into the near-IR or
unavailable from optical spectroscopy.
To correct \Ha\ in these 33 galaxies, we use \Hb\ fluxes and assume that \Ha\ is
three times stronger.
Of course, higher dust reddening will yield stronger \Ha\ corrections. We adopt a
minimum correction of $A(\Hae)$ = 0.12 mag.

\begin{figure*}
  \epsscale{1.1}
  \plotone{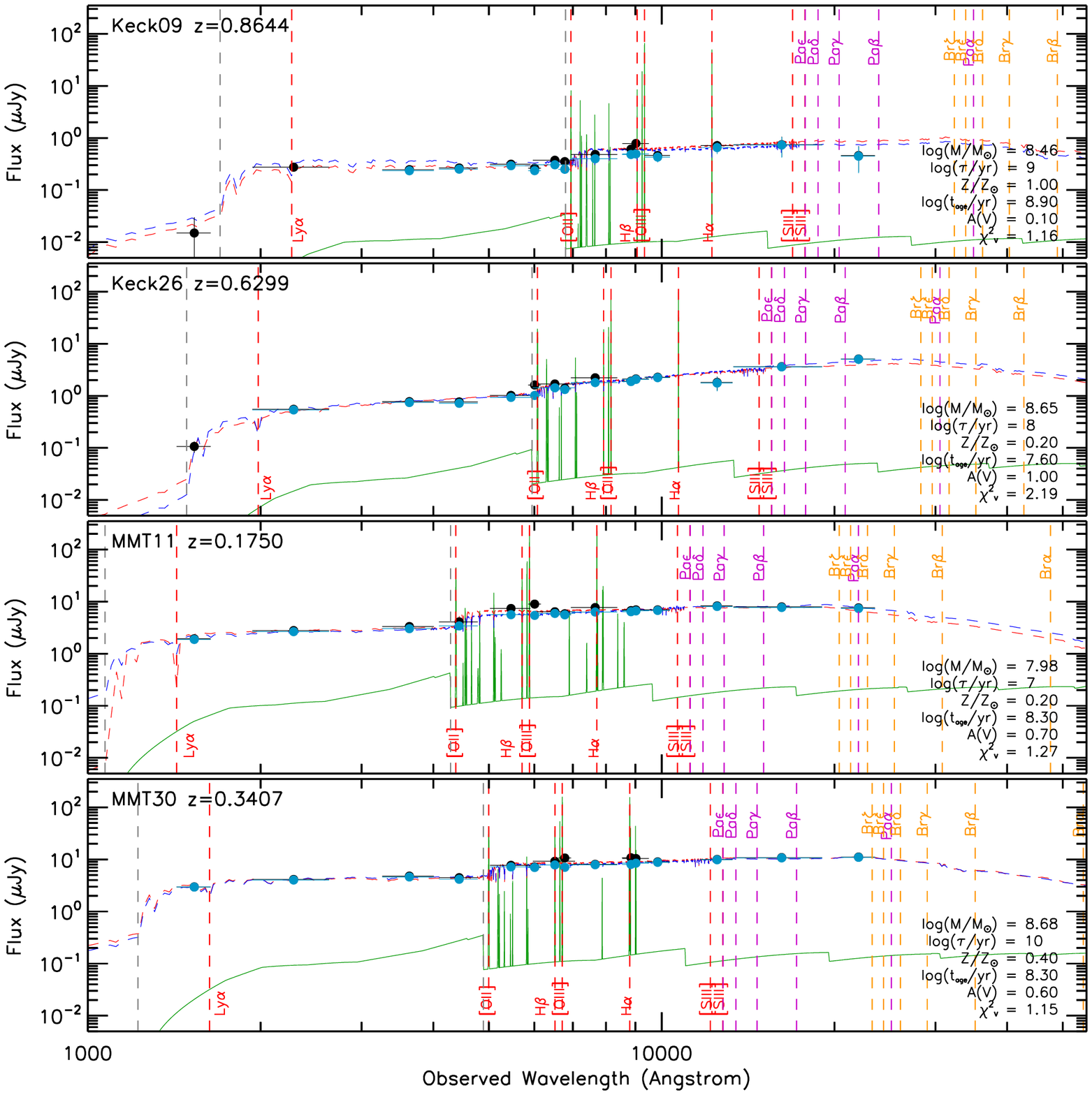}
  \caption{SEDs and the best-fitting stellar population results from modeling the SED in
    four galaxies: Keck09, Keck26, MMT11, and MMT30. The observed fluxes are shown by the
    black circles while the light blue circles illustrate the corrections for nebular
    continuum and line emission. Error bars along the $x$-axis demonstrate the FWHM of
    the filters while those along the $y$-axis are 1$\sigma$. The SEDs for the nebular
    emission are illustrated by the green solid lines. The best $\chi^2$ fits to the
    original and nebular emission corrected SEDs are shown by the red and blue dashed
    lines, respectively. The SED-fitting results after nebular emission corrections are
    reported on the right. Gray vertical lines indicate the locations of the Balmer and
    Lyman continuum breaks, while red, purple, and orange vertical lines show the
    location of various redshifted nebular emission lines. When correcting for nebular
    emission, the $\chi^2$ values are reduced by a factor of 2.7--4.3 in these four
    cases.}
  \label{fig:SED_fit}
\end{figure*}

In addition, we correct the photometry for nebular continuum emission from free--free,
free--bound, and two-photon emission.
To estimate the nebular contributions toward the total light, we generate spectral
synthesis models from Starburst99. Our models assume a constant SFR, a \cite{kro01}
IMF\footnote{This is similar to a \cite{cha03} IMF.}, Geneva stellar evolutionary
models, and $Z$/\zsun\ = 0.2, 0.4, and 1.0 (depending on the gas-phase metallicity; see
Equations~(\ref{eqn18})--(\ref{eqn20})). The nebular continuum emission is then scaled
by the dust-corrected \Ha\ or \Hb\ luminosity and reddened for dust attenuation by
assuming a \cite{car89} formalism with \EBVa\ estimates from Balmer decrements (see
Section~\ref{sec:dust}). These corrections are illustrated in Figure~\ref{fig:SED_fit}
for four of our \OIIIa-detected galaxies.
The black circles show the original broadband photometry, while the blue circles show
the corrected continuum fluxes after the removal of nebular continua and emission
lines. We also overlay the spectrum of the nebular emission.
Finally, both the original and corrected SEDs were fit with stellar population
synthesis models \citep{bc03} by the Fitting and Assessment of Synthetic Templates
\citep{kri09} code.  We use exponentially declining SFHs (i.e., $\tau$ models) similar
to previous fitting by \cite{ly11b} and many other groups.
We have chosen this SFH because: (1) broadband data are generally unable to distinguish
between more complicated SFHs (e.g., a constant SFR with a recent burst, which may be
more representative of our galaxies); (2) the inclusion of a recent burst in star
formation does not significantly alter stellar mass estimates, since the mass-to-light
ratio is set by the rest-frame optical-to-IR light from the older stellar population;
and (3) as we will later show, these fits that assume an exponentially declining SFH
are consistent with the data, with nearly unity $\chi^2_{\nu}$ values.


In these models, we adopt \cite{cal00} reddening ($A_V$ is free parameter which
ranged from 0 and 3 mag), inter-galactic medium attenuation following \cite{mad95},
$\tau$ values between 10$^7$ and 10$^{10}$ yr, and a \cite{cha03} IMF.
The only differences to our previous \OIIIa\ study \citepalias{ly14} is that we
use stellar atmosphere models with abundances consistent with the gas-phase metallicity:
\begin{eqnarray}
  Z_{\star} &=& 0.004,~(\OHm \leq 8.17),\label{eqn18}\\
  Z_{\star} &=& 0.008,~(8.17<\OHm \leq 8.39),\\
  Z_{\star} &=& 0.02 \equiv Z_{\sun},~(8.39<\OHm).\label{eqn20}
\end{eqnarray}
The latter assumption differs for our previous study \citepalias{ly14}, where
$Z_{\star} = 0.004$ was used for all galaxies.

\begin{figure*}
  \epsscale{0.56}
  \plotone{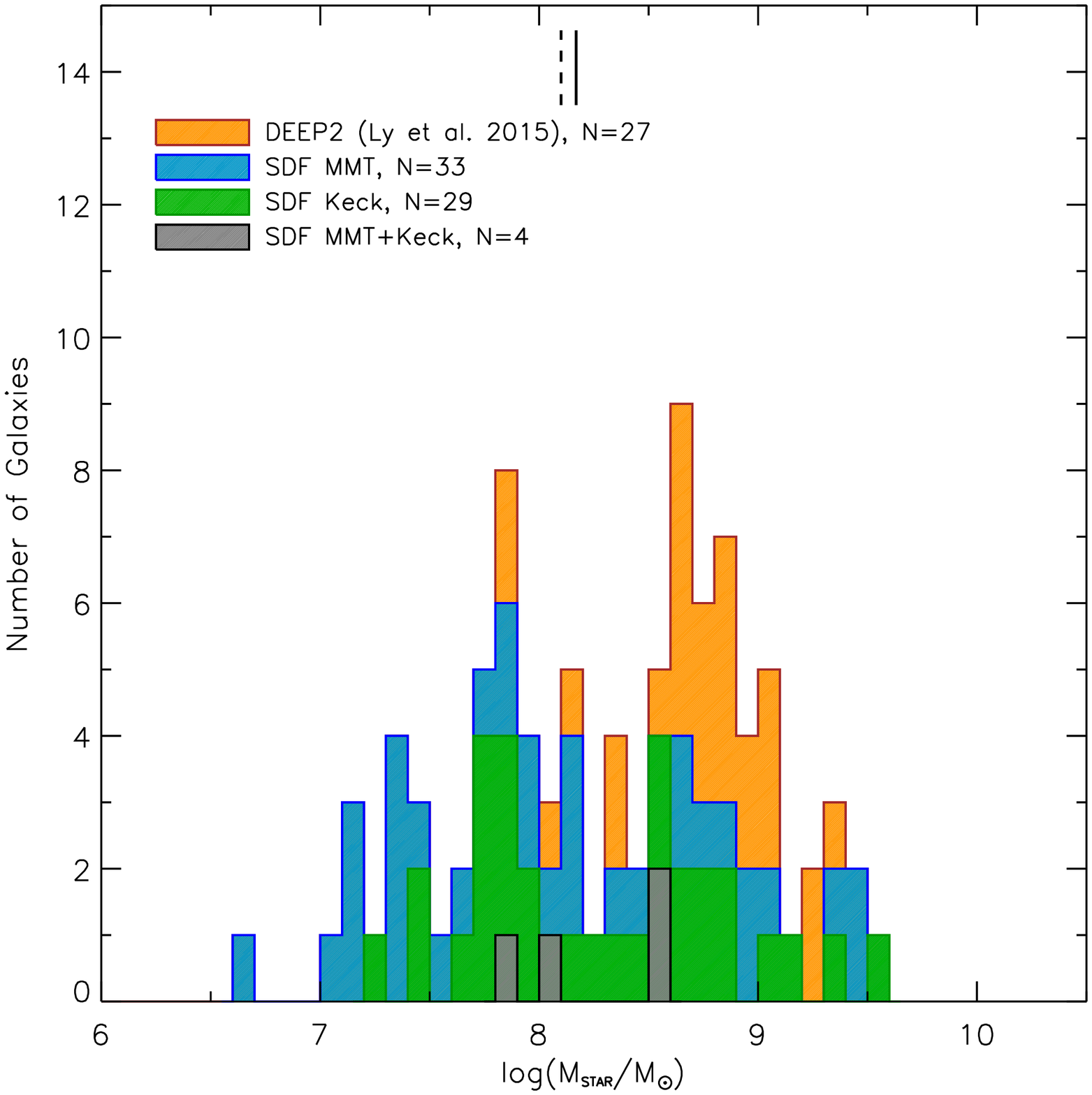}
  \plotone{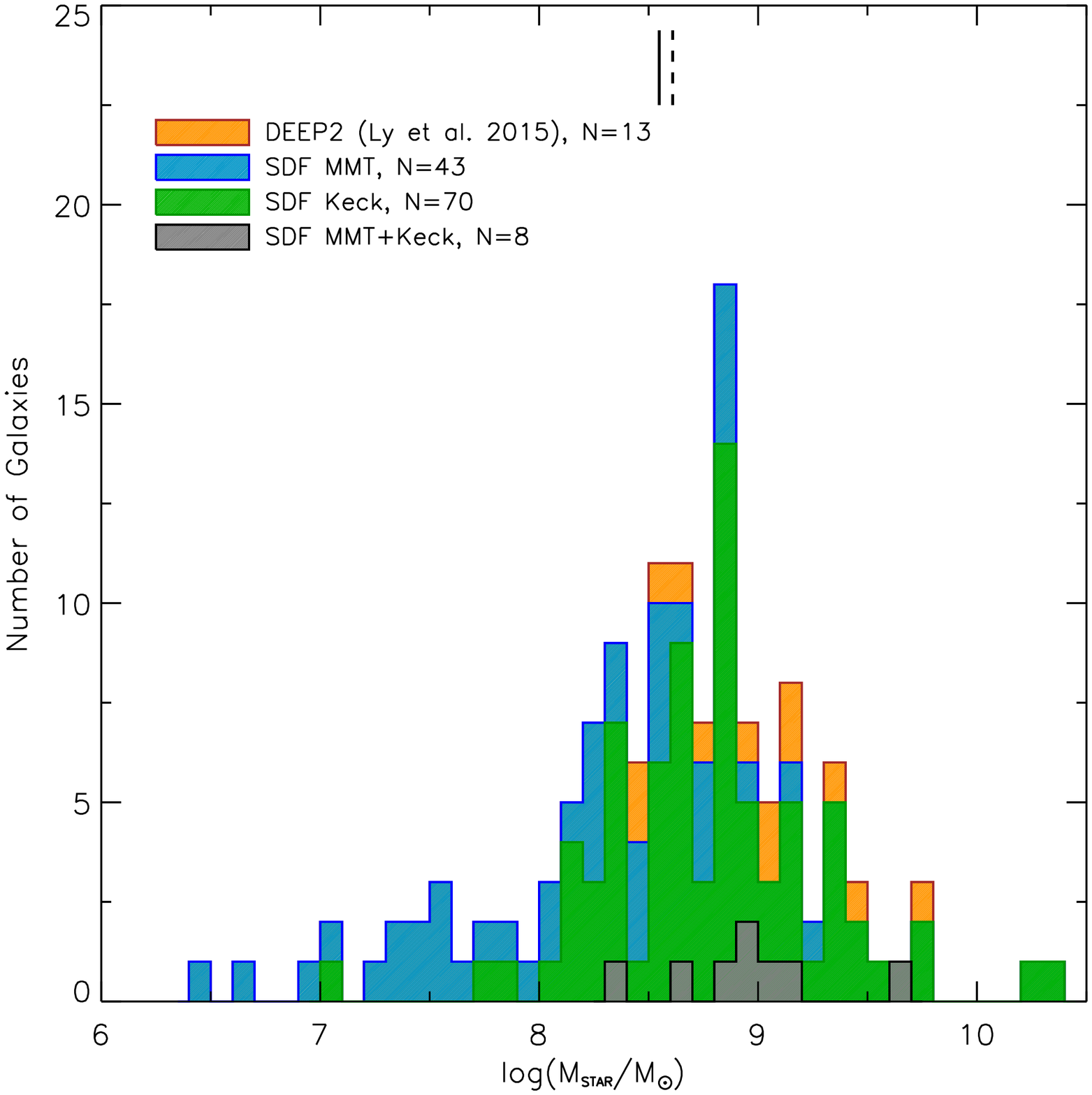}
  \caption{Distribution in stellar mass for the \OIIIa-detected (left) and
    \OIIIa-non-detected (right) samples determined from SED modeling (see
    Section~\ref{sec:SED}). Our SDF samples are shown in light blue (MMT), green (Keck),
    and gray (MMT+Keck). The average (median) stellar mass for the SDF samples is
    shown by the solid (dashed) black lines. Also overlaid in orange is the \cite{ly15}
    DEEP2 sample.}
  \label{fig:mass_dist}
\end{figure*}

Our SED-fitting results are illustrated in Figure~\ref{fig:mass_dist} and summarized in
Table~\ref{tab:SFR_Mass} for the \OIIIa-detected sample and
Table~\ref{tab:SFR_Mass_nondet} for the \OIIIa-non-detected sample. We find that these
galaxies are typically low-mass systems (median of \MassM\ \Msun, average of \MassA\ \Msun),
with a stellar mass distribution that extends from $4.3\times10^6$ to $3.9\times10^9$ \Msun.
The estimated light-weighted stellar ages of $t_{\rm age}=10^7$--10$^{9.9}$ yr (average of
10$^{\AgeA}$ yr) suggest that these galaxies formed most of their stars only recently,
which explains why they were detected by our emission-line flux sample.

We combine our dust-corrected SFRs (Section~\ref{sec:SFR}) and stellar mass determinations
to locate our galaxies on the \Mstar--SFR relation in Figure~\FMS\ of \citetalias{MACTII}.
We emphasize that the inverse of the sSFR is consistent with the stellar ages derived
from SED fitting, which are provided in Table~\ref{tab:SFR_Mass}. That is, these galaxies
could have formed most of their stars by maintaining the measured SFRs over their lifetimes.


\subsection{ISM Gas Densities}

The gas or electron density ($n_e$) is critical for understanding the multi-phased ISM,
specifically characterizing the neutral, ionized, and molecular gas components of the ISM.
In addition, the strength of nebular emission lines, such as hydrogen emission lines (e.g.,
\Ha, \Hb) emitted from the recombination of ionized hydrogen gas and collisionally excited
forbidden transition emission lines (e.g., \OIII, \OII), is proportional to $n_e^2$.

\begin{figure*}
  \epsscale{1.1}
  \plottwo{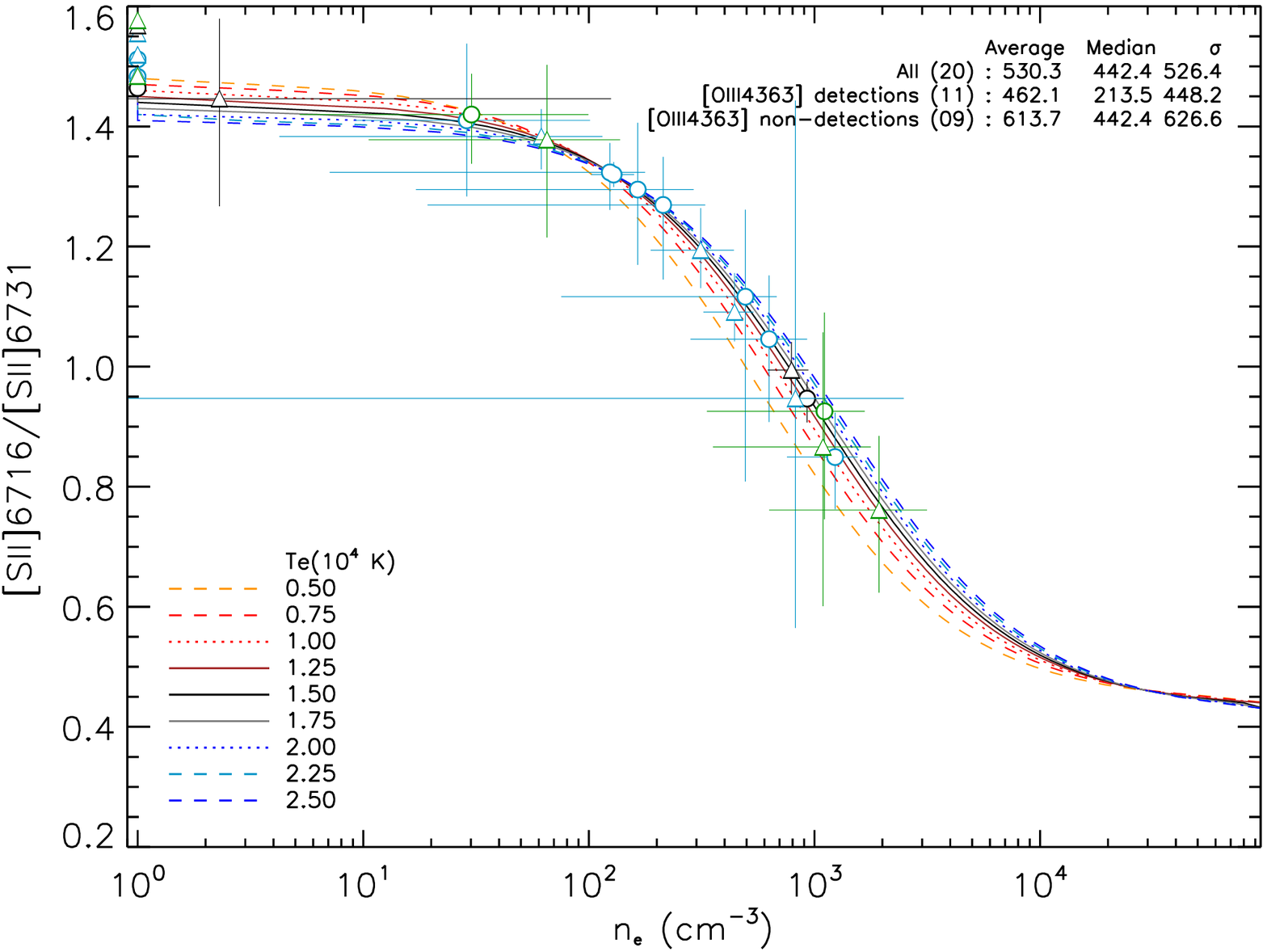}{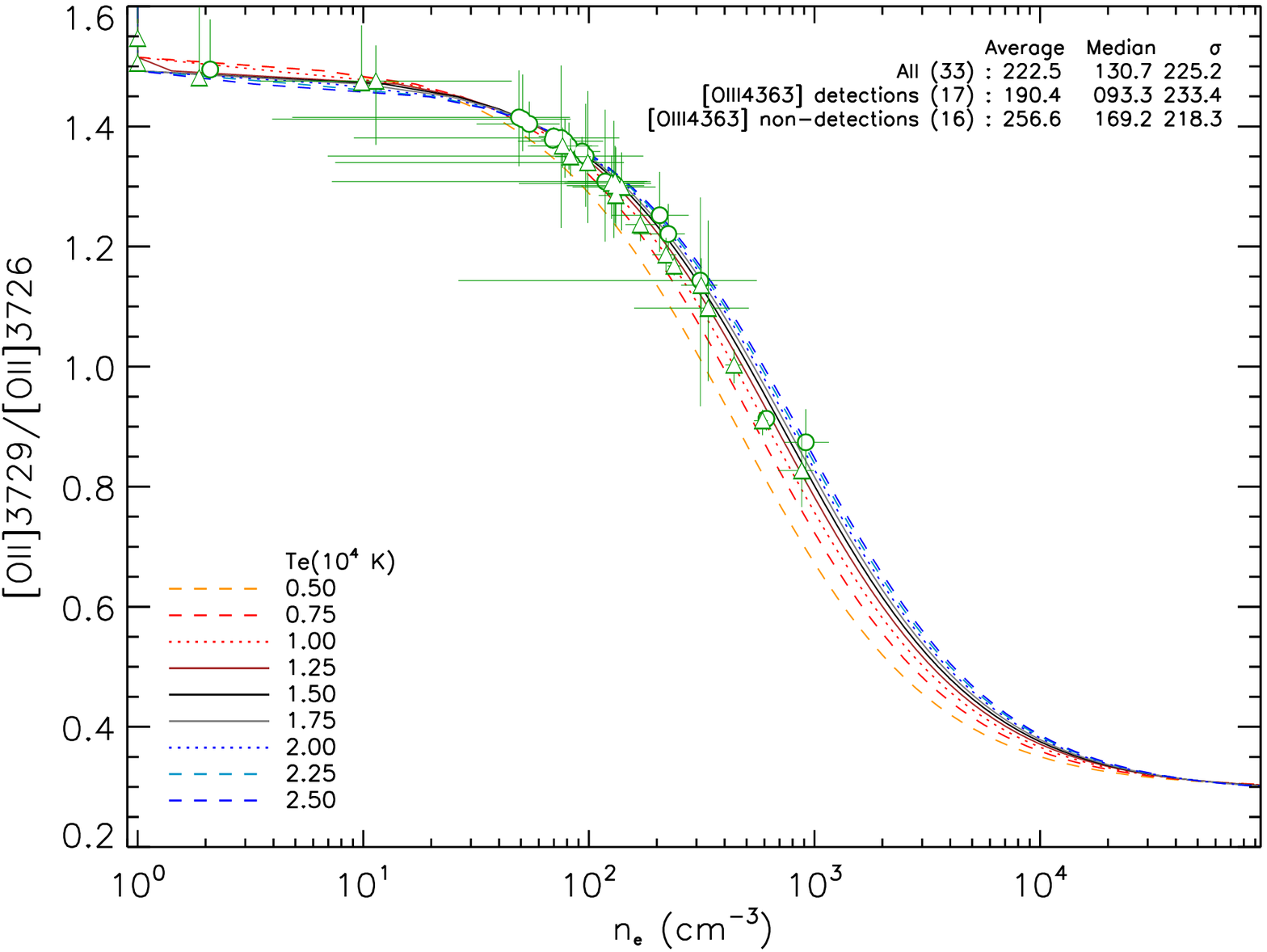}
  \caption{The electron densities derived from two emission-line doublet ratios,
    \SII$\lambda$6716/6731 (left) and \OII$\lambda$3729/3726 (right). The observed doublet
    ratios for SDF galaxies are shown on the $y$-axes, along with their uncertainties.
    These are then translated into {\it inferred} electron densities on the $x$-axes.
    This transformation is illustrated as a function of electron temperature (indicated
    by various color and line types from 5,000 to 25,000 K). For consistency, we use
    the \Te\ derived in Section~\ref{sec:Te}. Average and median densities are reported
    in the upper right of the figure. The derived electron densities are relatively
    high compared to integral measurements from SDSS; however, they are consistent with
    individual \textsc{H ii} regions.  Galaxies with doublet ratios at or beyond the
    low-density limit are plotted at $n_e = 1$ cm$^{-3}$. These electron densities are
    provided in Table~\ref{tab:metals}.}
  \label{fig:ne}
\end{figure*}

To estimate the electron density of the ISM, we use the flux ratios of the
\SII\,$\lambda\lambda$6716,6731 and \OII\,$\lambda\lambda$3726,3729 doublets. These ratios
are illustrated in Figure~\ref{fig:ne} and the inferred densities are tabulated in
Table~\ref{tab:metals}.
To compute electron densities, we use an IDL-based routine that is based on the
\textsc{iraf/temden} package, but with improvements with new atomic data
\citep{berg15,cro15}.\footnote{Developed by John Moustakas, Brian Moore, and
  Kevin Croxall. The code called \textit{impro} is available here:
  \url{https://github.com/moustakas/impro}.}
For consistency, we account for the $n_e$ dependence on \Te\ for each individual galaxy
(see the curves of different colors and line styles in Figure~\ref{fig:ne}).
The average (median) $n_e$ derived from \SII\ measurements are 462 (214) cm$^{-3}$ for
the \OIIIa-detected sample and 614 (443) cm$^{-3}$ for the \OIIIa-non-detected sample.
Likewise, the \OII\ measurements yield averages (medians) of 190 (93) cm$^{-3}$ for the
\OIIIa-detected sample and 257 (169) cm$^{-3}$ for the \OIIIa-non-detected
sample.\footnote{Averages and medians exclude galaxies with doublet ratios beyond the
  low-density limit.}
While these gas densities are higher than the integrated measurements from SDSS
\citep[$\sim$10 cm$^{-3}$;][]{hay15}, they are consistent with the typical densities
(10$^2$--10$^4$ cm$^{-3}$) of individual \textsc{H ii} regions. The high gas densities
suggest that the filling factor of ionized gas is large, encompassing most of the
galaxies in our sample.
We note that we are unable to compare directly the electron densities derived from \SII\
and \OII\ line ratios since \SII\ measurements are only available for lower redshift
galaxies and the resolved \OII\ measurements are from Keck, which are only available
at higher redshift.


\section{CONCLUSIONS}
\label{sec:End}

This paper (Paper I) describes the large spectroscopic survey, called ``\Sname,'' that we
have conducted. Using the MMT/Hectospec and Keck/DEIMOS spectrographs, we have obtained
optical spectra for $\approx$1900 star-forming galaxies. These galaxies are pre-selected
from narrowband or intermediate-band imaging in the SDF, which detects redshifted nebular
emission lines at redshifts between 0.07 and 1.61. The focus of \Sname\ is to obtain
robust measurements of the gas metallicity and other properties (e.g., gas density,
ionization parameter) of the interstellar gas from deep rest-frame optical spectroscopy.

\Sname\ is unique from previous spectroscopic surveys because our spectroscopy has
obtained measurements of the weak \OIIIa\ nebular emission for a large number of galaxies.
The strength of the \OIIIa\ line is set by the electron temperature for the ionized gas.
Since the gas temperature is regulated by the metal content---collisionally excited metal
emission lines enable the gas to cool---an inverse relationship exists between gas-phase
oxygen abundance and \OIIIa\ line strength. Specifically, \Sname\ has measured the \OIIIa\
line in \Ntot\ galaxies (\Ndet\ galaxies have detections at S/N $\geq$ 3, referred
to as the \OIIIa-detected sample, and \Nrel\ galaxies have reliable lower limits on
metallicity, referred to as the \OIIIa-non-detected sample). Combined with
multi-wavelength imaging data that span rest-frame UV to near-IR, we describe how we
determine dust attenuation, metallicity, SFR, and stellar mass for these galaxies in the
\OIIIa-detected and \OIIIa-non-detected samples. \citetalias{MACTII} of the \Sname\ survey
utilizes the \OIIIa-based metallicities to study, for the first time, the evolution of the
stellar mass--gas metallicity relation and its secondary dependence on SFR over 8 billion
years ($z\lesssim1$) using a temperature-sensitive metallicity diagnostics.

\acknowledgements
We thank the anonymous referee for comments that improved the paper.
The DEIMOS data presented herein were obtained at the W.M. Keck Observatory, which is
operated as a scientific partnership among the California Institute of Technology, the
University of California, and the National Aeronautics and Space Administration (NASA).
The Observatory was made possible by the generous financial support of the W.M. Keck
Foundation.
The authors wish to recognize and acknowledge the very significant cultural role and
reverence that the summit of Mauna Kea has always had within the indigenous Hawaiian
community. We are most fortunate to have the opportunity to conduct observations from
this mountain.
Hectospec observations reported here were obtained at the MMT Observatory, a joint
facility of the Smithsonian Institution and the University of Arizona.
A subset of MMT telescope time was granted by NOAO, through the NSF-funded Telescope
System Instrumentation Program (TSIP).
We gratefully acknowledge NASA's support for construction, operation, and science
analysis for the \GALEX\ mission.
This research is supported by an appointment to the NASA Postdoctoral Program at the
Goddard Space Flight Center, administered by Oak Ridge Associated Universities and
Universities Space Research Association through contracts with NASA.
CL is supported by NASA Astrophysics Data Analysis Program grant NNH14ZDA001N.
We thank Tohru Nagao and Brett Andrews for discussions that improve the paper.

{\it Facilities:} \facility{Subaru (Suprime-Cam)}, \facility{MMT (Hectospec)},
\facility{Keck:II (DEIMOS)}, \facility{\textit{GALEX}},
\facility{Mayall (MOSAIC, NEWFIRM)}, \facility{UKIRT (WFCAM)}


\clearpage
\input tab4.final.tex


\clearpage
\input tab5.final.tex


\clearpage
\LongTables
\input tab6.final.tex


\clearpage
\LongTables
\input tab7.final.tex


\input tab8.final.tex


\clearpage
\input tab9.final.tex


\clearpage
\LongTables
\input tab10.final.tex


\clearpage
\LongTables
\input tab11.final.tex


\input tab12.final.tex


\clearpage
\LongTables
\input tab13.final.tex


\clearpage
\LongTables
\input tab14.final.tex


\clearpage
\LongTables
\input tab15.final.tex


\clearpage
\LongTables
\input tab16.final.tex


\clearpage
\LongTables
\input tab17.final.tex

\end{document}

%% file: tab1.final.tex
\begin{deluxetable*}{lccccccrrrrrr}
  \tabletypesize{\scriptsize}
  \tablewidth{0pc}
  \tablecaption{Summary of Filters, Emission-line Samples, and Spectroscopic Samples for the SDF}
  \tablehead{
    Filter&
    \colhead{$\lambda_c$}&
    \colhead{FWHM}&
    \colhead{$m_{\rm lim}(3\sigma)$}&
    \colhead{Area}&
    \colhead{$N$}&
    \colhead{$N_{\rm total}$}&
    \multicolumn{3}{c}{$N_{\rm target}$}&
    \multicolumn{3}{c}{$N_{\rm specz}$}\\
    \cline{8-10}
    \cline{11-13}
    \colhead{}&
    \colhead{(\AA)}&
    \colhead{(\AA)}&
    \colhead{(mag)}&
    \colhead{(arcmin$^2$)}&
    \colhead{}&
    \colhead{}&
    \colhead{MMT} & \colhead{Keck} & \colhead{Total} &
    \colhead{MMT} & \colhead{Keck} & \colhead{Total}\\
    \colhead{(1)}&\colhead{(2)}&\colhead{(3)}& \colhead{(4)}&\colhead{(5)}&\colhead{(6)}&
    \colhead{(7)}&\colhead{(8)}&\colhead{(9)}&\colhead{(10)}&\colhead{(11)}&\colhead{(12)}&\colhead{(13)}}
  \startdata
  IA598 & 6007 &  303 & 26.79 & 870.4 &  118097 &\pa641 &  48 &   28 &   71 &  45 &   12 &   52\\
  IA679 & 6780 &  340 & 27.39 & 870.4 &  139585 &\pa790 & 110 &  124 &  193 & 101 &  105 &  165\\
  NB704 & 7046 &  100 & 26.71 & 870.4 &  123123 &  1695 & 279 &  226 &  423 & 253 &  191 &  362\\
  NB711 & 7111 &\pa72 & 26.07 & 870.4 &\pa97632 &  1480 & 131 &  188 &  276 & 115 &  160 &  232\\
  NB816 & 8150 &  120 & 26.90 & 870.4 &  133273 &  1602 & 221 &  209 &  381 & 166 &  152 &  270\\
  NB921 & 9196 &  132 & 26.71 & 870.4 &  119541 &  2361 & 277 &  530 &  710 & 237 &  458 &  598\\
  NB973 & 9755 &  200 & 25.69 & 788.7 &\pa84786 &  1243 &  87 &  171 &  222 &  79 &  110 &  153\\\hline
  Total &\ldots&\ldots&\ldots &\ldots &  \ldots &  \Nem & 845 & 1313 & 1911 & 708 & 1031 & 1493\\
  \vspace{-3mm}
  \enddata
  \label{tab:SDF_sample}
  \tablecomments{(1): Name of narrowband or intermediate-band filter. (2): Central
    wavelength of filter, $\lambda_c$. (3): Full width at half maximum of filter.
    (4): 3$\sigma$ limiting imaging depth, $m_{\rm lim}(3\sigma)$. (5): Surveyed area.
    (6): Number of sources detected in each image mosaic, $N$. (7) Number of narrowband
    or intermediate-band excess emitters, $N_{\rm total}$. (8)--(10): Number of galaxies
    with targeted MMT and/or Keck spectra, $N_{\rm target}$. (11)--(13): Number of
    galaxies with robust spectroscopic redshift, $N_{\rm specz}$. As discussed in
    Section~\ref{sec:NBem}, some excess emitters are selected by more than one filter.
    Thus the total number of galaxies for each column (last row) is less than the sum.}
\end{deluxetable*}

%% file: tab2.final.tex
\begin{deluxetable*}{lccccccc}
  \tabletypesize{\scriptsize}
  \tablewidth{0pc}
  \tablecaption{Comparison between MMT and Keck Spectroscopic Emission-line Fluxes}
  \tablehead{
    \colhead{}&
    \colhead{}&
    \multicolumn{3}{c}{Before Median-based Correction}&
    \multicolumn{3}{c}{After Median-based Correction}\\
    \cline{3-5}
    \cline{6-8}
    Mask Name&
    \colhead{$N$}&
    \colhead{Median}&
    \colhead{Average}&
    \colhead{$\sigma$}&
    \colhead{Median}&
    \colhead{Average}&
    \colhead{$\sigma$}\\
    \colhead{}&
    \colhead{}&
    \colhead{(dex)}&
    \colhead{(dex)}&
    \colhead{(dex)}&
    \colhead{(dex)}&
    \colhead{(dex)}&
    \colhead{(dex)}\\
    \colhead{(1)}&
    \colhead{(2)}&
    \colhead{(3)}&
    \colhead{(4)}&
    \colhead{(5)}&
    \colhead{(6)}&
    \colhead{(7)}&
    \colhead{(8)}}
  \startdata
  SDFM01 & \pa1 &  +0.011 &   +0.016 &  0.041 & \pa0.000 &   +0.006 &  0.041\\
  SDFM02 & \pa2 &  +0.044 &   +0.049 &  0.159 & \pa0.000 &  --0.034 &  0.081\\
  SDFM04 & \pa0 &  \ldots &   \ldots & \ldots &   \ldots &   \ldots & \ldots\\
  SDFM06 & \pa1 & --0.037 &  --0.095 &  0.097 & \pa0.000 &  --0.059 &  0.097\\
  SDFM07 & \pa3 & --0.009 &   +0.139 &  0.265 & \pa0.000 &   +0.018 &  0.110\\
  SDFM10 & \pa1 &  +0.029 &   +0.036 &  0.025 & \pa0.000 &   +0.007 &  0.025\\
  SDFM11 & \pa8 &  +0.077 &   +0.062 &  0.222 & \pa0.000 &  --0.001 &  0.187\\
  SDFM22 & \pa1 &  +0.012 &   +0.012 & \ldots & \pa0.000 & \pa0.000 & \ldots\\
  SDFM24 & \pa2 &  +0.218 &   +0.253 &  0.415 & \pa0.000 &   +0.068 &  0.317\\
  SDFM25 & \pa2 & --0.074 &  --0.125 &  0.084 & \pa0.000 &  --0.025 &  0.090\\
  SDFM27 & \pa1 & --0.050 &  --0.099 &  0.146 & \pa0.000 &  --0.049 &  0.146\\
  SDFM28 & \pa0 &  \ldots &   \ldots & \ldots &   \ldots &   \ldots & \ldots\\
  SDFM30 & \pa9 & --0.010 &   +0.059 &  0.311 & \pa0.000 &  --0.041 &  0.164\\
  SDFM31 &   24 &  +0.032 & \pa0.000 &  0.204 & \pa0.000 &  --0.036 &  0.119\\
  SDFM32 &   14 & --0.001 &   +0.017 &  0.248 & \pa0.000 &  --0.044 &  0.141\\
  SDFM33 &   26 &  +0.052 &   +0.110 &  0.232 & \pa0.000 &  --0.008 &  0.108\\
  SDFM34 &   26 & --0.038 &  --0.004 &  0.137 & \pa0.000 &  --0.038 &  0.122\\
  SDFM35 &   25 &  +0.005 &   +0.014 &  0.164 & \pa0.000 &  --0.030 &  0.111\\
  SDFM36 &   30 &  +0.042 &   +0.072 &  0.213 & \pa0.000 &  --0.025 &  0.133\\
  SDFM37 &   22 &  +0.046 &   +0.078 &  0.210 & \pa0.000 &   +0.001 &  0.165\\
  SDFM38 &   22 &  +0.108 &   +0.140 &  0.221 & \pa0.000 &  --0.039 &  0.142\\
  SDFM39 &   23 & --0.030 &   +0.042 &  0.252 & \pa0.000 &  --0.047 &  0.164\\
  Stacked&   23 &  \ldots &   \ldots & \ldots &  --0.003 &  --0.028 &  0.194\\
  \vspace{-3mm}
  \enddata
  \tablecomments{(1): Keck/DEIMOS mask name. (2): Number of galaxies with both
    MMT and Keck spectra. (3)--(5): Median, average, and dispersion for the difference
    between MMT and Keck spectroscopic fluxes \textit{before} a median-based
    correction is applied on individual galaxies. (6)--(8):
    Median, average, and dispersion for the difference between MMT and Keck
    spectroscopic fluxes \textit{after} a median-based correction is applied
    on individual galaxies. The median-based correction is discussed in
    Section~\ref{sec:f_calib}. Differences in fluxes are provided in units of
    dex. An illustration of this MMT--Keck comparison is provided in
    Figure~\ref{fig:MMTKeck}.}
  \label{tab:MMTKeck_flux}
\end{deluxetable*}

%% file: tab3.final.tex
\begin{deluxetable}{lcccccc}
  \tabletypesize{\scriptsize}
  \tablewidth{0pc}
  \tablecaption{Comparison between Spectroscopic and Narrowband Excess Fluxes}
  \tablehead{
    \colhead{}&
    \multicolumn{3}{c}{Individual}&
    \multicolumn{3}{c}{Stacked}\\
    \cline{2-4}
    \cline{5-7}
    Filter&
    \colhead{$N$}&
    \colhead{Median}&
    \colhead{$\sigma$}&
    \colhead{$N$}&
    \colhead{Median}&
    \colhead{$\sigma$}\\
    \colhead{}&
    \colhead{}&
    \colhead{(dex)}&
    \colhead{(dex)}&
    \colhead{}&
    \colhead{(dex)}&
    \colhead{(dex)}\\
    \colhead{(1)}&
    \colhead{(2)}&
    \colhead{(3)}&
    \colhead{(4)}&
    \colhead{(5)}&
    \colhead{(6)}&
    \colhead{(7)}}
  \startdata
  \multicolumn{7}{c}{MMT/Hectospec Spectroscopy}\\
  NB704 &   433 & --0.005 & 0.179 &   158 & \pa0.000 & 0.176\\
  NB816 &   237 & --0.044 & 0.173 & \pa88 &  --0.044 & 0.144\\
  NB921 & \pa90 &  +0.036 & 0.282 & \pa25 &   +0.061 & 0.139\\\hline
  \multicolumn{7}{c}{Keck/DEIMOS Spectroscopy}\\
  NB704 &   147 & --0.078 & 0.204 & \ldots & \ldots & \ldots\\
  NB816 & \pa82 & --0.040 & 0.221 & \ldots & \ldots & \ldots\\
  NB921 &   385 & --0.053 & 0.189 & \ldots & \ldots & \ldots\\
  NB973 & \pa73 &  +0.041 & 0.244 & \ldots & \ldots & \ldots\\
  \vspace{-3mm}
  \enddata
  \tablecomments{(1): Filter name. (2)--(7): Number of galaxies ($N$)
    with detections of nebular emission lines that are responsible for
    the narrowband excess, and median and dispersion difference
    between spectroscopic fluxes and narrowband excess fluxes.
    Comparisons are made against individual spectra (Columns 2--4),
    as well as, stacked spectra (Columns 5--7).}
  \label{tab:f_calib}
\end{deluxetable}

%% file: tab4.final.tex
\begin{landscape}
\begin{deluxetable*}{llllccclc}
  \tabletypesize{\scriptsize}
  \tablewidth{0pc}
  \tablecaption{Summary of MMT \OIIIa-detected Spectroscopic Sample}
  \tablehead{
    Merged ID&
    \colhead{ID}&
    \colhead{Name}&
    \colhead{Line Sel.}&
    \colhead{R.A.}&
    \colhead{Decl.}&
    \colhead{\zspec}&
    \colhead{Obs. Dates (UT)}&
    \colhead{$t_{\rm int}$}\\
    \colhead{}&
    \colhead{}&
    \colhead{}&
    \colhead{}&
    \colhead{(hr)}&
    \colhead{(deg)}&
    \colhead{}&
    \colhead{}&
    \colhead{(minutes)}\\
    \colhead{(1)}&
    \colhead{(2)}&
    \colhead{(3)}&
    \colhead{(4)}&
    \colhead{(5)}&
    \colhead{(6)}&
    \colhead{(7)}&
    \colhead{(8)}&
    \colhead{(9)}}
  \startdata
  \ldots & MMT01 & NB816--140623                               &          \OIII & 13:25:16.868 & +27:39:06.920 & 0.6377 &     2008 Apr 14, 2014 Feb 27, Mar 25, 29--31 & 580.0\\
  \ldots & MMT02 & NB711--064628                               &          \OIII & 13:23:39.133 & +27:32:52.705 & 0.4327 &         2008 Mar 13, 2014 Feb 27--28, Mar 29 & 585.0\\
  \ldots & MMT03\TA & NB973--104154                            &            \Ha & 13:23:39.170 & +27:31:47.336 & 0.4809 &     2008 Mar 13, 2014 Feb 28, Mar 25, 29--31 & 605.0\\
  MK01   & MMT04 & IA679--112491\_NB704--088982\_NB921--126525 & \OIII+\Hb, \Ha & 13:24:46.633 & +27:34:56.981 & 0.3932 &     2008 Mar 13, Apr 11, 2014 Feb 28, Mar 29 & 485.0\\
  \ldots & MMT05 & IA679--031637                               &      \OIII+\Hb & 13:25:03.365 & +27:17:23.770 & 0.3846 &             2008 Mar 13, 2014 Feb 27, Mar 29 & 340.0\\
  \ldots & MMT07 & IA679--062450\_NB704--049936\_NB921--079428 & \OIII+\Hb, \Ha & 13:24:06.943 & +27:24:01.763 & 0.3894 &     2008 Mar 13, Apr 11, 2014 Mar 25, 30--31 & 480.0\\
  \ldots & MMT08 & NB816--081644                               &          \OIII & 13:23:42.968 & +27:26:35.592 & 0.6332 &     2008 Apr 10, 2014 Feb 28, Mar 25, 29--31 & 565.0\\
  \ldots & MMT09 & NB973--094500                               &       \Ha+\SII & 13:23:49.797 & +27:28:35.316 & 0.4787 &     2008 Apr 10, 2014 Feb 27, Mar 25, 29--31 & 540.0\\
\enddata
  \label{tab:MMT_source_summary}
  \tablecomments{(1): Abbreviated MMT+Keck merged name. (2): Abbreviated name for MMT \OIIIa-detected galaxy. (3): Full narrowband and/or intermediate-band
    excess emitter name. (4): Emission line(s) responsible for narrowband and/or intermediate-band excess. (5): R.A. in units of hours, minutes, and seconds.
    (6): Decl. in units of degrees, arcminutes, and arcseconds. (7): Spectroscopic redshift. (8): UT observation dates.
    (9): On-source integration time in units of minutes.}
  \tablenotetext{1}{This source is likely a LINER (see Section \SAGN\ of \citetalias{MACTII}).}
\end{deluxetable*}
\clearpage
\end{landscape}

%% file: tab5.final.tex
\begin{landscape}
\begin{deluxetable*}{llllccclr}
  \tabletypesize{\scriptsize}
  \tablewidth{0pc}
  \tablecaption{Summary of Keck \OIIIa-detected Spectroscopic Sample}
  \tablehead{
    Merged ID&
    \colhead{ID}&
    \colhead{Name}&
    \colhead{Line Sel.}&
    \colhead{R.A.}&
    \colhead{Decl.}&
    \colhead{\zspec}&
    \colhead{Obs. Dates (UT)}&
    \colhead{$t_{\rm int}$}\\
    \colhead{}&
    \colhead{}&
    \colhead{}&
    \colhead{}&
    \colhead{(hr)}&
    \colhead{(deg)}&
    \colhead{}&
    \colhead{}&
    \colhead{(min.)}\\
    \colhead{(1)}&
    \colhead{(2)}&
    \colhead{(3)}&
    \colhead{(4)}&
    \colhead{(5)}&
    \colhead{(6)}&
    \colhead{(7)}&
    \colhead{(8)}&
    \colhead{(9)}}
  \startdata
\ldots & Keck01 & IA679--079866\_NB711--049857                               &        \OII, \NeIII & 13:25:11.938 & +27:27:31.204 & 0.8390 &              2004 Apr 23 & 120.0\\
\ldots & Keck02 & NB816--070113                                              &               \OIII & 13:24:34.911 & +27:24:10.195 & 0.6230 & 2004 Apr 23, 2014 May 02 & 117.6\\
\ldots & Keck03 & IA679--077341                                              &         \OII+\NeIII & 13:23:53.540 & +27:27:13.008 & 0.7906 & 2004 Apr 23, 2015 Mar 17 & 217.6\\
\ldots & Keck04 & NB704--087569                                              &                \OII & 13:23:43.510 & +27:34:20.980 & 0.8830 & 2008 May 02, 2015 Mar 17 & 230.0\\
\ldots & Keck05 & NB921--078003                                              &               \OIII & 13:24:43.656 & +27:23:34.861 & 0.8354 & 2009 Apr 28, 2014 May 02 & 300.0\\
\ldots & Keck06\TA & NB704--060432                                           &              \NeIII & 13:24:58.623 & +27:26:40.468 & 0.8237 & 2009 Apr 28, 2014 May 02 & 300.0\\
\ldots & Keck07 & IA679--066277                                              &                \OII & 13:24:57.971 & +27:24:44.603 & 0.8568 & 2009 Apr 28, 2014 May 02 & 120.0\\
\ldots & Keck08 & NB711--035903\_NB973--073302                               &         \OII, \OIII & 13:24:32.827 & +27:22:11.509 & 0.9200 & 2014 May 02, 2015 Mar 19 & 211.2\\
\enddata
  \label{tab:Keck_source_summary}
  \tablecomments{(1): Abbreviated MMT+Keck merged name. (2): Abbreviated name for Keck \OIIIa-detected galaxy. (3): Full narrowband and/or intermediate-band
    excess emitter name. (4): Emission line(s) responsible for narrowband and/or intermediate-band excess. (5): R.A. in units of hours, minutes, and seconds.
    (6): Decl. in units of degrees, arcminutes, and arcseconds. (7): Spectroscopic redshift. (8): UT observation dates.
    (9): On-source integration time in units of minutes.}
  \tablenotetext{1}{This galaxy was previously targeted for its NB921 excess flux due to \OIII. It was not identified as a NB921 emitter in the
    re-selection because of its faintness. The NB704 excess is due to \NeIII\,$\lambda$3869.}
\end{deluxetable*}
\clearpage
\end{landscape}

%% file: tab6.final.tex
\begin{landscape}
\begin{deluxetable*}{p{1.0cm}cccccccccccc}
  \tabletypesize{\scriptsize}
  \tablewidth{0pc}
  \tablecaption{Strong Nebular Emission Lines for the \OIIIa-detected Sample}
  \tablehead{
    \colhead{}&
    \colhead{\OII}&
    \colhead{\NeIII}&
    \colhead{\NeIII, H$\epsilon$}&
    \colhead{H$\delta$}&
    \colhead{H$\gamma$}&
    \colhead{\OIII}&
    \colhead{\Hb}&
    \colhead{\OIII}&
    \colhead{\OIII}&
    \colhead{\Ha}&
    \colhead{\SII}&
    \colhead{\SII}\\
    ID&
    \colhead{3727}&
    \colhead{3869}&
    \colhead{3967, 3970}&
    \colhead{4101}&
    \colhead{4340}&
    \colhead{4363}&
    \colhead{4861}&
    \colhead{4959}&
    \colhead{5007}&
    \colhead{6563}&
    \colhead{6716}&
    \colhead{6731}\\
    \colhead{(1)}&
    \colhead{(2)}&
    \colhead{(3)}&
    \colhead{(4)}&
    \colhead{(5)}&
    \colhead{(6)}&
    \colhead{(7)}&
    \colhead{(8)}&
    \colhead{(9)}&
    \colhead{(10)}&
    \colhead{(11)}&
    \colhead{(12)}&
    \colhead{(13)}}
  \startdata
 MK01  &     2.40$\pm$0.01\TA &     0.41$\pm$0.01\TA &     0.20$\pm$0.01\TA &     0.26$\pm$0.01\TA &     0.51$\pm$0.01\TA &   0.069$\pm$0.007\TA &    21.74$\pm$0.08\TB &              1.53\TB &              4.77\TB &              3.04\TB &     0.28$\pm$0.01\TB &     0.30$\pm$0.01\TB\\
 MK02  &     1.73$\pm$0.01\TA &     0.34$\pm$0.01\TA &              0.22\TA &              0.23\TA &              0.39\TB &   0.033$\pm$0.003\TB &    68.00$\pm$0.17\TB &              1.47\TB &              4.54\TB &  2.43$\pm$0.01\TC\TB &              0.13\TB &              0.09\TB\\
 MK03  &     0.64$\pm$0.03\TA &     0.50$\pm$0.03\TA &     0.23$\pm$0.03\TA &     0.21$\pm$0.03\TA &     0.45$\pm$0.03\TA &   0.123$\pm$0.019\TB &     4.91$\pm$0.07\TB &     2.35$\pm$0.01\TB &     6.65$\pm$0.01\TB &     2.75$\pm$0.02\TB &           $<$0.05\TB &           $<$0.05\TB\\
 MK04  &     1.68$\pm$0.03\TA &     0.39$\pm$0.02\TA &     0.23$\pm$0.02\TA &     0.27$\pm$0.01\TB &     0.53$\pm$0.01\TB &   0.074$\pm$0.006\TB &    19.96$\pm$0.09\TB &           1.95\TC\TB &              5.92\TB &              3.06\TB &     0.16$\pm$0.01\TB &            \ldots\TD\\
MMT01  &        1.23$\pm$0.02 &        0.40$\pm$0.02 &        0.22$\pm$0.02 &     0.20$\pm$0.02\TC &        0.40$\pm$0.02 &      0.152$\pm$0.022 &       10.17$\pm$0.30 &        1.92$\pm$0.03 &        5.61$\pm$0.02 &            \ldots\TD &            \ldots\TD &            \ldots\TD\\
MMT02  &        2.82$\pm$0.03 &        0.25$\pm$0.02 &              $<$0.05 &        0.24$\pm$0.02 &        0.38$\pm$0.02 &      0.042$\pm$0.014 &        7.55$\pm$0.15 &        1.04$\pm$0.02 &        2.90$\pm$0.02 &            \ldots\TD &            \ldots\TD &            \ldots\TD\\
MMT03  &        3.68$\pm$0.06 &        0.24$\pm$0.04 &              $<$0.00 &        0.26$\pm$0.02 &        0.46$\pm$0.02 &      0.102$\pm$0.028 &        5.71$\pm$0.17 &        0.51$\pm$0.04 &        1.58$\pm$0.03 &            \ldots\TD &            \ldots\TD &            \ldots\TD\\
\enddata
  \label{tab:em_lines_strong}
  \tablecomments{(1): Abbreviated name for MMT/Keck \OIIIa-detected galaxy. (2)--(13): Emission-line fluxes from shorter to longer wavelengths.
    All fluxes are observed values, prior to any correction for dust attenuation. All line fluxes, except for \Hb,
    are reported relative to the \Hb\ intensity. The \Hb\ observed fluxes are in units of 10$^{-17}$ erg s$^{-1}$ cm$^{-2}$.
    Upper limits are provided for emission lines undetected at 3$\sigma$.}
  \tablenotetext{1}{Measurements are from MMT/Hectospec.}
  \tablenotetext{2}{Measurements are from Keck/DEIMOS.}
  \tablenotetext{3}{Measurements may be affected by OH night skyline contamination. This is determined if at least half of the
    line falls on a region with any OH skyline emission. This does not account for the strength of the emission line.}
  \tablenotetext{4}{Measurements are unavailable due to limited spectral coverage.}
  \tablenotetext{5}{The \Hb\ line fell inside a CCD gap. The reported flux here is derived from \Hg\ assuming zero reddening.}
  \tablenotetext{6}{\OIII\,$\lambda$5007 are unavailable due to limited spectral coverage. We adopt a flux that is 3.1 times the 4959\AA\ flux.}
  \tablenotetext{7}{\OII\ fluxes were unavailable. We adopt a relation between \OII/\Hb\ and \OIII/\Hb\ (see Figure~\ref{fig:OIIOIII}).}
\end{deluxetable*}
\clearpage
\end{landscape}

%% file: tab7.final.tex
\begin{landscape}
\begin{deluxetable}{p{1.0cm}ccccccccccccc}
  \tabletypesize{\scriptsize}
  \tablewidth{0pc}
  \tablecaption{Weak Nebular Emission Lines for the \OIIIa-detected Sample}
  \tablehead{
    \colhead{}&
    \colhead{H10}&
    \colhead{H9}&
    \colhead{H$\zeta$, He \textsc{i}}&
    \colhead{He \textsc{i}}&
    \colhead{He \textsc{i}}&
    \colhead{[{\rm O}~\textsc{i}]}&
    \colhead{\NII}&
    \colhead{\NII}&
    \colhead{He \textsc{i}}&
    \colhead{He \textsc{i}}&
    \colhead{[{\rm Ar}~\textsc{iii}]}&
    \colhead{[{\rm O}~\textsc{ii}]}&
    \colhead{[{\rm O}~\textsc{ii}]}\\
    ID&
    \colhead{3798}&
    \colhead{3835}&
    \colhead{3888, 3889}&
    \colhead{4471}&
    \colhead{5876}&
    \colhead{6300}&
    \colhead{6548}&
    \colhead{6583}&
    \colhead{6678}&
    \colhead{7065}&
    \colhead{7135}&
    \colhead{7319}&
    \colhead{7330}\\
    \colhead{(1)}&
    \colhead{(2)}&
    \colhead{(3)}&
    \colhead{(4)}&
    \colhead{(5)}&
    \colhead{(6)}&
    \colhead{(7)}&
    \colhead{(8)}&
    \colhead{(9)}&
    \colhead{(10)}&
    \colhead{(11)}&
    \colhead{(12)}&
    \colhead{(13)}&
    \colhead{(14)}}
  \startdata
 MK01  &           $<$0.50\TB &           $<$0.09\TB &     0.17$\pm$0.01\TB &     0.04$\pm$0.00\TA &  0.10$\pm$0.00\TD\TA &  0.11$\pm$0.00\TD\TA &              0.03\TA &              0.10\TA &           $<$0.06\TA &            \ldots\TC &            \ldots\TC &            \ldots\TC &            \ldots\TC\\
 MK02  &           $<$0.01\TB &           $<$0.11\TB &     0.13$\pm$0.01\TB &     0.03$\pm$0.00\TA &     0.12$\pm$0.00\TA &     0.06$\pm$0.00\TA &           $<$0.01\TA &     0.15$\pm$0.01\TA &           $<$0.01\TA &            \ldots\TC &            \ldots\TC &            \ldots\TC &            \ldots\TC\\
 MK03  &           $<$0.15\TB &           $<$1.91\TB &     0.14$\pm$0.03\TB &           $<$0.07\TB &     0.10$\pm$0.01\TA &  0.04$\pm$0.01\TD\TA &           $<$0.04\TA &           $<$0.08\TA &     0.08$\pm$0.01\TA &            \ldots\TC &            \ldots\TC &            \ldots\TC &            \ldots\TC\\
 MK04  &            \ldots\TC &            \ldots\TC &     0.15$\pm$0.02\TB &     0.04$\pm$0.01\TA &     0.13$\pm$0.00\TA &  0.02$\pm$0.00\TD\TA &              0.04\TA &              0.12\TA &     0.06$\pm$0.01\TA &            \ldots\TC &            \ldots\TC &            \ldots\TC &            \ldots\TC\\
MMT01  &              $<$0.06 &              $<$0.13 &        0.12$\pm$0.02 &              $<$0.07 &            \ldots\TC &            \ldots\TC &            \ldots\TC &            \ldots\TC &            \ldots\TC &            \ldots\TC &            \ldots\TC &            \ldots\TC &            \ldots\TC\\
MMT02  &              $<$0.31 &              $<$0.01 &              $<$0.06 &              $<$0.05 &  0.15$\pm$0.02\TD\TA &              $<$0.36 &            \ldots\TC &            \ldots\TC &            \ldots\TC &            \ldots\TC &            \ldots\TC &            \ldots\TC &            \ldots\TC\\
\enddata
  \label{tab:em_lines_weak}
  \tablecomments{(1): Abbreviated name for MMT/Keck \OIIIa-detected galaxy. (2)--(14): Emission-line fluxes from shorter to longer wavelengths.
    All fluxes are observed values, prior to any correction for dust attenuation. All line fluxes, except for \Hb,
    are reported relative to the \Hb\ intensity. Upper limits are provided for emission lines undetected at 3$\sigma$.}
  \tablenotetext{1}{Measurements are from Keck/DEIMOS.}
  \tablenotetext{2}{Measurements are from MMT/Hectospec.}
  \tablenotetext{3}{Measurements are unavailable due to limited spectral coverage.}
  \tablenotetext{4}{Measurements are affected by OH night skyline contamination. This is determined if at least half of the
    line falls on a region with any OH skyline emission. This does not account for the strength of the emission line.}
\end{deluxetable}
\clearpage
\end{landscape}

%% file: tab8.final.tex
\begin{deluxetable*}{lccccc}
  \tabletypesize{\scriptsize}
  \tablewidth{0pc}
  \tablecaption{Flux Selection for \OIIIa-non-detected Samples}
  \tablehead{
    \colhead{Redshift}&
    \colhead{Flux(\OIII)}&
    \colhead{$N_z$}&
    \colhead{$N_{\OIII}$}&
    \colhead{$N_{\rm ND}$}&
    \colhead{$N_{\rm final}$}\\
    \colhead{(1)}&
    \colhead{(2)}&
    \colhead{(3)}&
    \colhead{(4)}&
    \colhead{(5)}&
    \colhead{(6)}}
  \startdata
  \multicolumn{6}{c}{MMT/Hectospec Selection}\\
  $z\leq0.3$           & $1.67\times10^{-16}$ ($1.86\times10^{-16}$) &   101 &   100 &  22 &   22 (21)\\
  $0.3 < z \leq 0.5$   & $1.59\times10^{-16}$ ($1.91\times10^{-16}$) &   389 &   378 &  38 &   28 (19)\\
  $0.5 < z \leq 1.0$   & $2.58\times10^{-16}$ ($3.28\times10^{-16}$) &   148 &   122 &   7 & \pa1 (\pa0)\\\hline
  \multicolumn{6}{c}{Keck/DEIMOS Selection}\\
  $0.10 < z \leq 0.30$ & $1.30\times10^{-16}$ ($1.50\times10^{-16}$) & \pa26 & \pa25 &   1 & \pa0 (\pa0)\\
  $0.30 < z \leq 0.50$ & $6.95\times10^{-17}$ ($8.11\times10^{-17}$) &   255 &   252 &  41 &   29 (24)\\
  $0.50 < z \leq 0.70$ & $4.73\times10^{-17}$ ($5.88\times10^{-17}$) &   123 &   119 &  32 &   21 (16)\\
\enddata
  \tablecomments{(1): Redshift range. (2): \OIII$\lambda$5007 emission-line flux limit in
    units of erg s$^{-1}$ cm$^{-2}$. The first values are the initial flux limits for sample
    selection. The values in parentheses are stricter flux limits that are later adopted for
    the \MZ\ relation analysis in \citetalias{MACTII}. (3): Number of galaxies with MMT/Keck
    spectra within redshift range, $N_z$. (4): Number of $N_z$ galaxies with \OIII\ measurements,
    $N_{\OIII}$. (5): Number of $N_{\OIII}$ galaxies that meet the \OIII\ flux limit, $N_{\rm ND}$.
    (6): Number of $N_{\rm ND}$ galaxies that remains after visual examination, $N_{\rm final}$
    (see Section~\ref{sec:reliable}). The values in parentheses are those with the stricter
    \OIII\ flux limits.}
  \label{tab:nondet_sel}
\end{deluxetable*}

%% file: tab9.final.tex
\begin{landscape}
\begin{deluxetable*}{llllccclr}
  \tabletypesize{\scriptsize}
  \tablewidth{0pc}
  \tablecaption{Summary of MMT \OIIIa-non-detected Spectroscopic Sample}
  \tablehead{
    Merged ID&
    \colhead{ID}&
    \colhead{Name}&
    \colhead{Line Sel.}&
    \colhead{R.A.}&
    \colhead{Decl.}&
    \colhead{\zspec}&
    \colhead{Obs. Dates (UT)}&
    \colhead{$t_{\rm int}$}\\
    \colhead{}&
    \colhead{}&
    \colhead{}&
    \colhead{}&
    \colhead{(hr)}&
    \colhead{(deg)}&
    \colhead{}&
    \colhead{}&
    \colhead{(min.)}\\
    \colhead{(1)}&
    \colhead{(2)}&
    \colhead{(3)}&
    \colhead{(4)}&
    \colhead{(5)}&
    \colhead{(6)}&
    \colhead{(7)}&
    \colhead{(8)}&
    \colhead{(9)}}
  \startdata
    MK05 & MMT12\TA & NB816--112403                               &           \OIII & 13:25:21.779 & +27:33:15.694 & 0.6404 &         2008 Apr 10, 2014 Feb 27--28, Mar 29 & 545.0\\
  \ldots & MMT40    & NB816--020228                               &             \Ha & 13:25:00.341 & +27:14:10.575 & 0.2294 &     2008 Apr 10, 2014 Feb 27, Mar 25, 29--31 & 540.0\\
  \ldots & MMT41    & NB816--042325                               &             \Ha & 13:24:33.171 & +27:18:15.605 & 0.2396 &     2008 Mar 13, 2014 Feb 27, Mar 25, 29--31 & 580.0\\
  \ldots & MMT42    & NB816--049386                               &             \Ha & 13:23:58.502 & +27:19:42.156 & 0.2485 &     2008 Apr 11, 2014 Feb 27, Mar 25, 29--31 & 580.0\\
  \ldots & MMT43    & NB816--053897                               &             \Ha & 13:23:44.341 & +27:20:41.263 & 0.2412 &         2008 Apr 14, 2014 Feb 27--28, Mar 29 & 585.0\\
  \ldots & MMT44    & IA598--070330\_NB816--063052                &  \OIII+\Hb, \Ha & 13:25:11.510 & +27:22:46.493 & 0.2328 &         2008 Apr 10, 2014 Feb 27--28, Mar 29 & 545.0\\
  \ldots & MMT45    & NB816--067584                               &             \Ha & 13:23:41.276 & +27:23:43.814 & 0.2472 &     2008 Apr 14, 2014 Feb 28, Mar 25, 29--31 & 605.0\\
  \ldots & MMT46    & NB816--068175                               &             \Ha & 13:24:39.064 & +27:23:49.369 & 0.2474 &         2008 Apr 11, 2014 Feb 27--28, Mar 29 & 585.0\\
\enddata
  \label{tab:MMT_source_summary_nondet}
  \tablecomments{(1): Abbreviated MMT+Keck merged name. (2): Abbreviated name for MMT \OIIIa-non-detected galaxy. (3): Full narrowband and/or intermediate-band
    excess emitter name. (4): Emission line(s) responsible for narrowband and/or intermediate-band excess. (5): R.A. in units of hours, minutes, and seconds.
    (6): Decl. in units of degrees, arcminutes, and arcseconds. (7): Spectroscopic redshift. (8): UT observation dates.
    (9): On-source integration time in units of minutes.}
  \tablenotetext{1}{Excluded from \OIIIa-non-detected sample with stricter \OIII$\lambda$5007 flux limit for \MZ\ relation analyses.}
\end{deluxetable*}
\clearpage
\end{landscape}

%% file: tab10.final.tex
\begin{landscape}
\begin{deluxetable*}{llllccclr}
  \tabletypesize{\scriptsize}
  \tablewidth{0pc}
  \tablecaption{Summary of Keck \OIIIa-non-detected Spectroscopic Sample}
  \tablehead{
    Merged ID&
    \colhead{ID}&
    \colhead{Name}&
    \colhead{Line Sel.}&
    \colhead{R.A.}&
    \colhead{Decl.}&
    \colhead{\zspec}&
    \colhead{Obs. Dates (UT)}&
    \colhead{$t_{\rm int}$}\\
    \colhead{}&
    \colhead{}&
    \colhead{}&
    \colhead{}&
    \colhead{(hr)}&
    \colhead{(deg)}&
    \colhead{}&
    \colhead{}&
    \colhead{(minutes)}\\
    \colhead{(1)}&
    \colhead{(2)}&
    \colhead{(3)}&
    \colhead{(4)}&
    \colhead{(5)}&
    \colhead{(6)}&
    \colhead{(7)}&
    \colhead{(8)}&
    \colhead{(9)}}
  \startdata
\ldots & Keck034    & NB973--066061                               &           \Ha & 13:24:37.504 & +27:20:03.909 & 0.4830 & 2015 Mar 19, 2015 Mar 26 & 231.2\\
\ldots & Keck035    & NB704--089683\_NB921--127544                &           \Ha & 13:24:44.854 & +27:34:53.163 & 0.3924 &          2015 Mar 16, 26 & 268.7\\
\ldots & Keck036\TA & NB711--044949\_NB973--085127                & \Hb, \Ha+\SII & 13:25:08.009 & +27:25:43.799 & 0.4638 & 2004 Apr 23, 2014 May 02 & 240.0\\
\ldots & Keck037    & NB816--031351                               &         \OIII & 13:24:27.480 & +27:16:08.479 & 0.6338 & 2015 Mar 19, 2015 Mar 26 & 260.0\\
\ldots & Keck038    & NB816--044249                               &         \OIII & 13:24:47.966 & +27:18:42.144 & 0.6361 & 2015 Mar 19, 2015 Mar 26 & 231.2\\
\ldots & Keck039    & NB816--055523                               &         \OIII & 13:25:20.409 & +27:21:08.928 & 0.6381 & 2004 Apr 23, 2015 Mar 26 & 260.0\\
\ldots & Keck040\TA & NB816--062942                               &         \OIII & 13:25:22.943 & +27:22:36.976 & 0.6350 & 2009 Apr 27, 2015 Mar 26 & 140.0\\
\ldots & Keck041    & NB921--116861                               &         \OIII & 13:25:16.212 & +27:32:31.769 & 0.8414 &              2004 Apr 24 & 153.2\\
\enddata                                                      
  \label{tab:Keck_source_summary_nondet}
  \tablecomments{(1): Abbreviated MMT+Keck merged name. (2): Abbreviated name for Keck \OIIIa-non-detected galaxy. (3): Full narrowband
      and/or intermediate-band excess emitter name. (4): Emission line(s) responsible for narrowband and/or intermediate-band excess.
      (5): R.A. in units of hours, minutes, and seconds. (6): Decl. in units of degrees, arcminutes, and arcseconds.
      (7): Spectroscopic redshift. (8): UT observation dates. (9): On-source integration time in units of minutes.}
  \tablenotetext{1}{Excluded from \OIIIa-non-detected sample with stricter \OIII$\lambda$5007 flux limit for \MZ\ relation analyses.}
\end{deluxetable*}
\clearpage
\end{landscape}

%% file: tab11.final.tex
\newcommand{\pc}{\phm{$^c$}}

\begin{deluxetable*}{lccccccc}
  \tabletypesize{\scriptsize}
  \tablewidth{0pc}
  \tablecaption{Balmer Decrement Measurements and Derived Dust Reddening for \OIIIa-detected Sample}
  \tablehead{
    ID&
    \colhead{\Ha/\Hb}&
    \colhead{\Hg/\Hb}&
    \colhead{EW(\Hb)}&
    \colhead{EW(\Hg)}&
    \colhead{EW(\Hd)}&
    \colhead{Source}&
    \colhead{\EBVa}\\
    \colhead{}&
    \colhead{}&
    \colhead{}&
    \colhead{(\AA)}&
    \colhead{(\AA)}&
    \colhead{(\AA)}&
    \colhead{}&
    \colhead{(mag)}\\
    \colhead{(1)}&
    \colhead{(2)}&
    \colhead{(3)}&
    \colhead{(4)}&
    \colhead{(5)}&
    \colhead{(6)}&
    \colhead{(7)}&
    \colhead{(8)}}
  \startdata
  MK01 & 3.04$^{+0.01}_{-0.01}$ & 0.51$^{+0.01}_{-0.01}$ & 0.0 &       3.3 & 2.1       & \Ha/\Hb & 0.06\\[1.5mm]
  MK02 & 2.43$^{+0.01}_{-0.01}$ & 0.39$^{+0.00}_{-0.00}$ & 0.0 &       1.0 & 1.6       & \Ha/\Hb & 0.00\TA\\[1.5mm]
  MK03 & 2.75$^{+0.04}_{-0.04}$ & 0.45$^{+0.03}_{-0.02}$ & 0.0 &       1.0 & 2.0       & \Ha/\Hb & 0.00$\pm$0.01\TA\\[1.5mm]
  MK04 & 3.06$^{+0.02}_{-0.02}$ & 0.53$^{+0.01}_{-0.01}$ & 0.0 &       0.0 & 2.0       & \Ha/\Hb & 0.07\\[1.5mm]
 MMT01 &               \ldots & 0.40$^{+0.02}_{-0.02}$ & 0.0 &       1.0 & 2.0       & \Hg/\Hb & 0.30$\pm$0.09\\[1.5mm]
 MMT02 &               \ldots & 0.38$^{+0.02}_{-0.02}$ & 2.8 &       2.0 & 3.8       & \Hg/\Hb & 0.39$\pm$0.12\\[1.5mm]
 MMT03 &               \ldots & 0.46$^{+0.03}_{-0.03}$ & 1.7 &       2.4 & 2.4       & \Hg/\Hb & 0.02$^{+0.09}_{-0.12}$\\[1.5mm]
 MMT05 & 2.79$^{+0.03}_{-0.02}$ & 0.49$^{+0.03}_{-0.02}$ & 0.0 &       0.0 & 0.0       & \Ha/\Hb & 0.00$\pm$0.01\TA\\
\enddata
  \label{tab:balmer}
  \tablecomments{(1): Abbreviated name for MMT/Keck \OIIIa-detected galaxy. (2): \Ha/\Hb\ Balmer decrement.
    (3): \Hg/\Hb\ Balmer decrement. (4): Equivalent width of stellar absorption correction for \Hb.
    (5): Equivalent width of stellar absorption correction for \Hg. (6): EW of stellar absorption correction
    for \Hd. (7): Balmer decrement used to determine nebular reddening. (8): Derived color excess for dust
    reddening in the nebular gas.}
  \tablenotetext{1}{Negative reddening was seen. We adopt \EBVa\ = 0.0.}
  \tablenotetext{2}{The continuum is too weak to measure the EW of stellar absorption correction.}
  \tablenotetext{3}{\Hg\ is affected by an OH skyline.}
  \tablenotetext{4}{\Hb\ affected by a gap between the blue and red CCDs.}
  \tablenotetext{5}{The \Hg\ and \Hd\ fluxes suggest no reddening.}
\end{deluxetable*}
\clearpage

%% file: tab12.final.tex
\begin{deluxetable*}{lccccccc}
  \tabletypesize{\scriptsize}
  \tablewidth{0pc}
  \tablecaption{Stacked Balmer Decrement Measurements for \OIIIa-non-detected Sample}
  \tablehead{
    \colhead{}&
    \colhead{}&
    \multicolumn{3}{c}{MMT}&
    \multicolumn{3}{c}{Keck}\\
    \cline{3-5}
    \cline{6-8}
    \colhead{$\log(L[\Hbe]/{\rm erg~s}^{-1})$}&
    \colhead{Redshift}&
    \colhead{\Ha/\Hb}&
    \colhead{\Hg/\Hb}&
    \colhead{\Hd/\Hb}&
    \colhead{\Ha/\Hb}&
    \colhead{\Hg/\Hb}&
    \colhead{\Hd/\Hb}\\
    \colhead{(1)}&
    \colhead{(2)}&
    \colhead{(3)}&
    \colhead{(4)}&
    \colhead{(5)}&
    \colhead{(6)}&
    \colhead{(7)}&
    \colhead{(8)}}
  \startdata
  \multicolumn{8}{c}{MMT Sample}\\
  38.452--40.127 & 0.035--0.346 & 3.60$\pm$0.07 (12) & 0.50$\pm$0.02 (15) & 0.25$\pm$0.02 (14) &             \ldots &             \ldots &             \ldots\\
  40.129--40.460 & 0.206--0.467 & 3.80$\pm$0.14 (10) & 0.40$\pm$0.01 (17) & 0.26$\pm$0.03 (14) & 5.83$\pm$0.14 (02) &             \ldots &             \ldots\\
  40.468--41.126 & 0.359--0.640 & 2.40$\pm$0.07 (05) & 0.47$\pm$0.01 (16) & 0.26$\pm$0.02 (14) & 3.07$\pm$0.02 (10) & 0.64$\pm$0.03 (07) & 0.17$\pm$0.02 (02)\\\hline
  \multicolumn{8}{c}{Keck Sample}\\
  40.072--40.518 & 0.369--0.633 & 2.52$\pm$0.32 (06) & 0.34$\pm$0.03 (15) & 0.26$\pm$0.05 (15) & 3.46$\pm$0.06 (13) & 0.63$\pm$0.04 (18) & 0.14$\pm$0.03 (06)\\
  40.525--40.783 & 0.405--0.676 &             \ldots & 0.46$\pm$0.03 (13) & 0.25$\pm$0.02 (13) & 5.67$\pm$0.11 (05) & 0.57$\pm$0.03 (19) & 0.28$\pm$0.03 (14)\\
  40.800--41.048 & 0.407--0.907 &             \ldots &             \ldots &             \ldots &             \ldots & 0.56$\pm$0.02 (20) & 0.30$\pm$0.01 (19)\\
\enddata
  \label{tab:balmer_nondet}
  \tablecomments{(1): Range in \Hb\ luminosity for each $\log(L[\Hbe])$ bin. (2): Range in redshift. (3): \Ha/\Hb\ Balmer
    decrement from MMT spectra. (4): \Hg/\Hb\ Balmer decrement from MMT spectra. (5): \Hd/\Hb\ Balmer decrement from MMT spectra.
    (6): \Ha/\Hb\ Balmer decrement from Keck spectra. (7): \Hg/\Hb\ Balmer decrement from Keck spectra. (8): \Hd/\Hb\ Balmer
    decrement from Keck spectra. Values within parentheses for Columns (3)--(8) indicate the number of galaxies used to compute
    the Balmer decrements.}
\end{deluxetable*}

%% file: tab13.final.tex
\newcommand{\phc}{\phantom{$<$}}
\begin{landscape}
\begin{deluxetable*}{lccccccccccccc}
  \tabletypesize{\scriptsize}
  \tablewidth{0pc}
  \tablecaption{Emission-line Ratios, Electron Densities and Temperatures, and
    Gas-phase Metallicities for \OIIIa-detected Galaxies}
  \tablehead{
    ID&
    \multicolumn{2}{c}{$\log{\left(\frac{\OIII}{\OIII\ \lambda4363}\right)}$}&
    \multicolumn{2}{c}{$\log{(T_e/{\rm K})}$}&
    \multicolumn{2}{c}{$\log{\left(\frac{\OII}{\Hbe}\right)}$}&
    \colhead{$\log{\left(\frac{\OIII}{\Hbe}\right)}$}&
    \multicolumn{2}{c}{$\log{\left(\frac{{\rm O}^+}{{\rm H}^+}\right)}$}&
    \colhead{$\log{\left(\frac{{\rm O}^{++}}{{\rm H}^+}\right)}$}&
    \multicolumn{2}{c}{$12+\log{\left(\frac{{\rm O}}{{\rm H}}\right)}$}&
    \colhead{$n_e$}\\
    \cline{2-3}
    \cline{4-5}
    \cline{6-7}
    \cline{9-10}
    \cline{12-13}
    \colhead{}&
    \colhead{(dex)}&
    \colhead{(dex)}&
    \colhead{(dex)}&
    \colhead{(dex)}&
    \colhead{(dex)}&
    \colhead{(dex)}&
    \colhead{(dex)}&
    \colhead{(dex)}&
    \colhead{(dex)}&
    \colhead{(dex)}&
    \colhead{(dex)}&
    \colhead{(dex)}&
    \colhead{(cm$^{-3}$)}\\
    \colhead{(1)}&
    \colhead{(2)}&
    \colhead{(3)}&
    \colhead{(4)}&
    \colhead{(5)}&
    \colhead{(6)}&
    \colhead{(7)}&
    \colhead{(8)}&
    \colhead{(9)}&
    \colhead{(10)}&
    \colhead{(11)}&
    \colhead{(12)}&
    \colhead{(13)}&
    \colhead{(14)}
  }
  \startdata
   MK01 & 1.96$^{+0.04}_{-0.04}$ & 1.94$^{+0.04}_{-0.04}$ & 4.11$^{+0.01}_{-0.02}$ & 4.11$^{+0.01}_{-0.02}$ & +0.38$^{+0.00}_{-0.00}$ & +0.41$^{+0.00}_{-0.00}$ & 0.80$^{+0.00}_{-0.00}$ & -4.08$^{+0.07}_{-0.07}$ & -4.07$^{+0.08}_{-0.07}$ & -4.12$^{+0.05}_{-0.05}$ & 8.21$^{+0.06}_{-0.06}$ & 8.21$^{+0.06}_{-0.06}$ & 927$^{+153}_{-99}$\\[1.5mm]
   MK02 & 2.27$^{+0.03}_{-0.03}$ & 2.27$^{+0.03}_{-0.03}$ & 4.00$^{+0.01}_{-0.01}$ & 4.00$^{+0.01}_{-0.01}$ & +0.24$^{+0.00}_{-0.00}$ & +0.24$^{+0.00}_{-0.00}$ & 0.78$^{+0.00}_{-0.00}$ & -3.80$^{+0.05}_{-0.06}$ & -3.80$^{+0.05}_{-0.06}$ & -3.77$^{+0.04}_{-0.04}$ & 8.51$^{+0.03}_{-0.05}$ & 8.51$^{+0.03}_{-0.05}$ & $<$1\\[1.5mm]
   MK03 & 1.87$^{+0.06}_{-0.07}$ & 1.87$^{+0.06}_{-0.07}$ & 4.15$^{+0.03}_{-0.03}$ & 4.15$^{+0.03}_{-0.03}$ & -0.20$^{+0.03}_{-0.02}$ & -0.20$^{+0.03}_{-0.02}$ & 0.95$^{+0.00}_{-0.01}$ & -4.79$^{+0.08}_{-0.14}$ & -4.79$^{+0.08}_{-0.14}$ & -4.05$^{+0.06}_{-0.08}$ & 8.02$^{+0.08}_{-0.09}$ & 8.02$^{+0.08}_{-0.09}$ & \ldots\\[1.5mm]
   MK04 & 2.03$^{+0.03}_{-0.03}$ & 2.01$^{+0.03}_{-0.03}$ & 4.08$^{+0.01}_{-0.01}$ & 4.09$^{+0.01}_{-0.01}$ & +0.22$^{+0.01}_{-0.01}$ & +0.26$^{+0.01}_{-0.01}$ & 0.90$^{+0.00}_{-0.00}$ & -4.13$^{+0.05}_{-0.06}$ & -4.12$^{+0.05}_{-0.05}$ & -3.94$^{+0.04}_{-0.04}$ & 8.29$^{+0.04}_{-0.05}$ & 8.28$^{+0.05}_{-0.05}$ & \ldots\\[1.5mm]
  MMT01 & 1.69$^{+0.05}_{-0.07}$ & 1.62$^{+0.05}_{-0.07}$ & 4.23$^{+0.03}_{-0.03}$ & 4.28$^{+0.04}_{-0.03}$ & +0.10$^{+0.01}_{-0.02}$ & +0.23$^{+0.07}_{-0.05}$ & 0.88$^{+0.01}_{-0.01}$ & -4.77$^{+0.09}_{-0.13}$ & -4.75$^{+0.08}_{-0.14}$ & -4.44$^{+0.08}_{-0.08}$ & 7.79$^{+0.08}_{-0.09}$ & 7.73$^{+0.09}_{-0.09}$ & \ldots\\[1.5mm]
  MMT02 & 1.97$^{+0.15}_{-0.13}$ & 1.88$^{+0.13}_{-0.12}$ & 4.10$^{+0.05}_{-0.07}$ & 4.14$^{+0.06}_{-0.08}$ & +0.45$^{+0.01}_{-0.01}$ & +0.66$^{+0.03}_{-0.08}$ & 0.60$^{+0.01}_{-0.01}$ & -3.99$^{+0.27}_{-0.28}$ & -3.94$^{+0.27}_{-0.22}$ & -4.40$^{+0.18}_{-0.16}$ & 8.19$^{+0.24}_{-0.20}$ & 8.19$^{+0.23}_{-0.24}$ & \ldots\\
\enddata
  \label{tab:metals}
  \tablecomments{(1): Abbreviated name for MMT/Keck \OIIIa-detected galaxy. (2)--(3): Flux ratio of \OIII$\lambda\lambda$4959,5007 to \OIIIa.
    (4)--(5): Electron temperature from \OIIIa\ (i.e., ${\rm O}^{++}$). (6)--(7): Flux ratio of \OII$\lambda$3727/\Hb.
    (8): Flux ratio of \OIII$\lambda\lambda$4959,5007/\Hb. (9)--(10): Gas-phase abundance of ${\rm O}^+/{\rm H}^+$.
    (11): Gas-phase abundance of ${\rm O}^{++}/{\rm H}^+$. (12)--(13): Gas-phase oxygen abundance.
    (14): Electron density in units of cm$^{-3}$. Columns (3), (5), (7), (10), and (13) include dust attenuation corrections
    (see Section~\ref{sec:dust}). Uncertainties are reported at 68\% confidence levels.}
  \tablenotetext{1}{This source is likely a LINER, so metallicity determinations are not trustworthy (see Section \SAGN\ of \citetalias{MACTII}).}
  \tablenotetext{2}{\OII\ fluxes were unavailable. We adopt a relation between \OII/\Hb\ and \OIII/\Hb\ (see Figure~\ref{fig:OIIOIII}).}
\end{deluxetable*}
\clearpage
\end{landscape}

%% file: tab14.final.tex
\begin{landscape}
\begin{deluxetable*}{lccccccccccccc}
  \tabletypesize{\scriptsize}
  \tablewidth{0pc}
  \tablecaption{Emission-line Ratios, Electron Densities and Temperatures, and
      Gas-Phase Metallicities for \OIIIa-non-detected Galaxies}
  \tablehead{
    ID&
    \multicolumn{2}{c}{$\log{\left(\frac{\OIII}{\OIII\ \lambda4363}\right)}$}&
    \multicolumn{2}{c}{$\log{(T_e/{\rm K})}$}&
    \multicolumn{2}{c}{$\log{\left(\frac{\OII}{\Hbe}\right)}$}&
    \colhead{$\log{\left(\frac{\OIII}{\Hbe}\right)}$}&
    \multicolumn{2}{c}{$\log{\left(\frac{{\rm O}^+}{{\rm H}^+}\right)}$}&
    \colhead{$\log{\left(\frac{{\rm O}^{++}}{{\rm H}^+}\right)}$}&
    \multicolumn{2}{c}{$12+\log{\left(\frac{{\rm O}}{{\rm H}}\right)}$}&
    \colhead{$n_e$}\\
    \cline{2-3}
    \cline{4-5}
    \cline{6-7}
    \cline{9-10}
    \cline{12-13}
    \colhead{}&
    \colhead{(dex)}&
    \colhead{(dex)}&
    \colhead{(dex)}&
    \colhead{(dex)}&
    \colhead{(dex)}&
    \colhead{(dex)}&
    \colhead{(dex)}&
    \colhead{(dex)}&
    \colhead{(dex)}&
    \colhead{(dex)}&
    \colhead{(dex)}&
    \colhead{(dex)}&
    \colhead{(cm$^{-3}$)}\\
    \colhead{(1)}&
    \colhead{(2)}&
    \colhead{(3)}&
    \colhead{(4)}&
    \colhead{(5)}&
    \colhead{(6)}&
    \colhead{(7)}&
    \colhead{(8)}&
    \colhead{(9)}&
    \colhead{(10)}&
    \colhead{(11)}&
    \colhead{(12)}&
    \colhead{(13)}&
    \colhead{(14)}}
  \startdata
   MK05    & $>$1.98 & $>$1.87 & $<$4.10 & $<$4.15 & +0.51$^{+0.00}_{-0.01}$ & +0.71$^{+0.07}_{-0.02}$ & 0.64$^{+0.01}_{-0.01}$ & $>$-3.92 & $>$-3.86 & $>$-4.37 & $>$8.25 & $>$8.26 & \ldots\\[1.5mm]
   MK06    & $>$2.35 & $>$2.30 & $<$3.97 & $<$3.99 & +0.43$^{+0.00}_{-0.01}$ & +0.54$^{+0.00}_{-0.01}$ & 0.66$^{+0.00}_{-0.00}$ & $>$-3.50 & $>$-3.47 & $>$-3.85 & $>$8.68 & $>$8.68 & 2$^{+123}_{-2}$\\[1.5mm]
   MK07    & $>$2.00 & $>$1.94 & $<$4.09 & $<$4.11 & +0.44$^{+0.02}_{-0.01}$ & +0.58$^{+0.02}_{-0.03}$ & 0.61$^{+0.02}_{-0.01}$ & $>$-3.95 & $>$-3.91 & $>$-4.30 & $>$8.24 & $>$8.24 & \ldots\\[1.5mm]
   MK08    & $>$1.90 & $>$1.89 & $<$4.13 & $<$4.14 & +0.40$^{+0.02}_{-0.01}$ & +0.42$^{+0.02}_{-0.02}$ & 0.70$^{+0.01}_{-0.01}$ & $>$-4.14 & $>$-4.13 & $>$-4.28 & $>$8.10 & $>$8.10 & $<$1\\[1.5mm]
   MK09    & $>$2.26 & $>$2.26 & $<$4.00 & $<$4.00 & +0.36$^{+0.01}_{-0.00}$ & +0.37$^{+0.00}_{-0.01}$ & 0.45$^{+0.00}_{-0.00}$ & $>$-3.69 & $>$-3.69 & $>$-4.11 & $>$8.45 & $>$8.45 & 791$^{+148}_{-170}$\\[1.5mm]
   MK10    & $>$2.19 & $>$2.15 & $<$4.02 & $<$4.04 & +0.41$^{+0.01}_{-0.02}$ & +0.51$^{+0.02}_{-0.02}$ & 0.83$^{+0.01}_{-0.01}$ & $>$-3.72 & $>$-3.69 & $>$-3.85 & $>$8.54 & $>$8.54 & \ldots\\[1.5mm]
   MK11    & $>$2.38 & $>$2.38 & $<$3.96 & $<$3.96 & +0.48$^{+0.01}_{-0.00}$ & +0.49$^{+0.01}_{-0.01}$ & 0.60$^{+0.00}_{-0.00}$ & $>$-3.41 & $>$-3.41 & $>$-3.82 & $>$8.73 & $>$8.73 & $<$1\\[1.5mm]
   MK12    & $>$1.93 & $>$1.87 & $<$4.12 & $<$4.15 & +0.55$^{+0.01}_{-0.01}$ & +0.68$^{+0.02}_{-0.02}$ & 0.63$^{+0.01}_{-0.01}$ & $>$-3.94 & $>$-3.91 & $>$-4.37 & $>$8.22 & $>$8.22 & $<$1\\
\enddata
  \label{tab:metals_nondet}
  \tablecomments{(1): Abbreviated name for MMT/Keck \OIIIa-non-detected galaxy. (2)--(3): Lower limit on flux ratio of \OIII$\lambda\lambda$4959,5007 to \OIIIa.
    (4)--(5): Upper limit on electron temperature from \OIIIa\ (i.e., ${\rm O}^{++}$). (6)--(7): Flux ratio of \OII$\lambda$3727/\Hb.
    (8): Flux ratio of \OIII$\lambda\lambda$4959,5007/\Hb. (9)--(10): Lower limit on gas-phase abundance of ${\rm O}^+/{\rm H}^+$.
    (11): Lower limit on gas-phase abundance of ${\rm O}^{++}/{\rm H}^+$. (12)--(13): Lower limit on gas-phase oxygen abundance.
    (14): Electron density in units of cm$^{-3}$. Columns (3), (5), (7), (10), and (13) include dust attenuation corrections
  (see Section~\ref{sec:dust}). Uncertainties are reported at 68\% confidence levels.}
  \tablenotetext{1}{Excluded from \OIIIa-non-detected sample with stricter \OIII$\lambda$5007 flux limit for \MZ\ relation analyses.}
  \tablenotetext{2}{\OII\ fluxes were unavailable. We adopt a relation between \OII/\Hb\ and \OIII/\Hb\ (see Figure~\ref{fig:OIIOIII}).}
\end{deluxetable*}
\clearpage
\end{landscape}

%% file: tab15.final.tex
\begin{deluxetable*}{lccccccccc}
  \tabletypesize{\scriptsize}
  \tablewidth{0pc}
  \tablecaption{Luminosities, Stellar Properties, and Star Formation Rates for the \OIIIa-detected Sample}
  \tablehead{
    ID&
    \colhead{\MB}&
    \multicolumn{2}{c}{$\log{\left[\frac{{\rm SFR}(\Hae)}{M_{\sun}/{\rm yr}}\right]}$}&
    \multicolumn{2}{c}{$\log{\left[\frac{{\rm SFR}(\Hbe)}{M_{\sun}/{\rm yr}}\right]}$}&
    \colhead{$\log{(M_{\star}/M_{\sun})}$}&
    \colhead{$Z_{\star}/Z_{\sun}$\TA}&
    \colhead{$\log{(t_{\rm age}/{\rm yr})}$}&
    \colhead{$A_V$}\\
    \cline{3-4}
    \cline{5-6}
    \colhead{}&
    \colhead{}&
    \colhead{Obs.}&
    \colhead{De-red.}&
    \colhead{Obs.}&
    \colhead{De-red.}&
    \colhead{}&
    \colhead{}&
    \colhead{}&
    \colhead{}\\\hline
    \colhead{}&
    \colhead{(mag)}&
    \colhead{(dex)}&
    \colhead{(dex)}&
    \colhead{(dex)}&
    \colhead{(dex)}&
    \colhead{(dex)}&
    \colhead{}&
    \colhead{(dex)}&
    \colhead{(mag)}\\
    \colhead{(1)}&
    \colhead{(2)}&
    \colhead{(3)}&
    \colhead{(4)}&
    \colhead{(5)}&
    \colhead{(6)}&
    \colhead{(7)}&
    \colhead{(8)}&
    \colhead{(9)}&
    \colhead{(10)}}
  \startdata
  MK01 & -18.98 &  +0.04 & +0.10$^{+0.00}_{-0.00}$ & \ldots & \ldots                  & 8.50$^{+0.11}_{-0.10}$ & 0.4 & 7.90$^{+0.42}_{-0.46}$ & 0.7$^{+0.2}_{-0.2}$\\[1mm]
  MK02 & -20.23 &  +0.70 & +0.70$^{+0.00}_{-0.00}$ & \ldots & \ldots                  & 8.53$^{+0.30}_{-0.00}$ & 1.0 & 7.10$^{+0.82}_{-0.07}$ & 0.7$^{+0.2}_{-0.2}$\\[1mm]
  MK03 & -17.61 &  -0.62 & -0.62$^{+0.02}_{-0.01}$ & \ldots & \ldots                  & 7.88$^{+0.07}_{-0.06}$ & 0.2 & 8.70$^{+0.16}_{-0.40}$ & 0.0$^{+0.1}_{-0.0}$\\[1mm]
  MK04 & -18.70 &  +0.12 & +0.18$^{+0.01}_{-0.00}$ & \ldots & \ldots                  & 8.07$^{+0.03}_{-0.00}$ & 0.4 & 7.50$^{+0.24}_{-0.06}$ & 0.4$^{+0.2}_{-0.0}$\\[1mm]
 MMT01 & -18.67 & \ldots & \ldots                  &  +0.09 & +0.53$^{+0.13}_{-0.14}$ & 8.10$^{+0.13}_{-0.00}$ & 0.2 & 7.70$^{+0.63}_{-0.15}$ & 0.0$^{+0.4}_{-0.0}$\\[1mm]
 MMT02 & -19.39 & \ldots & \ldots                  &  -0.35 & +0.21$^{+0.18}_{-0.17}$ & 9.09$^{+0.03}_{-0.11}$ & 0.4 & 8.70$^{+0.25}_{-0.40}$ & 0.8$^{+0.1}_{-0.2}$\\[1mm]
 MMT03 & -20.15 & \ldots & \ldots                  &-0.48\TB&-0.45$^{+0.13}_{-0.18}$\TB& 9.43$^{+0.10}_{-0.11}$ & 0.2 & 8.70$^{+0.34}_{-0.42}$ & 0.9$^{+0.2}_{-0.2}$\\[1mm]
 MMT05 & -16.43 &  -0.58 & -0.58$^{+0.01}_{-0.01}$ & \ldots & \ldots                  & 7.05$^{+0.09}_{-0.14}$ & 0.2 & 7.50$^{+0.49}_{-0.37}$ & 0.1$^{+0.5}_{-0.1}$\\
\enddata
  \label{tab:SFR_Mass}
  \tablecomments{(1): Abbreviated name for MMT/Keck \OIIIa-detected galaxy. (2): Rest-frame absolute $B$-band magnitude. (3)--(4): \Ha\ star formation rate.
    (5)--(6): \Hb\ star formation rate. (7): Stellar mass from SED fitting. (8): Assumed stellar metallicity in SED-fitting model. (9): Light-weighted
    stellar age from SED fitting. (10): Extinction at rest-frame 5500 \AA\ from SED fitting. All values assume a \cite{cha03} IMF. Dust
    attenuation for H$\alpha$ and H$\beta$ SFRs are from Balmer decrements, following the \cite{car89} reddening formalism. The SED
    fitting assumes \cite{cal00} reddening for the stellar continuum. Uncertainties are reported for 68\% confidence levels.}
  \tablenotetext{1}{The stellar metallicity is assumed to be one of three values, 0.2, 0.4, or 1.0 $Z_{\sun}$, based on the dust-corrected
    gas-phase metallicity (see Equations~(\ref{eqn18})--(\ref{eqn20})).}
  \tablenotetext{2}{This source is likely a LINER, so SFR estimates are not trustworthy.}
\end{deluxetable*}
\clearpage

%% file: tab16.final.tex
\begin{deluxetable*}{lccccccccc}
  \tabletypesize{\scriptsize}
  \tablewidth{0pc}
  \tablecaption{Luminosities, Stellar Properties, and SFRs for the \OIIIa-non-detected Sample}
  \tablehead{
    ID&
    \colhead{\MB}&
    \multicolumn{2}{c}{$\log{\left[\frac{{\rm SFR}(\Hae)}{M_{\sun}/{\rm yr}}\right]}$}&
    \multicolumn{2}{c}{$\log{\left[\frac{{\rm SFR}(\Hbe)}{M_{\sun}/{\rm yr}}\right]}$}&
    \colhead{$\log{(M_{\star}/M_{\sun})}$}&
    \colhead{$Z_{\star}/Z_{\sun}$\TA}&
    \colhead{$\log{(t_{\rm age}/{\rm yr})}$}&
    \colhead{$A_V$}\\
    \cline{3-4}
    \cline{5-6}
    & & \colhead{Obs.}&\colhead{De-red.} &\colhead{Obs.}&\colhead{De-red.}&
    & & & \\\hline
    \colhead{(1)}&\colhead{(2)}&\colhead{(3)}&\colhead{(4)}&\colhead{(5)}&
    \colhead{(6)}&\colhead{(7)}&\colhead{(8)}&\colhead{(9)}&\colhead{(10)}}
  \startdata
MK05       & -19.83 & \ldots & \ldots                  &  +0.08 & +0.73$^{+0.14}_{-0.11}$ & 8.86$^{+0.08}_{-0.29}$ & 0.4 & 7.90$^{+0.37}_{-0.90}$ & 0.9$^{+0.3}_{-0.2}$\\[1mm]
MK06       & -19.42 &  +0.29 & +0.50$^{+0.01}_{-0.01}$ & \ldots & \ldots                  & 9.05$^{+0.04}_{-0.10}$ & 1.0 & 8.60$^{+0.22}_{-0.44}$ & 0.7$^{+0.2}_{-0.1}$\\[1mm]
MK07       & -18.91 &  -0.26 & -0.01$^{+0.04}_{-0.04}$ & \ldots & \ldots                  & 8.91$^{+0.02}_{-0.10}$ & 0.4 & 9.10$^{+0.10}_{-0.33}$ & 0.3$^{+0.2}_{-0.1}$\\[1mm]
MK08       & -18.27 &  -0.58 & -0.55$^{+0.02}_{-0.03}$ & \ldots & \ldots                  & 8.60$^{+0.01}_{-0.17}$ & 0.2 & 9.00$^{+0.16}_{-0.53}$ & 0.3$^{+0.3}_{-0.1}$\\[1mm]
MK09       & -20.32 &  +0.13 & +0.13$^{+0.01}_{-0.01}$ & \ldots & \ldots                  & 9.61$^{+0.00}_{-0.12}$ & 1.0 & 8.80$^{+0.09}_{-0.45}$ & 0.8$^{+0.1}_{-0.1}$\\[1mm]
MK10       & -18.22 &  -0.47 & -0.28$^{+0.03}_{-0.03}$ & \ldots & \ldots                  & 8.35$^{+0.10}_{-0.01}$ & 1.0 & 8.40$^{+0.30}_{-0.13}$ & 0.2$^{+0.2}_{-0.0}$\\[1mm]
MK11       & -19.46 &  +0.07 & +0.09$^{+0.01}_{-0.01}$ & \ldots & \ldots                  & 9.13$^{+0.03}_{-0.08}$ & 1.0 & 8.90$^{+0.13}_{-0.20}$ & 0.5$^{+0.1}_{-0.2}$\\[1mm]
MK12       & -18.82 &  -0.42 & -0.16$^{+0.02}_{-0.02}$ & \ldots & \ldots                  & 8.94$^{+0.10}_{-0.01}$ & 0.4 & 8.90$^{+0.30}_{-0.10}$ & 0.7$^{+0.1}_{-0.2}$\\
\enddata
  \label{tab:SFR_Mass_nondet}
  \tablecomments{(1): Abbreviated name for MMT/Keck \OIIIa-non-detected galaxy. (2): Rest-frame absolute $B$-band magnitude. (3)--(4): \Ha\ SFR.
    (5)--(6): \Hb\ SFR. (7): Stellar mass from SED fitting. (8): Assumed stellar metallicity in SED-fitting model. (9): Light-weighted
    stellar age from SED fitting. (10): Extinction at rest-frame 5500 \AA\ from SED fitting. All values assume a \cite{cha03} IMF. Dust
    attenuation for H$\alpha$ and H$\beta$ SFRs are from Balmer decrements, following the \cite{car89} reddening formalism. The SED
    fitting assumes \cite{cal00} reddening for the stellar continuum. Uncertainties are reported for 68\% confidence levels.}
  \tablenotetext{1}{The stellar metallicity is assumed to be one of three values, 0.2, 0.4, or 1.0 $Z_{\sun}$, based on
    the dust-corrected gas-phase metallicity (see Equations~(\ref{eqn18})--(\ref{eqn20})).}
  \tablenotetext{2}{Excluded from \OIIIa-non-detected sample with stricter \OIII$\lambda$5007 flux limit for \MZ\ relation analyses.}
\end{deluxetable*}
\clearpage

%% file: tab17.final.tex
\begin{deluxetable*}{lrrrrrrrrrrrrrrr}
  \tabletypesize{\tiny}
  \tablewidth{0pc}
  \tablecaption{SEDs for the \OIIIa-detected Sample}
  \tablehead{
    ID&
    \colhead{\FUV}&
    \colhead{\NUV}&
    \colhead{$U$}&
    \colhead{$B$}&
    \colhead{$V$}&
    \colhead{IA598}&
    \colhead{\Rc}&
    \colhead{IA679}&
    \colhead{$i$\arcmin}&
    \colhead{$z_b$}&
    \colhead{$z$\arcmin}&
    \colhead{$z_r$}&
    \colhead{$J$}&
    \colhead{$H$}&
    \colhead{$K$}\\
    \colhead{}&
    \colhead{($\mu$Jy)}&
    \colhead{($\mu$Jy)}&
    \colhead{($\mu$Jy)}&
    \colhead{($\mu$Jy)}&
    \colhead{($\mu$Jy)}&
    \colhead{($\mu$Jy)}&
    \colhead{($\mu$Jy)}&
    \colhead{($\mu$Jy)}&
    \colhead{($\mu$Jy)}&
    \colhead{($\mu$Jy)}&
    \colhead{($\mu$Jy)}&
    \colhead{($\mu$Jy)}&
    \colhead{($\mu$Jy)}&
    \colhead{($\mu$Jy)}&
    \colhead{($\mu$Jy)}\\
    \colhead{(1)}&
    \colhead{(2)}&
    \colhead{(3)}&
    \colhead{(4)}&
    \colhead{(5)}&
    \colhead{(6)}&
    \colhead{(7)}&
    \colhead{(8)}&
    \colhead{(9)}&
    \colhead{(10)}&
    \colhead{(11)}&
    \colhead{(12)}&
    \colhead{(13)}&
    \colhead{(14)}&
    \colhead{(15)}&
    \colhead{(16)}}
  \startdata
  MK01 &       1.72 &       2.85 &       3.23 &       3.03 &       4.96 &       4.56 &       7.19 &       7.89 &       6.09 &       7.26 &       7.02 &       5.87 &       6.15 &       7.04 &       7.23 \\
\ldots &  $\pm$0.06 &  $\pm$0.05 &  $\pm$0.04 &  $\pm$0.01 &  $\pm$0.01 &  $\pm$0.03 &  $\pm$0.01 &  $\pm$0.02 &  $\pm$0.01 &  $\pm$0.04 &  $\pm$0.03 &  $\pm$0.10 &  $\pm$0.14 &  $\pm$0.35 &  $\pm$0.22 \\
\ldots &     --0.02 &     --0.12 &     --0.21 &     --0.34 &     --0.84 &     --0.58 &     --2.18 &     --3.66 &     --1.31 &     --2.14 &     --1.85 &     --0.26 &     --0.19 &     --0.19 &     --0.17 \\[1.0mm]
  MK02 &       3.81 &       7.68 &       9.12 &       8.67 &      12.16 &       9.23 &      11.13 &       9.81 &      16.30 &       9.73 &      13.32 &      27.41 &      11.35 &      13.88 &      13.34 \\
\ldots &  $\pm$0.11 &  $\pm$0.15 &  $\pm$0.04 &  $\pm$0.01 &  $\pm$0.01 &  $\pm$0.02 &  $\pm$0.01 &  $\pm$0.03 &  $\pm$0.01 &  $\pm$0.04 &  $\pm$0.03 &  $\pm$0.10 &  $\pm$0.18 &  $\pm$0.35 &  $\pm$0.29 \\
\ldots &     --0.05 &     --0.40 &     --0.60 &     --0.89 &     --2.29 &     --0.99 &     --1.32 &     --0.40 &     --6.76 &     --0.82 &     --2.92 &    --10.01 &     --0.52 &     --0.41 &     --0.40 \\[1.0mm]
  MK03 &       0.55 &       1.35 &       0.62 &       0.57 &       0.95 &       1.13 &       1.42 &       0.90 &       1.57 &       1.00 &       1.26 &       1.19 &       0.74 &       0.92 &       0.64 \\
\ldots &  $\pm$0.05 &  $\pm$0.05 &  $\pm$0.04 &  $\pm$0.01 &  $\pm$0.01 &  $\pm$0.02 &  $\pm$0.01 &  $\pm$0.03 &  $\pm$0.01 &  $\pm$0.04 &  $\pm$0.03 &  $\pm$0.09 &  $\pm$0.35 &  $\pm$0.31 &  $\pm$0.27 \\
\enddata
  \label{tab:SED}
  \tablecomments{(1): Abbreviated name for MMT/Keck \OIIIa-detected galaxy. (2)--(16): Photometric fluxes from shorter
    to longer wavelengths. For each source, the observed fluxes are provided on the first row, the uncertainties in fluxes are
    provided on the second row, and the corrections for nebular emission (continuum and line) are provided in the third row
    (see Section~\ref{sec:SED} for nebular emission determinations).
    All fluxes and uncertainties are provided in units of $\mu$Jy. All uncertainties are reported at 68\% confidence levels.}
\end{deluxetable*}
\clearpage

%% file: Ly2015_MACT1.arxiv.v3.bbl
\begin{thebibliography}{}

\bibitem[Allende Prieto et al.(2001)]{prieto01}
  Allende Prieto, C., Lambert, D.~L., \& Asplund, M.\ 2001, \apjl, 556, L63 

\bibitem[Aller(1984)]{all84}
  Aller, L.~H.\ 1984, Astrophysics and Space Science Library, (Dordrecht: Reidel)

\bibitem[Andrews \& Martini(2013)]{and13}
  Andrews, B.~H., \& Martini, P.\ 2013, \apj, 765, 140

\bibitem[Baldwin et al.(1981)]{bal81}
  Baldwin, A., Phillips, M.~M., \& Terlevich, R.\ 1981, \pasp, 93, 817 

\bibitem[Berg et al.(2015)]{berg15}
  Berg, D.~A., Skillman, E.~D., Croxall, K.~V., et al.\ 2015, \apj, 806, 16

\bibitem[Bertin \& Arnouts(1996)]{ber96}
  Bertin, E., \& Arnouts, S.\ 1996, \aaps, 117, 393

\bibitem[Bressan et al.(1993)]{bre93}
  Bressan, A., Fagotto, F., Bertelli, G., \& Chiosi, C.\ 1993, \aaps, 100, 647 

\bibitem[Brinchmann et al.(2004)]{bri04}
  Brinchmann, J., Charlot, S., White, S.~D.~M., et al.\ 2004, \mnras, 351, 1151 

\bibitem[Bruzual \& Charlot(2003)]{bc03}
  Bruzual, G., \& Charlot, S.\ 2003, \mnras, 344, 1000

\bibitem[Calzetti et al.(2000)]{cal00}
  Calzetti, D., Armus, L., Bohlin, R.~C., Kinney, A.~L., Koornneef, J.,
  \& Storchi-Bergmann, T.\ 2000, \apj, 533, 682

\bibitem[Cardelli et al.(1989)]{car89}
  Cardelli, J.~A., Clayton, G.~C., \& Mathis, J.~S.\ 1989, \apj, 345, 245

\bibitem[Chabrier(2003)]{cha03}
  Chabrier, G.\ 2003, \pasp, 115, 763

\bibitem[Cooper et al.(2012)]{coo12}
  Cooper, M.~C., Newman, J.~A., Davis, M., Finkbeiner, D.~P., 
  \& Gerke, B.~F.\ 2012, Astrophysics Source Code Library, ascl:1203.003 

\bibitem[Cowie et al.(2016)]{cow16}
  Cowie, L.~L., Barger, A.~J., \& Songaila, A.\ 2016, \apj, 817, 57 

\bibitem[Cresci et al.(2012)]{cre12}
  Cresci, G., Mannucci, F., Sommariva, V., et al.\ 2012, \mnras, 421, 262 

\bibitem[Croxall et al.(2015)]{cro15}
  Croxall, K.~V., Pogge, R.~W., Berg, D.~A., Skillman, E.~D.,
  \& Moustakas, J.\ 2015, \apj, 808, 42
    
\bibitem[Dalcanton(2007)]{dal07}
  Dalcanton, J.~J.\ 2007, \apj, 658, 941 

\bibitem[Dav{\'e} et al.(2011)]{dave11}
  Dav{\'e}, R., Finlator, K., \& Oppenheimer, B.~D.\ 2011, \mnras, 416, 1354 

\bibitem[de los Reyes et al.(2015)]{rey15}
  de los Reyes, M.~A., Ly, C., Lee, J.~C., et al.\ 2015, \aj, 149, 79 

\bibitem[Dopita et al.(2016)]{dop16}
  Dopita, M.~A., Kewley, L.~J., Sutherland, R.~S., \& Nicholls, D.~C.\ 2016,
  \apss, 361, 61

\bibitem[Eddington(1913)]{edd13}
  Eddington, A.~S.\ 1913, \mnras, 73, 359 

\bibitem[Faber et al.(2003)]{fab03}
  Faber, S.~M., Phillips, A.~C., Kibrick, R.~I., et al.\ 2003,
  \procspie, 4841, 1657 

\bibitem[Fabricant et al.(2005)]{fab05}
  Fabricant, D., Fata, R., Roll, J., et al.\ 2005, \pasp, 117, 1411 

\bibitem[Fabricant et al.(2008)]{fab08}
  Fabricant, D.~G., Kurtz, M.~J., Geller, M.~J., et al.\ 2008,
  \pasp, 120, 1222

\bibitem[Fagotto et al.(1994a)]{fag94a}
  Fagotto, F., Bressan, A., Bertelli, G., \& Chiosi, C.\ 1994a, \aaps, 105, 29

\bibitem[Fagotto et al.(1994b)]{fag94b}
  Fagotto, F., Bressan, A., Bertelli, G., \& Chiosi, C.\ 1994b, \aaps, 104, 365

\bibitem[Fujita et al.(2003)]{fuj03}
  Fujita, S.~S., Ajiki, M., Shioya, Y., et al.\ 2003, \apjl, 586, L115 

\bibitem[Hayashi et al.(2015)]{hay15}
  Hayashi, M., Ly, C., Shimasaku, K., et al.\ 2015, \pasj, 67, 80

\bibitem[Hayashi et al.(2009)]{hay09}
  Hayashi, M., Motohara, K., Shimasaku, K., et al.\ 2009, \apj, 691, 140 

\bibitem[Hu et al.(2009)]{hu09}
  Hu, E.~M., Cowie, L.~L., Kakazu, Y., \& Barger, A.~J.\ 2009, \apj, 698, 2014 

\bibitem[Humphreys et al.(2011)]{hum11}
  Humphreys, R.~M., Beers, T.~C., Cabanela, J.~E., et al.\ 2011, \aj, 141, 131

\bibitem[Izotov et al.(2006a)]{izo06a}
  Izotov, Y.~I., Papaderos, P., Guseva, N.~G., Fricke, K.~J.,
  \& Thuan, T.~X.\ 2006, \aap, 454, 137 

\bibitem[Izotov et al.(2006b)]{izo06b}
  Izotov, Y.~I., Stasi{\'n}ska, G., Meynet, G., Guseva, N.~G.,
  \& Thuan, T.~X.\ 2006a, \aap, 448, 955

\bibitem[Kashikawa et al.(2006)]{kas06}
  Kashikawa, N., Shimasaku, K., Malkan, M.~A., et al.\ 2006, \apj, 648, 7 

\bibitem[Kashikawa et al.(2011)]{kas11}
  Kashikawa, N., Shimasaku, K., Matsuda, Y., et al.\ 2011, \apj, 734, 119 

\bibitem[Kashikawa et al.(2004)]{kas04}
  Kashikawa, N., Shimasaku, K., Yasuda, N., et al.\ 2004, \pasj, 56, 1011 

\bibitem[Kauffmann et al.(2003)]{kau03}
  Kauffmann, G., Heckman, T.~M., Tremonti, C., et al.\ 2003, \mnras, 346, 1055

\bibitem[Kennicutt(1998)]{ken98}
  Kennicutt, R.~C.\ 1998, \araa, 36, 189

\bibitem[Kochanek et al.(2012)]{koc12}
  Kochanek, C.~S., Eisenstein, D.~J., Cool, R.~J., et al.\ 2012, \apjs, 200, 8 

\bibitem[Kriek et al.(2009)]{kri09}
  Kriek, M., van Dokkum, P.~G., Labb{\'e}, I., Franx, M., Illingworth, G.~D.,
  Marchesini, D., \& Quadri, R.~F.\ 2009, \apj, 700, 221

\bibitem[Kroupa(2001)]{kro01}
  Kroupa, P.\ 2001, \mnras, 322, 231

\bibitem[Lee et al.(2012)]{lee12}
  Lee, J.~C., Ly, C., Spitler, L., et al.\ 2012, \pasp, 124, 782 

\bibitem[Leitherer et al.(1999)]{lei99}
  Leitherer, C., Schaerer, D., Goldader, J.~D.,
  et al.\ 1999, \apjs, 123, 3 

\bibitem[Lilly et al.(2013)]{lil13}
  Lilly, S.~J., Carollo, C.~M., Pipino, A., Renzini, A., \&
  Peng, Y.\ 2013, \apj, 772, 119 

\bibitem[Lilly et al.(2009)]{lil09}
  Lilly, S.~J., Le Brun, V., Maier, C., et al.\ 2009, \apjs, 184, 218 

\bibitem[Ly et al.(2011a)]{newha}
  Ly, C., Lee, J.~C., Dale, D.~A., et al.\ 2011a, \apj, 726, 109 

\bibitem[Ly et al.(2011b)]{ly11b}
  Ly, C., Malkan, M.~A., Hayashi, M., et al.\ 2011b, \apj, 735, 91 

\bibitem[Ly et al.(2007)]{ly07}
  Ly, C., Malkan, M.~A., Kashikawa, N., et al.\ 2007, \apj, 657, 738 (Ly07)

\bibitem[Ly et al.(2012a)]{HaSFR}
  Ly, C., Malkan, M.~A., Kashikawa, N., et al.\ 2012a, \apjl, 747, L16 

\bibitem[Ly et al.(2012b)]{OIIpop}
  Ly, C., Malkan, M.~A., Kashikawa, N., et al.\ 2012b, \apj, 757, 63 

\bibitem[Ly et al.(2014)]{ly14}
  Ly, C., Malkan, M.~A., Nagao, T., et al.\ 2014, \apj, 780, 122 

\bibitem[Ly et al.(2016)]{MACTII}
  Ly, C., Malkan, M.~A., Rigby, J.~R., Nagao, T.\ 2016, \apj, 828, 67

\bibitem[Ly et al.(2009)]{ly09}
  Ly, C., Malkan, M.~A., Treu, T., et al.\ 2009, \apj, 697, 1410 

\bibitem[Ly et al.(2015)]{ly15}
  Ly, C., Rigby, J., Cooper, M., \& Yan, R.\ 2015, \apj, 805, 45

\bibitem[Madau(1995)]{mad95}
  Madau, P.\ 1995, \apj, 441, 18

\bibitem[Markwardt(2009)]{mar09}
  Markwardt, C.~B.\ 2009, ADASS XVIII, 411, 251 

\bibitem[Martin et al.(2005)]{mar05}
  Martin, D.~C., Fanson, J., Schiminovich, D., et al.\ 2005, \apjl, 619, L1 

\bibitem[Miyazaki et al.(2002)]{miy02}
  Miyazaki, S., Komiyama, Y., Sekiguchi, M., et al.\ 2002, \pasj, 54, 833 

\bibitem[Moustakas et al.(2011)]{mou11}
  Moustakas, J., Zaritsky, D., Brown, M., et al.\ 2011, \apj,
  submitted (arXiv:1112.3300)

\bibitem[Nagao et al.(2008)]{nag08}
  Nagao, T., Sasaki, S.~S., Maiolino, R., et al.\ 2008, \apj, 680, 100 

\bibitem[Newman et al.(2013)]{new13}
  Newman, J.~A., Cooper, M.~C., Davis, M., et al.\ 2013, \apjs, 208, 5 

\bibitem[Nicholls et al.(2014)]{nic14}
  Nicholls, D.~C., Dopita, M.~A., Sutherland, R.~S., Jerjen, H., 
  \& Kewley, L.~J.\ 2014, \apj, 790, 75

\bibitem[Nicholls et al.(2013)]{nic13}
  Nicholls, D.~C., Dopita, M.~A., Sutherland, R.~S., Kewley, L.~J., 
  \& Palay, E.\ 2013, \apjs, 207, 21

\bibitem[Oke(1974)]{oke74}
  Oke, J.~B.\ 1974, \apjs, 27, 21

\bibitem[Osterbrock \& Ferland(2006)]{ost06}
  Osterbrock, D.~E., \& Ferland, G.~J.\ 2006, Astrophysics of Gaseous
  Nebulae and Active Galactic Nuclei, 2nd.~ed.~by D.E.~Osterbrock and
  G.J.~Ferland.~Sausalito, CA: University Science Books, 2006 

\bibitem[Ouchi et al.(2004)]{ouc04}
  Ouchi, M., Shimasaku, K., Okamura, S., et al.\ 2004, \apj, 611, 660 

\bibitem[Palay et al.(2012)]{pal12}
  Palay, E., Nahar, S.~N., Pradhan, A.~K., \& Eissner, W.\ 2012, \mnras, 423, L35

\bibitem[Probst et al.(2008)]{pro08}
  Probst, R.~G., George, J.~R., Daly, P.~N., Don, K.,
  \& Ellis, M.\ 2008, \procspie, 7014, 70142S 

\bibitem[Salim et al.(2007)]{sal07}
  Salim, S., Rich, R.~M., Charlot, S., et al.\ 2007, \apjs, 173, 267 

\bibitem[Sanders et al.(2015)]{san15}
  Sanders, R.~L., Shapley, A.~E., Kriek, M., et al.\ 2015, \apj, 799, 138

\bibitem[Steidel et al.(2014)]{ste14}
  Steidel, C.~C., Rudie, G.~C., Strom, A.~L., et al.\ 2014, \apj, 795, 165
  
\bibitem[Stetson(1987)]{ste87}
  Stetson, P.~B.\ 1987, \pasp, 99, 191

\bibitem[Tremonti et al.(2004)]{tre04}
  Tremonti, C.~A., Heckman, T.~M., Kauffmann, G., et al.\ 2004, \apj, 613, 898 

\bibitem[Veilleux \& Osterbrock(1987)]{vei87}
  Veilleux, S., \& Osterbrock, D.~E.\ 1987, \apjs, 63, 295 

\bibitem[Whitaker et al.(2014a)]{whi14}
  Whitaker, K.~E., Franx, M., Leja, J., et al.\ 2014a, \apj, 795, 104

\bibitem[Yagi et al.(2002)]{yagi02}
  Yagi, M., Kashikawa, N., Sekiguchi, M., Doi, M., Yasuda, N.,
  Shimasaku, K., \& Okamura, S.\ 2002, \aj, 123, 66 

\bibitem[York et al.(2000)]{york00}
  York, D.~G., Adelman, J., Anderson, J.~E., Jr., et al.\ 2000,
  \aj, 120, 1579

\bibitem[Zahid et al.(2011)]{zah11}
  Zahid, H.~J., Kewley, L.~J., \& Bresolin, F.\ 2011, \apj, 730, 137 

\end{thebibliography}
